\documentclass[prd,11pt,nofootinbib,preprint]{revtex4}
\usepackage[T1]{fontenc}
\usepackage{amsmath,amssymb}
\usepackage{url}
\usepackage{subfigure}
\usepackage{graphicx}
\usepackage{grffile}
\usepackage[usenames,dvipsnames]{color}
\usepackage{mathtools}
\usepackage{slashed}
\usepackage{array}
\usepackage{multirow}
\usepackage{xcolor}
\usepackage[colorlinks,citecolor=blue]{hyperref}
\usepackage{color}
\usepackage{cancel}
\usepackage{gensymb}
\usepackage{soul}

\def \met{\not \! E_T }

\def\a {\alpha}
\def\b {\beta}

\def\d {\delta}

\def\l {\lambda}

\def\bar {\overline}

\def\be {\begin{equation}}
\def\ee {\end{equation}}
\def\beq {\begin{equation}}
\def\eeq {\end{equation}}
\def\bea {\begin{eqnarray}}
\def\eea {\end{eqnarray}}

\newcommand{\besub}{\begin{subequations}}
\newcommand{\eesub}{\end{subequations}}

\def\beq{\begin{equation}}
\def\eeq{\end{equation}}
\def\barr{\begin{array}}
\def\earr{\end{array}}

\def\a{\alpha}

\def\b{\beta}

\def\d{\delta}

\def\l{\lambda}

\def\q2 {q^2}

\def\bt{\begin{table}}
\def\et{\end{table}}

\begin{document}

\title{Radiative $H^+_{1,2}W^- Z$ vertices in flavour conserving three Higgs-doublet scenarios}

\author{Nabarun Chakrabarty}
\email{nabarun.chakrabarty@visva-bharati.ac.in}
\affiliation{Department of Physics, Siksha Bhavana, Visva-Bharati,
Santiniketan, West Bengal 731235, India}
\author{Indrani Chakraborty}
\email{indrani.chakraborty@jiit.ac.in}
\affiliation{Department of Physics and Material Science and Engineering,
Jaypee Institute of Information Technology, A-10,
Sector-62, Noida 201307, Uttar Pradesh, India}

\vspace{5mm}

\begin{abstract} 
A detailed calculation of the radiatively induced $H_{1,2}^+ W^- Z$ vertices is carried out in the context of flavour conserving three Higgs doublet models (3HDMs). The Type-II, lepton specific and democratic versions of the 3HDM are chosen as representative cases and the \emph{alignment limit} is adopted. We arrange the amplitudes in UV-finite and gauge-invariant subsets for a sharper understanding of the underlying one-loop structure. Factoring-in various theoretical constraints and the ones from the $h \to \gamma \gamma$ signal strength and the $B \to X_s \gamma$ branching ratio, we compute the relevant form factors and compare them among the various 3HDM types taken. The results also indicate a sizeable increment ($\sim 100\%$) over the corresponding form factors in 2HDMs. In addition, we probe the $H_{1,2}^+ W^- Z$ vertices at the 14 TeV LHC using vector boson fusion (VBF). It is seen that production of $H_1^+$ via VBF and the subsequent $H_1^+ \to H_2^+ H_2/A_2$ decays can have $\sigma \times \text{BR}$ in the $\mathcal{O}$(0.1 fb) ball park. Therefore, such a cascade can shed light on the strength of the $H_1^+ W^- Z$ interaction and confirm the presence of two charged scalars thereby acting as a smoking gun signal of a 3HDM.    
\end{abstract} 
\maketitle

\section{Introduction}
\label{intro}

The discovery of the Higgs boson with a mass around 125 GeV at the Large Hadron Collider (LHC) \cite{CMS:2012qbp,ATLAS:2012yve}  has completed the particle content of Standard Model (SM). While the SM minimally a\text{ch}ounts for electroweak symmetry breaking (EWSB) and fermion mass generation, it leaves many fundamental questions unanswered on both theoretical and experimental fronts. This renders the SM far from being a complete theory thereby motivating extensions Beyond the SM (BSM).
However, the LHC has been consistently reporting that the interaction
strengths of the scalar with the SM fermions and gauge bosons are in good agreement with the corresponding SM predictions. Such results, along with a host of null outcomes pertaining to direct searches of additional degrees freedom, have collectively pushed a plethora of BSM scenarios at bay. 

That said, extended scalar sectors-particularly those with additional Higgs doublets-provide simplistic framework to address several of these shortcomings. Multi-Higgs doublet frameworks can be motivated from several perpectives. First, there is no restriction on the number of scalar doublets from fundamental principles, especially when the $\rho$-parameter is maintained at unity at the tree level in the presence of $SU(2)_L$ scalar doublets.  Secondly, such frameworks shut off flavour changing neutral currents (FCNC), predict additional CP-violation through the scalar potential that can eventually explain the observed matter-antimatter imbalance. Thirdly, the same setups can accommodate an \emph{alignment limit} wherein the interactions of one of the CP-even scalars, identified with the discovered boson, become identical to the corresponding SM ones. Though many interactions involving the non-standard scalars and gauge bosons vanish at an exact alignment, such scalars can still be probed at particle colliders primarily through their fermionic interactions, modulo the restrictions from direct search.

The most minimal and also the most widely studied multi-Higgs doublet setup is the two-Higgs doublet model (2HDM) \cite{Lee1973,Paschos:1976ay,GlashowWeinberg1977,Deshpande:1977rw,Branco2012,GunionHaber1986} and it can emerge in diverse theoretical contexts such as supersymmetry, axion models, flavor symmetries, and grand unification. FCNCs in this case can be annulled by imposing a single $\mathbb{Z}_2$ symmetry. A 2HDM is in fact
the smallest non-trivial $SU(2)_L$ multiplet to predict a singly charged
Higgs $H^+$. An $H^+$ can be searched at the LHC through it fermionic interactions. While an $H^+ \to t \overline{b}$ mode is useful to look
for an heavy $H^+$, an $H^+ \to \tau \nu_\tau$ channel becomes for suited to an apppriately light one. However, heavy QCD backgrounds can limit the observability of such modes thereby advocating the need to look at other channels, the bosonic decays $H^+ \to W^+ h,~W^+ Z,~W^-\gamma$, for instance. That said, the last two of the
aforementioned modes are prima facie more intriguing since
they arise only at one-loop in multi-Higgs doublet models~\cite{Grifols:1980uq}.
The absence of the $H^+ W^- Z(\gamma) $ coupling at the tree level is an
artefact of the isospin symmetry of the kinetic terms of the
Higgs sector. Since both these characteristics are, in general,
broken raditively through quantum effects from other sectors
that do not respect the custodial invariance, these vertices
are potentially induced at the one-loop level. It is therefore established that the strength of the $H^+ W^- Z(\gamma)$ interaction quantifies the
extent of custodial symmetry breaking in the embedding model. Moreover, the fact that a radiatively induced $H^+ W^- Z(\gamma)$ vertex carries a momentum-dependence distinguishes a multi-Higgs doublet setup from one featuring scalar triplets (since the tree-level $H^+ W^- Z$ interaction in case of the latter does not carry any momentum-dependence). The $H^+ W^- Z(\gamma)$ interaction at one-loop has been studied in various versions of the 2HDM in \cite{Kanemura2000HWZ,Kanemura2000Custodial,Aoki2011HWZ, Kanemura2015Review,Abbas:2018pfp} and in a scenario comprising two scalar doublets and a scalar iso-doublet color octet in \cite{Chakrabarty:2020msl}.

The next minimal multi-Higgs doublet framework is the three-Higgs doublet model (3HDM). Various aspects of the 3HDMs have also been studied in the recent past, a partial list of which is \cite{Ferreira2008,Machado2011,Aranda2012,Ivanov2012,Ivanov2013,
GonzalezFelipe2013a,GonzalezFelipe2013b,Keus2014,Aranda2013,
Ivanov2015,Das2014,Maniatis:2015kma,DasPhD2015,Maniatis2015O2,
Moretti2015,Chakrabarty2016,Merchand2017,EmmanuelCosta2016,
Yagyu2016,Bento2017,EmmanuelCosta2017,CamargoMolina2018,
Pramanick2018,Varzielas2019,Logan:2020mdz,Akeroyd:2021fpf,Kuncinas:2025mcn,Dey:2024epo,Keus:2025ova,Kuncinas:2024zjq,Batra:2025amk}. In this work, we study in detail the $H_1^+ W^- Z$ and $H_2^+ W^- Z$ vertices at one-loop in the context of flavour-conserving 3HDMs of Type-II, Type-X and Type-Z \footnote{The rationale behind picking the aforesaid variants is relegated to the following sections}. Discrete symmetry-breaking quadratic terms are allowed for phenomenological reasons and an exact alignment is demanded. It must be mentioned here that while \cite{Moretti:2015tva} investigated the $H^+ W^- Z(\gamma)$ interaction for the three Higgs doublet scenario, of the doublets was \emph{inert} and therefore the aforementioned interaction was forbidden for the inert charged scalar. The main features of the present study are outlined below.

\begin{itemize}

\item  We adopt a non-linear gauge where the unphysical $G^+ W^- Z$ vertex disappears. This makes a check of gauge invariance easier using Ward identity. Further, we divide the one-loop amplitudes into UV-finite and gauge invariant subsets.

\item Given the existing literature on 3HDM, we refrain from performing an exhaustive parameter space scan. We therefore adopt a simplified approach by adhering to the alignment limit and enforcing additional relations involving model parameters to satisfy electroweak precision constraints.   

\item We look for possible (anti)correlations between the form factors corresponding to $H_1^+$ and $H_2^+$. We also look to discern a possible enhancement of the form factors over 2HDM.

\item The $H^+_{1,2} W^- Z$ interaction is probed at the 14 TeV LHC through vector boson fusion (VBF) in this study. Further the decays 
$H_1^+ \to H^+_2 H_2,~H^+ A_2$ are investigated that can confirm the presence of two charged scalars. Therefore such a cascade can be used a smoking gun signal of a 3HDM.

\end{itemize}

This study is organised as follows. We review the flavour-conserving 3HDM types under a $\mathbb{Z}_2 \times \mathbb{Z}_2^\prime \times \mathbb{Z}_2^{\prime\prime}$ symmetry in section \ref{model}. In section \ref{one-loop}, the one-loop form factors are detailed and their UV-finiteness and gauge invariance are demonstrated. The relevant constraints are discussed in section \ref{constraints}. Numerical estimation of the form factors is carried out in section \ref{results} and collider signatures are elucidated in section \ref{collider}. We conclude in section \ref{summary}.

\section{Flavour conserving three Higgs doublet models}
\label{model}
\subsection{Scalar sector}
We introduce three scalar doublets $\phi_1, \phi_2$ and $\phi_3$ that are parametrized as :
\bea
\Phi_j = \begin{pmatrix}
\omega_j^+ \\
\frac{1}{\sqrt{2}} (v_j + h_j + i z_j ) 
\end{pmatrix},~~~ j = 1,2,3.
\eea
Here, $v_1,v_2$ and $v_3$ are the three vacuum expectation values (VEVs) satisfying $v_1^2 + v_2^2 + v_3^2 = v^2$ with $v$ = 246 GeV, and, $\omega_j^+,h_j$ and $z_j$ respectively denote the charged, CP-even and CP-odd components. One additionally defines $\tan \b$ = $\frac{v_2}{v_1}$ and $\tan \gamma$ = $\frac{\sqrt{v_1^2 + v_2^2}}{v_3}$. Thus, $v_1 = v \cos \beta \sin \gamma,~v_2 = v \sin \beta \sin \gamma,~v_3 = v \cos \gamma$.
Further, aiming to suppress FCNCs, the framework is endowed with a $\mathbb{Z}_2 \times \mathbb{Z}^{'}_2 \times \mathbb{Z}^{''}_2$ symmetry under which $\Phi_1,\Phi_2$ and $\Phi_3$ respectively are assigned the charges $(+,+,-)$, $(+,-,+)$ and $(-,+,+)$. The most general scalar potential consistent with the SM and discrete symmetries is therefore
\bea
V &=& m_{11}^2 \Phi_1^\dag \Phi_1 + m_{22}^2 \Phi_2^\dag \Phi_2 + m_{33}^2 \Phi_3^\dag \Phi_3 \nonumber \\
&& - [m_{12}^2 \Phi_1^\dag \Phi_2 + m_{13}^2 \Phi_1^\dag \Phi_3 + m_{23}^2 \Phi_2^\dag \Phi_3 + {\rm h.c.}  ] \nonumber \\
&& + \frac{1}{2} \lambda_1 (\Phi_1^\dag \Phi_1)^2 + \frac{1}{2} \lambda_2 (\Phi_2^\dag \Phi_2)^2 + \frac{1}{2} \lambda_3 (\Phi_3^\dag \Phi_3)^2 \nonumber \\
&& + \lambda_{12} (\Phi_1^\dag \Phi_1)(\Phi_2^\dag \Phi_2) + \lambda_{13} (\Phi_1^\dag \Phi_1)(\Phi_3^\dag \Phi_3) + \lambda_{23} (\Phi_2^\dag \Phi_2)(\Phi_3^\dag \Phi_3) \nonumber \\
&& + \lambda^{\prime}_{12} (\Phi_1^\dag \Phi_2)(\Phi_2^\dag \Phi_1) + \lambda^{\prime}_{13} (\Phi_1^\dag \Phi_3)(\Phi_3^\dag \Phi_1) + \lambda^{\prime}_{23} (\Phi_2^\dag \Phi_3)(\Phi_3^\dag \Phi_2) \nonumber \\
&& + \frac{1}{2} [\lambda^{''}_{12} (\Phi_1^\dag \Phi_2)^2 + \lambda^{\prime\prime}_{13} (\Phi_1^\dag \Phi_3)^2 + \lambda^{\prime\prime}_{23} (\Phi_2^\dag \Phi_3)^2 + {\rm h.c.}].
\label{scalar-pot}
\eea
All parameters above are chosen real and dimension-2 terms are included in Eq.\ref{scalar-pot} that break the discrete symmetry \emph{softly}. The physical mass spectrum is spanned by the charged scalars $H_1^+$ and $H_2^+$, the CP-odd neutral scalars $A_1$ and $A_2$, and, the CP-even neutral scalars $h,H_1$ and $H_2$. Of these, $h$ is chosen to be the observed Higgs. 

The tadpole conditions $\Big(\frac{\partial V}{\partial h_i}\Big)_{(v_1,v_2,v_3)} = 0,$ for $i$ = 1,2,3, lead to
\besub
\bea
m^2_{11} &=& \frac{1}{v_1}\Big(m_{12}^2 v_2 + m_{13}^2 v_3 - \frac{\l_1}{2}v_1^3 - \frac{\l_{12}}{2}v_1 v_2^2 - \frac{\l^\prime_{12}}{2}v_1 v_2^2 - \frac{\l^{\prime\prime}_{12}}{2}v_1 v_2^2 \nonumber \\
&&
 - \frac{\l_{13}}{2}v_1 v_3^2 - \frac{\l^\prime_{13}}{2}v_1 v_3^2 - \frac{\l^{\prime\prime}_{13}}{2}v_1 v_3^2 \Big), \\
m^2_{22} &=& \frac{1}{v_2}\Big(m_{12}^2 v_1 + m_{23}^2 v_3 - \frac{\l_2}{2}v_2^3 - \frac{\l_{12}}{2}v_2 v_1^2 - \frac{\l^\prime_{12}}{2}v_2 v_1^2 - \frac{\l^{\prime\prime}_{12}}{2}v_2 v_1^2 \nonumber \\
&&
 - \frac{\l_{23}}{2}v_2 v_3^2 - \frac{\l^\prime_{23}}{2}v_2 v_3^2 - \frac{\l^{\prime\prime}_{23}}{2}v_2 v_3^2 \Big), \\
m^2_{33} &=& \frac{1}{v_3}\Big(m_{13}^2 v_1 + m_{23}^2 v_2 - \frac{\l_3}{2}v_3^3 - \frac{\l_{13}}{2}v_3 v_1^2 - \frac{\l^\prime_{13}}{2}v_3 v_1^2 - \frac{\l^{\prime\prime}_{13}}{2}v_3 v_1^2 \nonumber \\
&&
 - \frac{\l_{23}}{2}v_3 v_2^2 - \frac{\l^\prime_{23}}{2}v_3 v_2^2 - \frac{\l^{\prime\prime}_{23}}{2}v_3 v_2^2 \Big).  
\eea
\eesub
The mass terms can subsequently be expressed as
\bea
V \supset \frac{1}{2}\begin{pmatrix}
h_1 & h_2 & h_3\end{pmatrix}
\mathcal{M}^2_{\text{even}}\begin{pmatrix}
h_1 \\
h_2 \\
h_3
\end{pmatrix} + \frac{1}{2}\begin{pmatrix}
z_1 & z_2 & z_3\end{pmatrix}
\mathcal{M}^2_{\text{odd}}\begin{pmatrix}
z_1 \\
z_2 \\
z_3
\end{pmatrix} + \begin{pmatrix}
\omega_1^- & \omega_2^- & \omega_3^-\end{pmatrix}
\mathcal{M}^2_{\text{ch}}\begin{pmatrix}
\omega_1^+ \\
\omega_2^+ \\
\omega_3^+
\end{pmatrix}
\eea
The mass matrices in the gauge basis, i.e., $\mathcal{M}^2_{\text{even}}$, $\mathcal{M}^2_{\text{even}}$ and $\mathcal{M}^2_{\text{ch}}$ are given in the Appendix. These matrices can be diagonalised using appropriate unitary tranformations connecting the gauge and physical bases. Therefore, one writes in the lines of \cite{Akeroyd:2021fpf,Chakraborti:2021bpy}
\besub
\bea
\begin{pmatrix}
h \\
H_1 \\
H_2
\end{pmatrix}
&=& \begin{pmatrix}
1 & 0 & 0 \\
0 & c_{\a_3} & s_{\a_3} \\
0 & -s_{\a_3} & c_{\a_3}
\end{pmatrix}
\begin{pmatrix}
c_{\a_2} & 0 & s_{\a_2} \\
0 & 1 & 0 \\
-s_{\a_2} & 0 & c_{\a_2}
\end{pmatrix}
\begin{pmatrix}
c_{\a_1} & s_{\a_1} & 0 \\
-s_{\a_1} & c_{\a_1} & 0 \\
0 & 0 & 1
\end{pmatrix}
\begin{pmatrix}
h_1 \\
h_2 \\
h_3
\end{pmatrix}, \\
\begin{pmatrix}
G^0 \\
A_1 \\
A_2
\end{pmatrix}
&=& \begin{pmatrix}
1 & 0 & 0 \\
0 & c_{\d_1} & -s_{\d_1} \\
0 & s_{\d_1} & c_{\d_1}
\end{pmatrix}
\begin{pmatrix}
s_{\gamma} & 0 & c_{\gamma} \\
0 & 1 & 0 \\
-c_{\gamma} & 0 & s_{\gamma}
\end{pmatrix}
\begin{pmatrix}
c_\b & s_\b & 0 \\
-s_\b & c_\b & 0 \\
0 & 0 & 1
\end{pmatrix}
\begin{pmatrix}
z_1 \\
z_2 \\
z_3
\end{pmatrix}, \\
\begin{pmatrix}
G^+ \\
H_1^+ \\
H_2^+
\end{pmatrix}
&=&  \begin{pmatrix}
1 & 0 & 0 \\
0 & c_{\d_2} & -s_{\d_2} \\
0 & s_{\d_2} & c_{\d_2}
\end{pmatrix}
\begin{pmatrix}
s_{\gamma} & 0 & c_{\gamma} \\
0 & 1 & 0 \\
-c_{\gamma} & 0 & s_{\gamma}
\end{pmatrix}
\begin{pmatrix}
c_\b & s_\b & 0 \\
-s_\b & c_\b & 0 \\
0 & 0 & 1
\end{pmatrix}
\begin{pmatrix}
\omega_1^+ \\
\omega_2^+ \\
\omega_3^+
\end{pmatrix},
\label{mass-to-flav}
\eea
\eesub
Here, $\alpha_{1,2,3}$ and $\delta_{1,2}$ are mixing angles that are taken real. To this end, we take stock of the independent parameters in the model. Having eliminated $m^2_{11},m^2_{22}$ and $m^2_{33}$ through the tadpole conditions, there are the three dimensionful parameters $m^2_{12},m^2_{13}$ and $m^2_{23}$, the mixing angles $\beta,\gamma$ and twelve quartic couplings leading to a total of seventeen independent parameters in the gauge basis. However, the  quartic couplings have been solved in terms of $\{\text{tan}\b,\text{tan}\gamma,\delta_1,\delta_2,\a_1,\a_2,\a_3,m_{12},m_{13},m_{23},
M_h,M_{H_1},M_{H_2},
M_{H_1^+},M_{H_2^+},M_{A_1},M_{A_2}\}$ prompting us take the latter set as the independent parameters.

\subsection{Yukawa sector}
\label{Yukawa-pot}
The possible Yukawa coupling schemes leading to flavour conservation in a 3HDM are listed in table~\ref{table_yuk}. Depending on the assignment of the charges, there can be five such types, namely (i) Type-I, (ii) Type-II, (iii) Type-X (lepton-specific), (iv) Type-Y (Flipped) and (v) Type-Z (democratic). It is seen that in Type-I 3HDM, up-type, down-type quarks and leptons only couple to scalar doublet $\phi_2$ to acquire masses, whereas in Type-Z or democratic 3HDM, down-type, up-type quarks and leptons receive their masses through individual couplings with $\Phi_1, \Phi_2$ and $\Phi_3$ respectively.
\begin{table}[htpb!]
\centering
\begin{tabular}{ |c|c|c|c|c|c| } 
\hline
Scalar doublet & Type-I & Type-II & Lepton-specific & Flipped & Democratic \\ 
\hline \hline 
$\Phi_1$ & $-$ & $d,l$ & $l$ & $d$ & $d$ \\
$\Phi_2$ & $u,d,l$ & $u$ & $u,d$ & $u,l$ & $u$ \\
$\Phi_3$ & $-$ & $-$ & $-$ & $-$ & $l$ \\ \hline
\end{tabular}
\label{table_yuk}
\end{table}
For instance, an assignment of the $\mathbb{Z}_2 \times \mathbb{Z}_2^\prime \times \mathbb{Z}_2^{\prime\prime}$ charges responsible for a Type-Z 3HDM is shown in table \ref{quantum-no}.
\begin{table}[htpb]
\centering
\begin{tabular}{ |c|c|c|c|c|c|c|c| } 
\hline
Symmetry & $\Phi_1$ & $\Phi_2$ & $\Phi_3$ & $Q_{L_i}, L_{L_i}$ & $u_R$ & $d_R$ & $l_R$ \\ 
\hline \hline 
$\mathbb{Z}_2$ & + & + & - & + & + & - & + \\
$\mathbb{Z}^{'}_2$ & + & - & + & + & - & + & + \\
$\mathbb{Z}^{''}_2$ & - & + & + & + & + & + & -\\ \hline
\end{tabular}
\caption{$\mathbb{Z}_2 \times \mathbb{Z}^{'}_2 \times \mathbb{Z}^{''}_2$ quantum numbers assigned to the doublets and fermions.}
\label{quantum-no}
\end{table}

Next, we parameterise the interactions of the neutral scalar mass eigenstates with the fermions as follows.
\bea
\mathcal{L}_Y = \sum_{f=u,d,l} \bigg[ -\frac{m_f}{v}  \Big\{ \sum_{\phi_S = h,H_1,H_2} y_f^{\phi_S}\bar{f} f \phi_S + \sum_{\phi_P = A_1,A_2} y_f^{\phi_P}\bar{f} i\gamma_5 f \phi_P \Big\} \bigg] 
\eea
Here, $m_f$ refers to the mass of the fermion $f$, and, $ y_f^{\phi_S}$ and $ y_f^{\phi_P}$ are the Yukawa scale factors corresponding to the CP-even and CP-odd scalars respectively. These scale factors are expressible in terms of the various mixing angles. We list the scale factors corresponding to $h$ for all the 3HDM types in table \ref{scale-fact} for illustration.
\begin{table}[htpb!]
\centering
\begin{tabular}{ |c|c|c|c|c|c| } 
\hline
 & Type-I & Type-II & Type-X & Type-Y & Type-Z \\ 
\hline \hline 
$y_u^h$ & $\frac{s_{\a_1}c_{\a_2}}{s_\b s_\gamma}$  & $\frac{s_{\a_1}c_{\a_2}}{s_\b s_\gamma}$ & $\frac{s_{\a_1} c_{\a_2}}{s_\b s_\gamma}$ & $\frac{s_{\a_1} c_{\a_2}}{s_\b s_\gamma}$ & $\frac{s_{\a_1} c_{\a_2}}{s_\b s_\gamma}$ \\
$y_d^h$ & $\frac{s_{\a_1}c_{\a_2}}{s_\b s_\gamma}$ & 
$\frac{c_{\a_1}c_{\a_2}}{c_\b s_\gamma}$ & $\frac{s_{\a_1} c_{\a_2}}{s_\b s_\gamma}$ & $\frac{c_{\a_1} c_{\a_2}}{c_\b s_\gamma}$ & $\frac{c_{\a_1} c_{\a_2}}{c_\b s_\gamma}$ \\
$y_l^h$ & $\frac{s_{\a_1}c_{\a_2}}{s_\b s_\gamma}$ & 
$\frac{c_{\a_1}c_{\a_2}}{c_\b s_\gamma}$ & $\frac{c_{\a_1} c_{\a_2}}{c_\b s_\gamma}$  & $\frac{s_{\a_1} c_{\a_2}}{s_\b s_\gamma}$ & $\frac{s_{\a_2}}{c_\gamma}$ \\ \hline
\end{tabular}
\caption{Scale factors corresponding to the various Yukawa interactions involving $h$.}
\label{scale-fact}
\end{table}

The interactions involving the charged eigenstates $H_1^+,~H_2^+$ and  the quarks are quoted as
\bea
\mathcal{L}_{\rm charged} =  \sum_{i=1}^2 \left[\frac{\sqrt{2} V_{ud}}{v} \bar{u}(m_u A_u^i P_L - m_d A_d^i P_R ) d H_i^+  + {\rm h.c.}\right]\label{Lcharged},
\eea
where $V_{ud}$ and $P_L(P_R)$ respectively denote the CKM matrix element
and the left (right) projection operator respectively. The expressions for $A_u^i$ and $A_d^i$ for the five 3HDM types taken are given in table \ref{Au_Ad}.
\begin{table}[htpb!]
\centering
\resizebox{17cm}{!}{
\begin{tabular}{ |c|c|c|c|c|c| } 
\hline
 & Type-I & Type-II & Type-X & Type-Y & Type-Z \\ 
\hline \hline 
$A_u^1$ & $\frac{c_\b c_{\d_2} + c_\gamma s_\b s_{\d_2}}{s_\b s_\gamma}$ & $\frac{c_\b c_{\d_2}+ c_\gamma s_\b s_{\d_2}}{s_\b s_\gamma}$ & $\frac{c_\b c_{\d_2}+ c_\gamma s_\b s_{\d_2}}{s_\b s_\gamma}$ & $\frac{c_\b c_{\d_2}+ c_\gamma s_\b s_{\d_2}}{s_\b s_\gamma}$  & $\frac{c_\b c_{\d_2}+ c_\gamma s_\b s_{\d_2}}{s_\b s_\gamma}$  \\ \hline \hline
$A_d^1$ & $\frac{c_\b c_{\d_2} + c_\gamma s_\b s_{\d_2}}{s_\b s_\gamma}$ &   $\frac{-c_{\d_2} s_\b + c_\b c_\gamma s_{\d_2}}{s_\gamma c_\b}$ & $\frac{c_\b c_{\d_2}+ c_\gamma s_\b s_{\d_2}}{s_\b s_\gamma}$ & $\frac{-c_{\d_2} s_\b + c_\b c_\gamma s_{\d_2}}{s_\gamma c_\b}$& $\frac{-c_{\d_2} s_\b + c_\b c_\gamma s_{\d_2}}{s_\gamma c_\b}$ \\ \hline \hline
$A_u^2$ & $\frac{-c_{\d_2} c_\gamma s_\b + c_\b s_{\d_2}}{s_\b s_\gamma}$  & $\frac{- c_{\d_2} c_\gamma s_\b+ c_\b s_{\d_2}}{s_\b s_\gamma}$ & $\frac{-c_{\d_2} c_\gamma s_\b + c_\b s_{\d_2}}{s_\b s_\gamma}$ & $\frac{-c_{\d_2} c_\gamma s_\b + c_\b s_{\d_2}}{s_\b s_\gamma}$ & $\frac{-c_{\d_2}c_\gamma s_\b + c_\b s_{\d_2}}{s_\b s_\gamma}$ \\ \hline \hline
$A_d^2$  & $\frac{-c_{\d_2} c_\gamma s_\b + c_\b s_{\d_2}}{s_\b s_\gamma}$  & $\frac{-c_\b c_{\d_2}c_\gamma - s_\b s_{\d_2}}{s_\gamma c_\b}$ & $\frac{-c_{\d_2} c_\gamma s_\b + c_\b s_{\d_2}}{s_\b s_\gamma}$ & $\frac{-c_\b c_{\d_2} c_\gamma - s_\b s_{\d_2}}{s_\gamma c_\b}$ & $\frac{-c_\b c_{\d_2}c_\gamma-s_\b s_{\d_2}}{s_\gamma c_\b}$ \\ \hline \hline
\end{tabular}}
\caption{Expressions of $A_u^i$ and $A_d^i$ for five types of 3HDMs.}
\label{Au_Ad}
\end{table}
We implement an exact alignment limit $\alpha_1 = \beta,~\alpha_2 = \frac{\pi}{2} - \gamma$ in the ensuing sections in which case couplings of $h$ to all SM particles become equal to the corresponding SM values.

\section{The $H_{1,2}^+ ~W^- Z$ vertex at one-loop}
\label{one-loop}

The amplitude for the decay $H_j^+ \to W^+ V$ ($j=1,2$) reads
\bea
\mathcal{M}_j = g M_W A_{jV}^{\mu\nu}\epsilon^*_{W\mu}\epsilon^*_{V\nu},
\eea
where 
\bea
A^{\mu \nu}_{jV} = g^{\mu \nu} F_{jV} + \frac{p^{\mu}_V p^{\nu}_W}{M_W^2} G_{jV}  + i \epsilon^{\mu \nu \rho \sigma} \frac{p_{V_{\rho}}~ p_{W_{\sigma}}}{M_W^2} H_{jV}. \label{vertex}
\eea
It is mentioned here that $F_{jV},G_{jV}$ and $H_{jV}$ are the form-factors corresponding to the different Lorentz structures, and, $p_W^\nu$, $p^\mu_V$ are the incoming momenta of $W^\pm$ and V. To this end, we also introduce nonlinear gauge-fixing functions as \cite{Fujikawa:1973qs,Bace:1975qi,Gavela:1981ri,Monyonko:1985kuv,Hernandez:1999xn,Hernandez-Sanchez:2004yid}
\bea
f^+ &=& (D_\mu^e + \frac{i g s_W^2}{c_W} Z_\mu) W^{+\mu} - i \zeta M_W G^+, \nonumber \\
f^Z &=& \partial_\mu  Z^\mu - \zeta M_Z G^0, \nonumber \\
f^A &=& \partial_\mu A^\mu \,.
\eea
In the above, $D_\mu^e$ and $\zeta$ are the electromagnetic covariant derivative and gauge parameter respectively. It is noted that $f^+$ transforms covariantly under the electromagnetic gauge group and is nonlinear. The gauge-fixing Lagrangian involving $f^+, f^Z$ and $f^A$ can be written as:
\bea
\mathcal{L} = - \frac{1}{\zeta} f^+ f^- - \frac{1}{2 \zeta} (f^Z)^2 - \frac{1}{2 \zeta} (f^A)^2 \,.
\eea
One observes that setting $\zeta=1$ removes the unphysical $G^+ W^- Z(\gamma)$ vertices from the framework. The same also entails that the Feynman rules for the $ Z(\gamma) W^+ W^- $ trilinear gauge vertices gets modified to
\besub
\bea
\Gamma^{\gamma W^+ W^-}_{\rho \nu \mu} (k,p,q) = - i e [g_{\mu \nu} (k-p - q)_{\rho} + g_{\nu \rho} (p-q)_\mu + g_{\rho \mu}(q-k + p)_\nu], \\
\Gamma^{Z W^+ W^-}_{\rho \nu \mu} (k,p,q) = - i g c_W [g_{\mu \nu} (k-p + \frac{s_W^2}{c_W^2} q)_{\rho} + g_{\nu \rho} (p-q)_\mu + g_{\rho \mu}(q-k-\frac{s_W^2}{c_W^2} p)_\nu].
\eea
\eesub
We are now in a position to discuss the contribution of a 3HDM to the form factors $F_{jZ},G_{jZ}$ and $H_{jZ}$ for $j=1,2$. A generic $H_j^+ W^- Z$ one-loop amplitude contains one power of a trilinear coupling at one vertex and two powers of gauge couplings from the remaining vertex (vertices), as  is shown in the past studies \cite{Kanemura2000HWZ,Kanemura2000Custodial,Aoki2011HWZ, Kanemura2015Review,Abbas:2018pfp}. This allows for arranging all the one-loop diagrams into UV-finite and gauge-invariant subsets. We display below, for instance, all diagrams featuring the $H_1-H_1^+-H_1^-$ trilinear interaction (denoted by $\l_{H_1 H_1^+ H_1^-}$) in an exact alignment. 
\begin{figure}[htpb!]{\centering
\subfigure[$E_1$]{
\includegraphics[height = 3 cm, width = 6 cm]{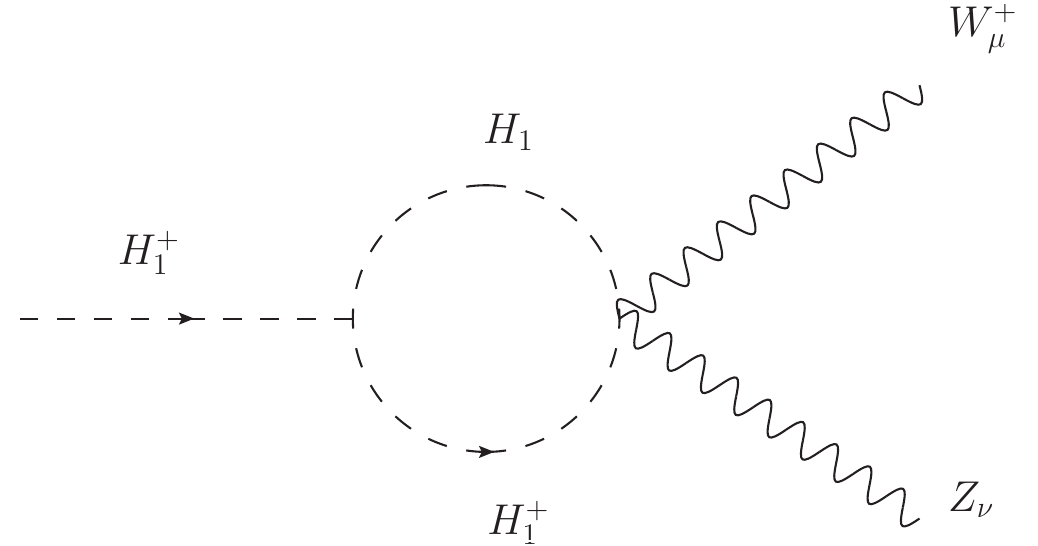}}\\ 
\subfigure[$L_1$]{
\includegraphics[height = 2 cm, width = 5.2 cm]{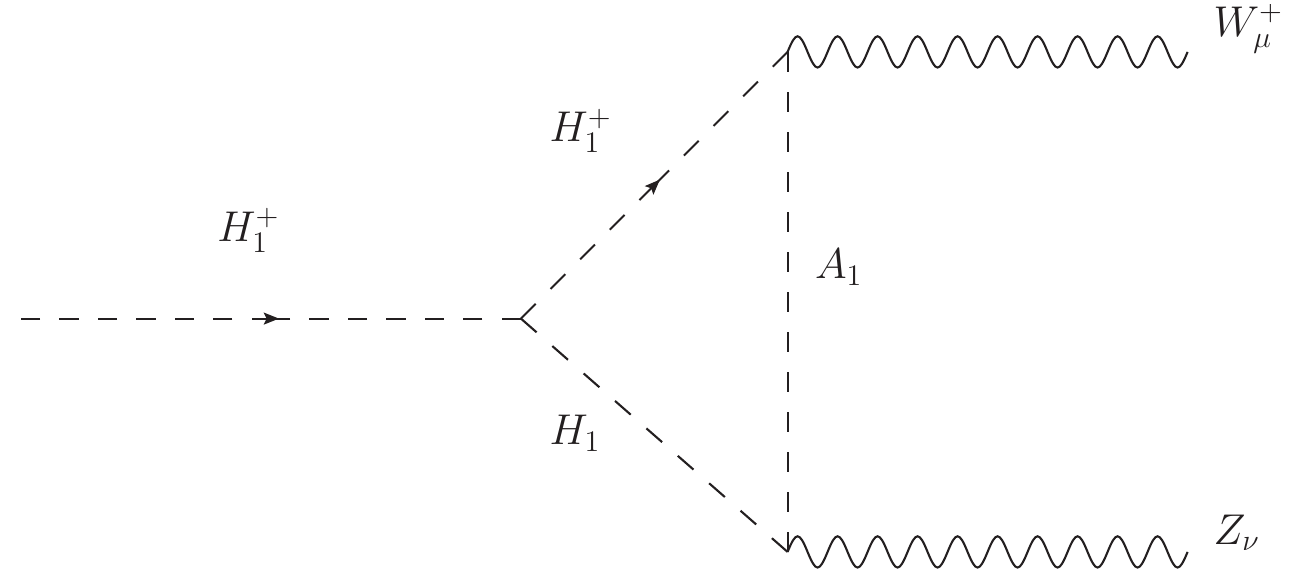}}
\subfigure[$M_1$]{
\includegraphics[height = 2 cm, width = 5.2 cm]{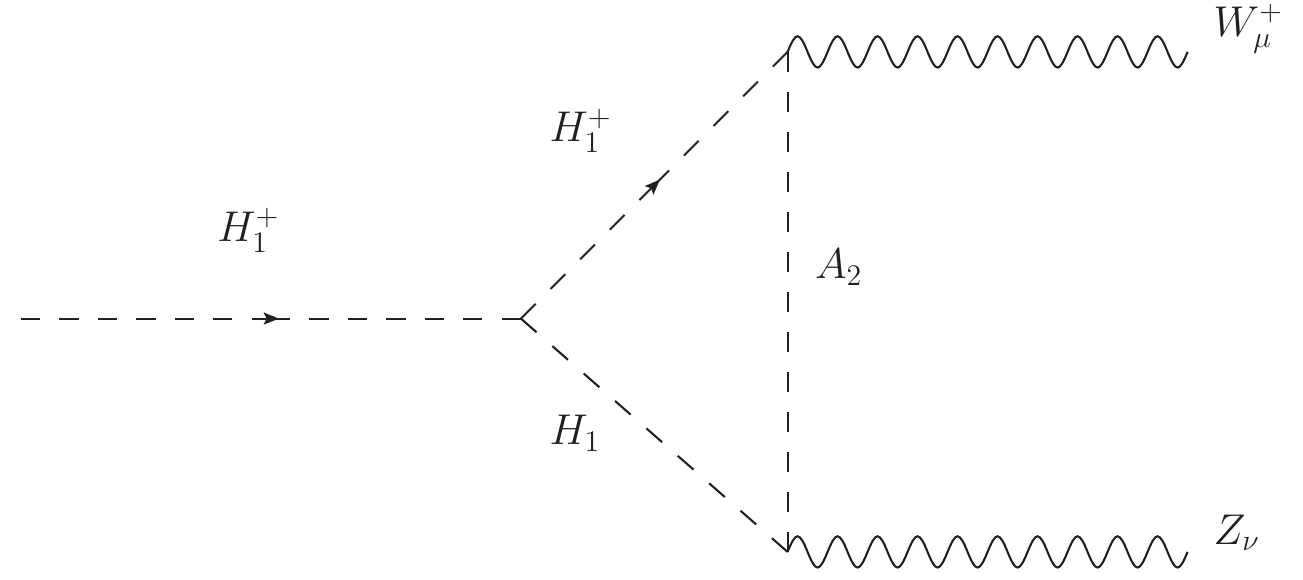}} 
\subfigure[$N_1$]{
\includegraphics[height = 2 cm, width = 5.2 cm]{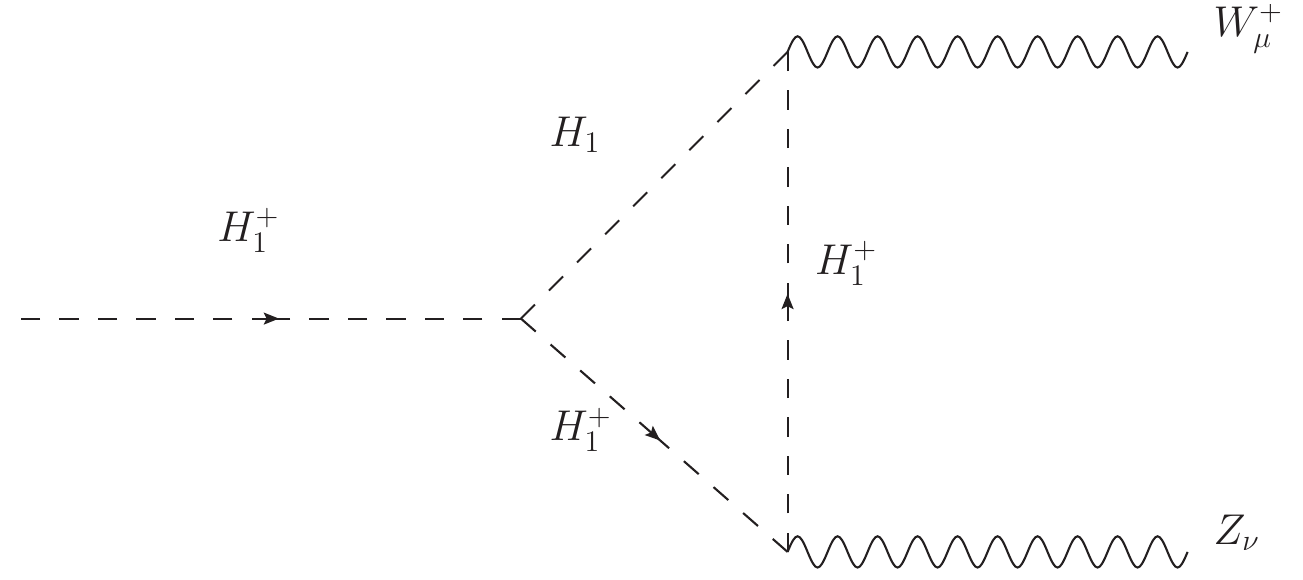}}\\
\subfigure[$E_1^\prime$]{
\includegraphics[height = 3 cm, width = 6 cm]{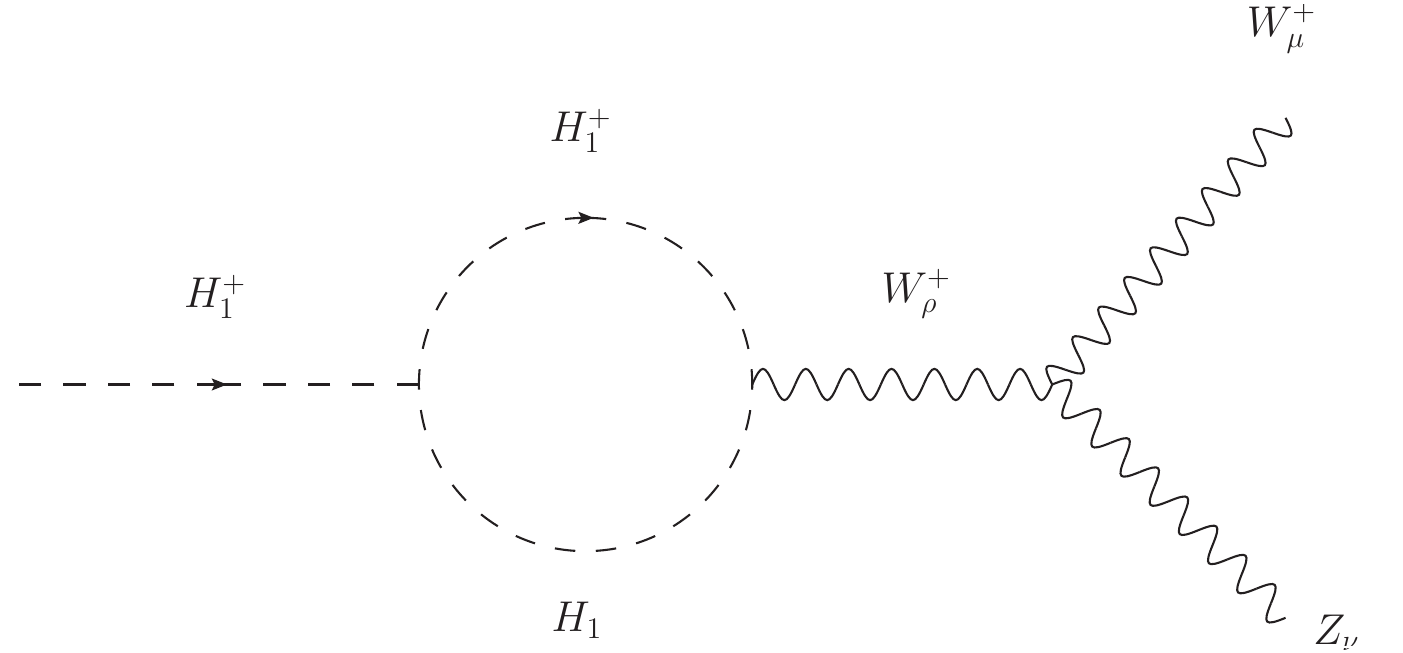}}}
\caption{}
\label{fig-1}
\end{figure}
We elucidate on the topology of each diagram. Diagrams $E_1$ and $E_1^\prime$ are bubble topologies containing a 4-point $S_i S_j W^+ Z$ vertex and a 3-point $W^+ W^- Z$ vertex respectively, where $S_i$ is a generic scalar. On the other hand, $L_1, M_1$ and $N_1$ are triangle topologies involving two $S_i S_j V$ vertices. The amplitudes corresponding to the aforesaid five diagrams are calculated in $d = 4 -\epsilon$ dimensions with $\mu$ being the dimensional regularisation scale,  and, expressed in terms of Passarino-Veltman functions \cite{tHooft:1978jhc} below. We also indicate the corresponding UV-divergent parts.
\besub
\bea
\big(F_{1Z}\big)_{E_1} &=& \frac{1}{16 \pi^2 v}~\frac{s_W^2}{c_W}~ c_{\alpha_3 + \delta_2}~ \lambda_{H_1^+ H_1^- H_1}~B_0(H_1^+; H_1^+, H_1) \nonumber \\
&\xRightarrow{\text{UV-divergent}}& 
\frac{1}{16 \pi^2 v}~\frac{s_W^2}{c_W}~ c_{\alpha_3 + \delta_2}~ \lambda_{H_1^+ H_1^- H_1}(\text{div}) \\
\big(F_{1Z}\big)_{L_1} &=&  \left(\frac{-2 \lambda_{H_1^+ H_1^- H_1}}{16 \pi^2 v ~c_W}\right) (c_{\delta_1-\delta_2}~ c_{\alpha_3+\delta_1}) C_{24}(W,Z,H_1^+;H_1^+,A_1,H_1) \nonumber \\
&\xRightarrow{\text{UV-divergent}}& \left(\frac{-2 \lambda_{H_1^+ H_1^- H_1}}{16 \pi^2 v ~c_W}\right) (c_{\delta_1-\delta_2}~ c_{\alpha_3+\delta_1}) \Big(\frac{\text{div}}{4}\Big) \\
\big(F_{1Z}\big)_{M_1} &=& \left(\frac{-2 \lambda_{H_1^+ H_1^- H_1}}{16 \pi^2 v ~c_W}\right) (s_{\delta_1-\delta_2}~ s_{\alpha_3+\delta_1}) C_{24}(W,Z,H_1^+;H_1^+,A_2,H_1) \nonumber \\
&\xRightarrow{\text{UV-divergent}}& \left(\frac{-2 \lambda_{H_1^+ H_1^- H_1}}{16 \pi^2 v ~c_W}\right) (s_{\delta_1-\delta_2}~ s_{\alpha_3+\delta_1})\Big(\frac{\text{div}}{4}\Big)  \\
\big(F_{1Z}\big)_{N_1} &=& \left(\frac{2 c_{2W}}{c_W}\right) \frac{\lambda_{H_1^+ H_1^- H_1}}{16 \pi^2 v} (c_{\alpha_3+\delta_2}) C_{24}(W,Z,H_1^+;H_1, H_1^+, H_1^+) \nonumber \\
&\xRightarrow{\text{UV-divergent}}& \left(\frac{2 c_{2W}}{c_W}\right) \frac{\lambda_{H_1^+ H_1^- H_1}}{16 \pi^2 v} (c_{\alpha_3+\delta_2}) \Big(\frac{\text{div}}{4}\Big), \\ 
\big(F_{1Z}\big)_{E_1^\prime} &=& \frac{1}{16 \pi^2 v}~\frac{s_W^2}{c_W}~ c_{\alpha_3 + \delta_2}~ \lambda_{H_1^+ H_1^- H_1}~[B_0(H_1^+;H_1^+, H_1)+2 B_1(H_1^+; H_1^+,H_1)]\nonumber \\
&\xRightarrow{\text{UV-divergent}}& 0. 
\eea
\label{amp}
\eesub
A quantity $P$ appearing in the arguments of the Passarino-Veltman functions in Eqs.(\ref{amp}) refers to the squared mass $M_P^2$.
We refer to the Appendix for expressions of the various Passarino-Veltman functions and dub $\frac{2}{\epsilon} - 4\pi + \gamma_E + \text{ln}(\mu^2)$ as the quantity "div". It can be straightforwardly checked that UV-divergent and $\mu$-dependent parts of $\big(F_{1Z}\big)_{E_1} + \big(F_{1Z}\big)_{L_1} + \big(F_{1Z}\big)_{M_1} + \big(F_{1Z}\big)_{N_1} + \big(F_{1Z}\big)_{E_1^\prime}$ vanishes

\begin{figure}[htpb!]{\centering
\includegraphics[height = 7 cm, width = 8 cm]{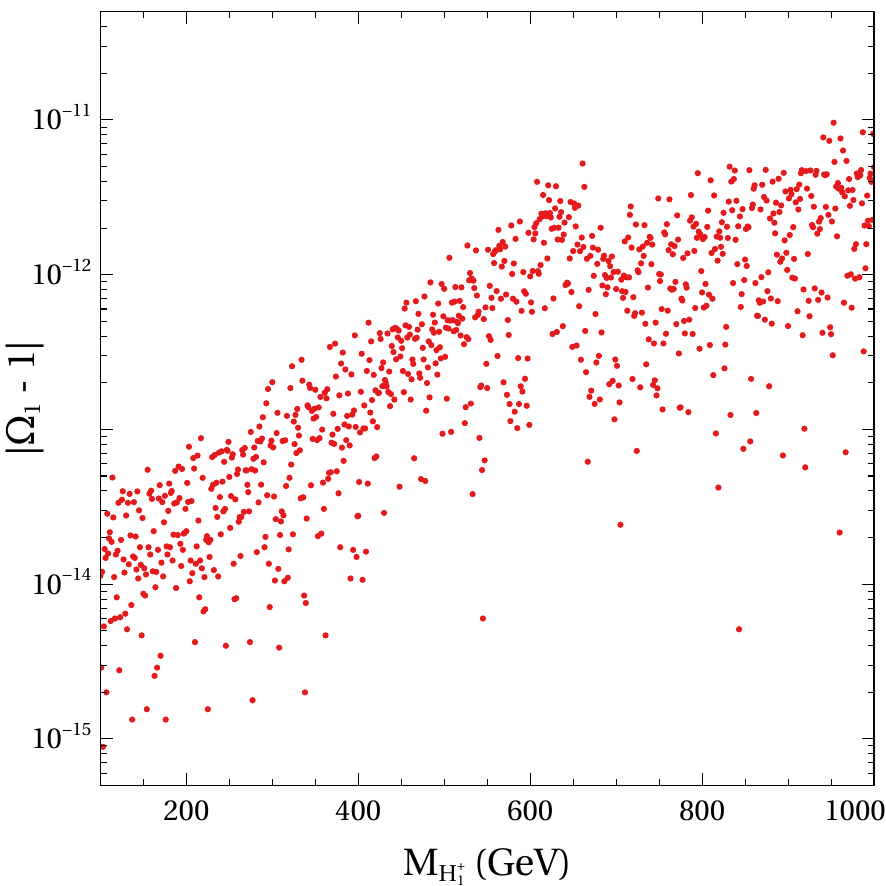}}
\caption{}
\label{ward}
\end{figure}

In addition to checking for UV-finiteness, gauge invariance of an $H_j^+ W^- \gamma$ vertex can serve as a useful check of the $H_j^+ W^- Z$ amplitude even if a detailed quantification of the former in a given model is not explicitly aimed at. The Ward
identity enforces $A^{\mu \nu}_{j\gamma}p_{\gamma \nu} = 0$ that ultimately leads to
\bea
F_{j\gamma} &=& \frac{G_{j\gamma}}{2}\Bigg(1 - \frac{M_{H^+_j}^2}{M^2_W} \Bigg).
\eea
The amplitudes featuring a given trilinear coupling would form a UV-finite subset in case of $H^+_j \to W^+ \gamma$ too. The diagrams corresponding to the $H_1-H_1^+-H_1^-$ trillinear coupling are named $R_1,~S_1$ and $R_1^\prime$ and are displayed below.
\begin{figure}[htpb!]{\centering
\subfigure[$R_1$]{
\includegraphics[height = 3 cm, width = 6 cm]{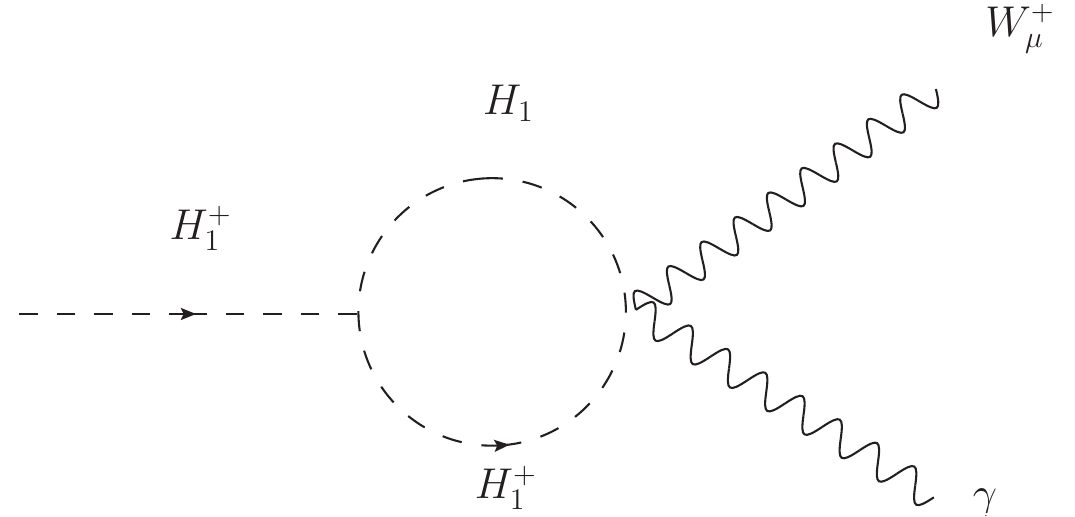}}~~~  
\subfigure[$S_1$]{
\includegraphics[height = 2 cm, width = 5.2 cm]{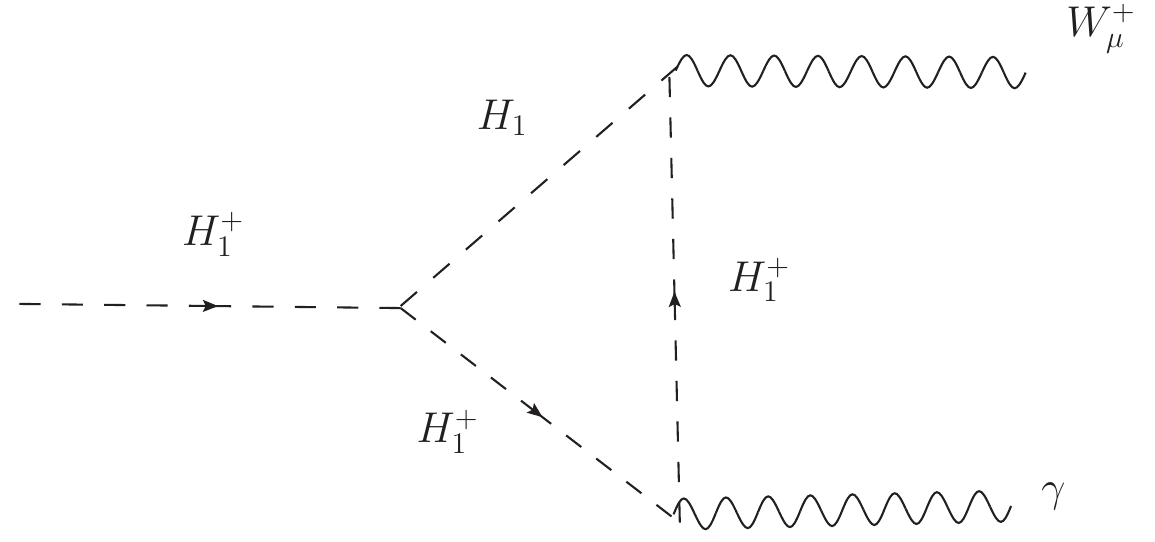}}~~~
\subfigure[$S_1^\prime$]{
\includegraphics[height = 3 cm, width = 6 cm]{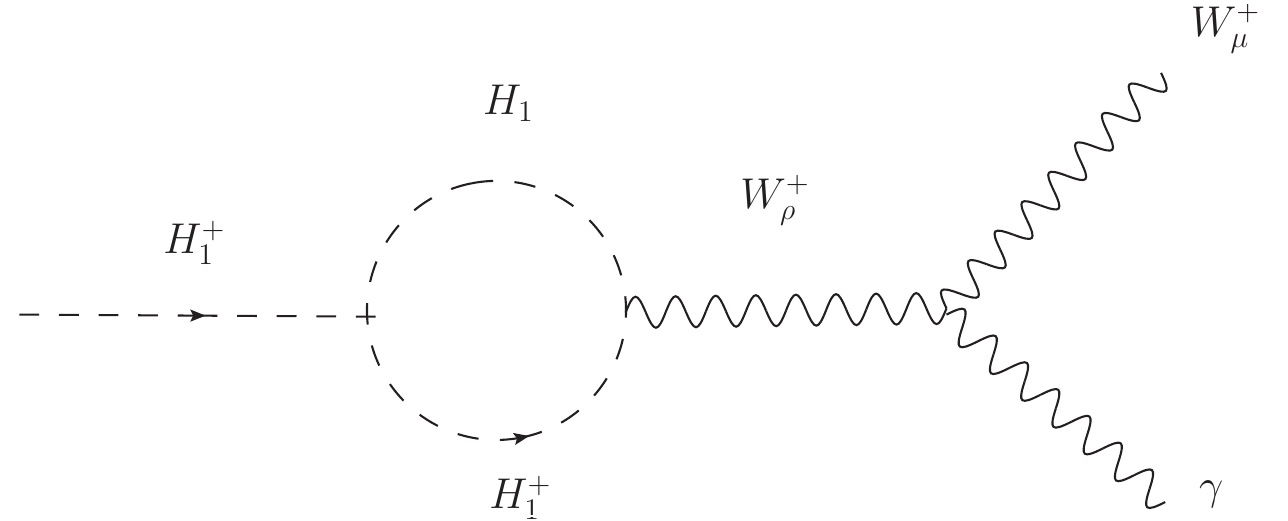}}}
\caption{}
\label{fig-gamma}
\end{figure}
We define the dimensionless quantity $\Omega_j = \frac{2 F_{j\gamma}}{G_{j\gamma}}\Bigg(1 - \frac{M_{H^+_j}^2}{M^2_W} \Bigg)^{-1}$ with a numerical cross-check in view. The aforementioned set of amplitudes then gives
\bea
\Omega_1 &=& -\Bigg\{\frac{B_0(H_1^+;H_1^+,H_1) - 2 C_{24}(H_1^+,W,0;H_1, H_1^+, H_1^+) + B_1(H_1^+;H_1^+,H_1)}{C_{12}(H_1^+,W,0;H_1, H_1^+, H_1^+) + C_{23}(H_1^+,W,0;H_1, H_1^+, H_1^+)}\Bigg\}\Bigg(1 - \frac{M_{H^+_1}^2}{M^2_W} \Bigg)^{-1}
\eea
We numerically evaluate $\Omega_1$ using the publicly available tool \texttt{LoopTools} \cite{Hahn:1998yk} with $M_{H^+_1}$ = 200 GeV and $M_{H_1}$ varied in the  [200 GeV,1 TeV] range. Treating $|\Omega_1 - 1|$ as a measure of breakdown of gauge invariance, we display the behaviour of the same \emph{w.r.t.} $M_{H_1}$ in Fig.\ref{ward}.
The deviation turns out to be tiny and is mostly an artefact of numerical accuracy. Thus, it is safely concluded that the $H_j^+ W^- \gamma$ amplitude stemming from $R_1,S_1$ and $S_1^\prime $ is consistent with Ward identity. We argue here that such an observation for $R_1,S_1$ and $S_1^\prime $ also serves as an additional check on the correctness of $E_1,L_1,M_1,N_1,E_1^\prime$, given the similarities of the two sets of amplitudes at one-loop.

In all, it is demonstrated that $H^+_j \to W^+ Z$ amplitudes featuring a given scalar trilinear interaction constitute a UV-finite and gauge invariant subset. Therefore, in view of the large number of amplitudes in this framework, we cluster them according to the trilinear couplings in Tables \ref{amp1_list} ($H_1^+$ decay) and \ref{amp2_list} ($H_2^+$ decay). 
\begin{center}
\begin{table}[htb!]
\begin{tabular}{|c|c|}\hline
~~~~Trilinear interaction ~~~~& ~~~~Diagrams~~~~ \\ \hline  
$H_1-H_1^+-H_1^-$ & $E_1,L_1,M_1,N_1,E_1^\prime$ \\  
$H_1-H_1^+-H_2^-$ & $E_2,L_2,M_2,N_2,E_2^\prime$ \\  
$H_2-H_1^+-H_1^-$ & $E_3,L_3,M_3,N_3,E_3^\prime$ \\  
$H_2-H_1^+-H_2^-$ & $E_4,L_4,M_4,N_4,E_4^\prime$ \\  
$A_1-H_1^+-H_2^-$ & $O_1,X_1,Y_1,Z_1,O_1^\prime$ \\  
$A_2-H_1^+-H_2^-$ & $O_2,X_2,Y_2,Z_2,O_2^\prime$ \\  \hline
 \end{tabular}
\caption{Serial numbers of scalar 1-loop amplitudes for $H_1^+ \to W^+ Z$}
\label{amp1_list}
\end{table}
\end{center}

\begin{center}
\begin{table}[htb!]
\begin{tabular}{|c|c|}\hline
~~~~Trilinear interaction ~~~~& ~~~~Diagrams~~~~ \\ \hline  
$H_1-H_1^+-H_1^-$ & $E_5,L_5,M_5,N_5,E_5^\prime$ \\  
$H_1-H_1^+-H_2^-$ & $E_6,L_6,M_6,N_6,E_6^\prime$ \\  
$H_2-H_1^+-H_1^-$ & $E_7,L_7,M_7,N_7,E_7^\prime$ \\  
$H_2-H_1^+-H_2^-$ & $E_8,L_8,M_8,N_8,E_8^\prime$ \\  
$A_1-H_1^+-H_2^-$ & $O_3,X_3,Y_3,Z_3,O_3^\prime$ \\  
$A_2-H_1^+-H_2^-$ & $O_4,X_4,Y_4,Z_4,O_4^\prime$ \\  \hline
 \end{tabular}
\caption{Serial numbers of scalar 1-loop diagrams for $H_2^+ \to W^+ Z$}
\label{amp2_list}
\end{table}
\end{center}

Fermionic contributions at one-loop to $H_j^+ \to W^+ Z$ are shown in the Appendix.
Next, the decay width for $H_j^+ \to W^+ Z$ reads
\bea
\Gamma (H_j^+ \rightarrow W^+ Z) = M_{H_j^+} \frac{\sqrt{\lambda(1, \omega_j, z_j)}}{16 \pi} \sum_{i = L,T} |M_{ii}|_j^2 \,,.
\eea
Here $\lambda (a,b,c) = (a-b-c)^2 - 4 abc,~ \omega_j = \frac{M_W^2}{M_{H_j^+}^2},~z = \frac{M_Z^2}{M_{H_j^+}^2}$. In addition, $L$ and $T$ refer to the longitudinal and transverse polarizations respectively. The longitudinal and transverse squared amplitudes are given by
\bea
|M_{LL}|_j^2 &=& \frac{g^2}{4 z_j}|(1-\omega_j-z_j) F_{jZ} + \frac{\lambda (1, \omega_j,z_j)}{2 \omega_j} G_{jZ}|^2 \,, \nonumber \\
|M_{TT}|_j^2 &=& g^2 (2 \omega_j |F_{jZ}|^2 + \frac{\lambda(1, \omega_j, z_j)}{2 \omega_j} |H_{jZ}|^2) \,.
\eea


\section{Constraints}
\label{constraints}
We parameter space of a flavour-conserving 3HDM is subjected to the following theoretical and experimental constraints.

\subsection{Theoretical Constraints}\label{theo_con}
A perturbative framework stipulates that the magnitude of the quartic couplings remains bounded from above at $4\pi$, that is $|\l_i| \leq 4\pi$.
The corresponding bound on a Yukawa coupling is $|y| \leq \sqrt{4\pi}$.
Next, the scalar potential remains bounded from below along various directions in field space if the following conditions are satisfied.
\bea
\l_1 > 0,~\l_2 > 0,~\l_3 > 0,
~\l_{12} + \sqrt{\l_1 \l_2} > 0,~\l_{23} + \sqrt{\l_2 \l_3} > 0, 
~\l_{13} + \sqrt{\l_1 \l_3} > 0 \,, \nonumber \\
\l_{12} + \l_{12}^{'} - |\l_{12}^{''}| + \sqrt{\l_1 \l_2} > 0, 
~\l_{23} + \l_{23}^{'} - |\l_{23}^{''}| + \sqrt{\l_2 \l_3} > 0, 
~\l_{13} + \l_{13}^{'} - |\l_{13}^{''}| + \sqrt{\l_1 \l_3} > 0. 
\eea
The theoretical constraint taken up next is perturbative unitarity. Various $2 \to 2$ scattering amplitudes involving longitudinal components of gauge bosons can mapped to ones involving the Goldstone bosons using the electroweak equivalence theorem~\cite{Lee:1977eg,Cornwall:1974km}. The latter amplitudes, expressible in terms of the scalar quartic couplings, can be arranged in the form of a matrix. The framework obeys perturbative unitarity if the eigenvalues of the aforementioned matrix have magnitudes not exceeding $8\pi$. For a 3HDM, 
one can construct the amplitudes using neutral and singly-charged two-particle states, which themselves comprise of the scalar fields $\omega_i^\pm, h_i, z_i$ (with $i=1,2,3$) in the gauge basis. This in turn leads to $30 \times 30$ and $18 \times 18$ scattering matrices in the bases of two-particle neutral and singly charged bases respectively~\cite{Bento:2022vsb}.

It is cumbersome to extract the eigenvalues analytically. Therefore, we adopt another approach for a 3HDM. The details are relegated in the Appendix. This new approach furnishes analytical forms for the following 18 eigenvalues.
\bea
\l_{12} \pm \l^\prime_{12},~\l_{12} \pm \l^{\prime\prime}_{12},~\l_{13} \pm \l^\prime_{13},~\l_{13} \pm \l^{\prime\prime}_{13},~\l_{23} \pm \l^\prime_{23},~\l_{23} \pm \l^{\prime\prime}_{23}, \nonumber \\
\l_{12} + 2\l^\prime_{12} \pm 3\l^{\prime\prime}_{12},~\l_{13} + 2\l^\prime_{13} \pm 3\l^{\prime\prime}_{13},~\l_{23} + 2\l^\prime_{23} \pm 3\l^{\prime\prime}_{23}
\eea  
Nine more eigenvalues are obtained from three $3 \times 3$  matrices as detailed in the Appendix.

\subsection{Experimental Constraints}
\subsubsection{$h \to \gamma\gamma$ signal strength}
The $h \to \gamma \gamma$ decay width can deviate from the corresponding SM value in spite of an exact alignment on account of the additional one-loop contributions coming from $H_{1,2}^+$. The corresponding amplitude and decay width read \cite{Djouadi:2005gi,Djouadi:2005gj}:
\besub
\bea
\mathcal{M}^{\text{3HDM}}_{h \to \gamma \gamma} &=& 
\sum_f N_f Q_f^2 y_f^h A_{1/2}\Big(\frac{M^2_h}{4 M^2_f}\Big)
 + f^h_{VV} A_1\Big(\frac{M^2_h}{4 M^2_W}\Big)\nonumber \\
&& 
 + \sum_{i=j}^2 \frac{\l_{h H_i^+ H_i^-} v}{2 M^2_{H_j^+}} A_0\Big(\frac{M^2_h}{4 M^2_{H_i^+}}\Big) \\ 
\Gamma^{\text{3HDM}}_{h \to \gamma \gamma} &=& \frac{G_F \a^2 M_h^3}{128 \sqrt{2} \pi^3} |\mathcal{M}^{\text{3HDM}}_{h \to \gamma \gamma}|^2,
\eea
\eesub
where $N_f$ and $Q_f$ respectively denote respectively color factor, charge of fermion $f$. Additionally, $G_F$ and $\alpha$ are the Fermi-constant and the QED fine-structure constant respectively. Moreover, $f^h_{VV} = c_\gamma s_{\alpha_2} + c_{\alpha_2} c_{\alpha_1 - \beta} s_\gamma$ is the scale factor corresponding to the $h-V-V$ coupling. One however notes $y_f^h = f_{VV}^h$ = 1 for alignment. The functions $A_0(x)$, $A_{1/2}(x)$ and $A_1(x)$ are given in \cite{Djouadi:2005gi,Djouadi:2005gj}. Given the production cross section of $h$ is unchanged \emph{w.r.t.} the SM for an exact alignment, the signal strength for the $h \to \gamma\gamma$ channel becomes
\bea
\mu_{\gamma \gamma} \simeq \frac{\Gamma_{\rm 3HDM}(h \to \gamma \gamma)}{\Gamma_{\rm exp}(h \to \gamma \gamma)} \,.
\label{h-to-gaga}
\eea
We have implemented this constraint at the 2$\sigma$-level following the latest measurements from CMS \cite{CMS:2023HiggsDiphoton} and ATLAS \cite{ATLAS:2022tnm}.

\subsubsection{Electroweak Precision Observables}
A multi-Higgs setup induces additional contributions to the oblique ($S,T,U$) parameters~\cite{Toussaint:1978zm,Peskin:1990zt,Peskin:1991sw}. The latest limits on $\Delta S$ and $\Delta T$ for $\Delta U$ = 0 are
\bea
\Delta S = 0.05 \pm 0.08,~\Delta T = 0.09 \pm 0.07,~\rho_{ST} = 0.92.
\eea
We have calculated the corresponding contributions coming from a $\mathbb{Z}_2 \times \mathbb{Z}_2^\prime \times \mathbb{Z}_2^{\prime\prime}$ 3HDM following~\cite{Grimus:2008nb} and imposed the aforementioned limit at 2$\sigma$.

\subsubsection{$B \to X_s \gamma$ decay}
\label{btosgamma-sub}
The most stringent flavour constraint for a multi-doublet framework that forbids tree level FCNCs arises from the $B \to X_s \gamma$ branching ratio. SM predicts the aforesaid ratio to be $(3.36 \pm 0.23) \times 10^{-4}$ while its experimental average is $(3.32 \pm 0.16) \times 10^{-4}$. NP contributions to the $b \to s \gamma$ amplitude thus get tightly constrained. Such contributions stem from $H_{1,2}^+$ in a 3HDM and modify the $C_{7,8}$ Wilson coefficients \emph{w.r.t.} the SM values. Following \cite{PhysRevLett.114.221801}, one has the following at a matching scale $\simeq$ 160 GeV
\bea
\delta C^{\text{eff}}_{7,8} = \sum_{j=1,2} \bigg[\frac{(A^j_u)^2}{3} F^{(1,2)}_{7,8}(x_j) - A_u^j A_d^j F^{(1,2)}_{7,8}(x_j)    \bigg].\label{C_78}
\eea
One notes that $x_j = M^2_t/M^2_{H^+_j}$ and 
\besub
\bea
F_7^{(1)}(x)
&=& \frac{x(7 - 5x - 8x^2)}{24 (x - 1)^3}
  + \frac{x^2(3x - 2)}{4 (x - 1)^4} \ln x, \\
F_8^{(1)}(x)
&=& \frac{x(2 + 5x - x^2)}{8 (x - 1)^3}
  - \frac{3x^2}{4 (x - 1)^4} \ln x. \\
F_7^{(2)}(x) &=& \frac{x(3 - 5x)}{12 (x - 1)^2}
             + \frac{x(3x - 2)}{6 (x - 1)^3} \ln x, \\
F_8^{(2)}(x) &=& \frac{x(3 - x)}{4 (x - 1)^2}
             - \frac{x}{2 (x - 1)^3} \ln x.
\eea
\eesub
The factors $A_u^j$ and $A_d^j$ are expressed in Table \ref{Au_Ad} for the 3HDM types considered. Following [36,38 of Avik], the
limits on the branching ratio translate into
\bea
-0.063 \leq \delta C_7^{\text{eff}} + 0.24~\delta C_8^{\text{eff}} \leq 0.073.
\eea

\section{Numerical scans and evaluation of the form factors}
\label{results}

This section is devoted to sampling the 3HDM parameter space $\emph{w.r.t.}$ the said constraints and evaluating the $H^+_{1,2} \to W^+ Z$ form factors numerically. And we choose the Type-II, lepton specific and democratic 3HDM in view of the difference in the Yukawa structures. 
Drawing analogies with the 2HDM, a few comments are in order. First, the stringent 
$M_{H^+} \geq$ 580 GeV limit from $B \to X_s \gamma$ for a Type-II 2HDM is expected to appropriately relax in the 3HDM, thanks to cancellations in the amplitude level. It is worthwhile to examine the form factors for the compartively lighter $H_1^+$ or $H_2^+$ in foresight. Secondly, the $B \to X_s \gamma$ branching fraction is identical in the Type-I and lepton specific 3HDM owing to the identical quark couplings in the two cases. We choose the latter in view of its potentially enhanced lepton Yukawas. We also keep in mind a possible scope to search for $H_{1,2}^+$ at colliders through their lepton couplings. Finally, the democratic 3HDM has also generated interest [DD, Moretti etc] on account of its enticing Yukawa structure  where the u-quark, b-quark and lepton masses come from three different scalar doublets. 

Before embarking on the numerical scans, it is pointed out that an elaborate sweep of the 3HDM parameter space becomes computationally hectic given the large number of free parameters. Some past studies [Coleppa etc] have carried out detailed investigations in this direction. We therefore follow a more economical approach in this paper. First, we implement an exact alignment limit as $\alpha_1 = \beta,~\alpha_2 = \frac{\pi}{2} -\gamma$. Secondly, we fix $M_{H_1} = M_{H^+_1},~M_{H_2} = M_{H^+_2},~\delta_1 = \delta_2,~\alpha_3 = -\delta_2$.
in order to ensure $\Delta T = 0$. The rest of the parameters are varied as under:
\bea
&& 0 \leq m_{12}, m_{23}, m_{13} \leq 500~ \rm{GeV}, \nonumber\\
&& 100~\rm{GeV} \leq M_{H_1^+}, M_{H_2^+}, M_{A_1}, M_{A_2} \leq 1000~\rm{GeV}, \nonumber \\
&& 1 \leq \tan \gamma, \tan \beta \leq 20,~ 0 \leq \delta_2 \leq 2 \pi.
\eea

\begin{figure}[htpb!]{\centering
\subfigure[]{
\includegraphics[height = 6.5 cm, width = 7 cm]{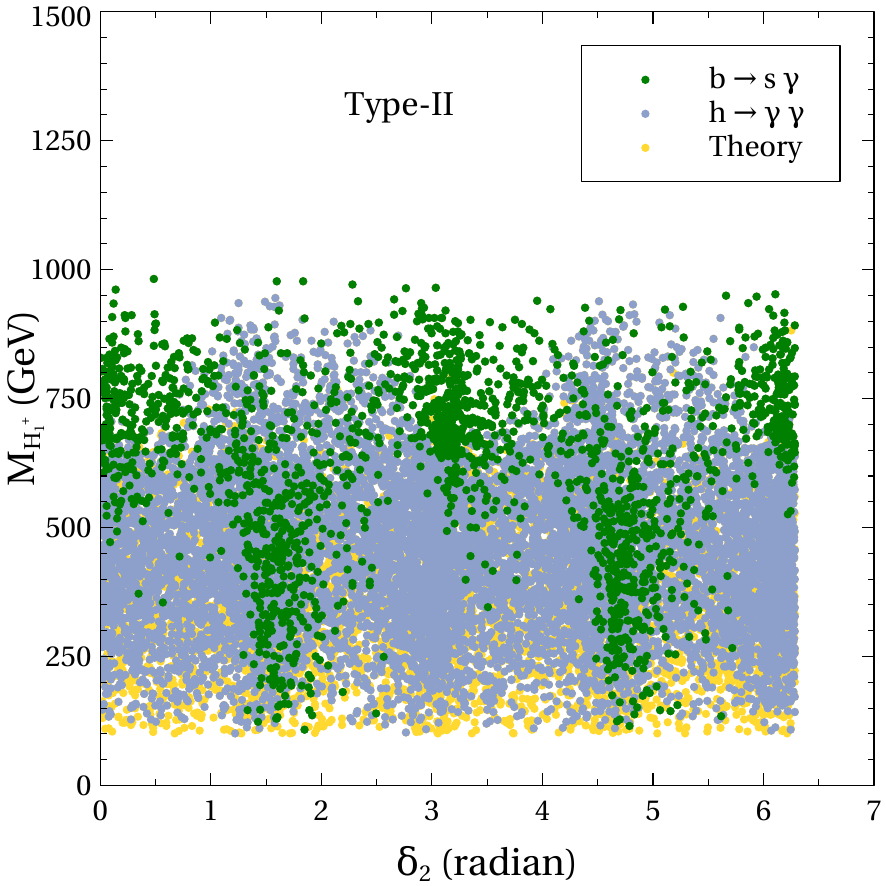}}
\subfigure[]{
\includegraphics[height = 6.5 cm, width = 7 cm]{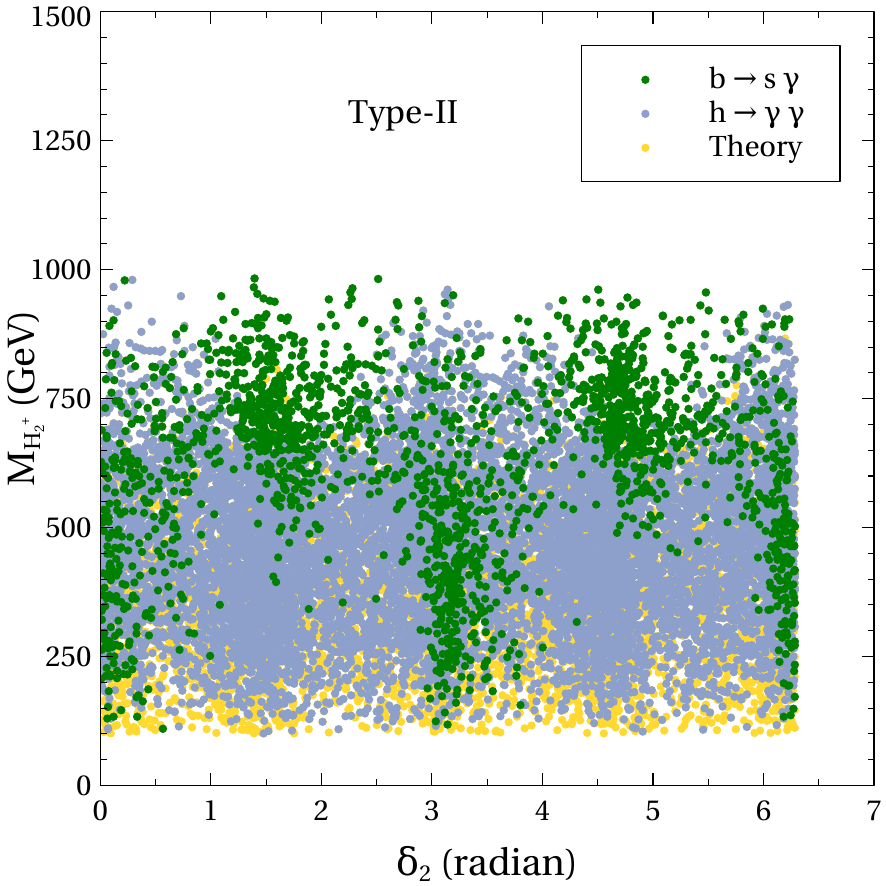}} \\
\subfigure[]{
\includegraphics[height = 6.5 cm, width = 7 cm]{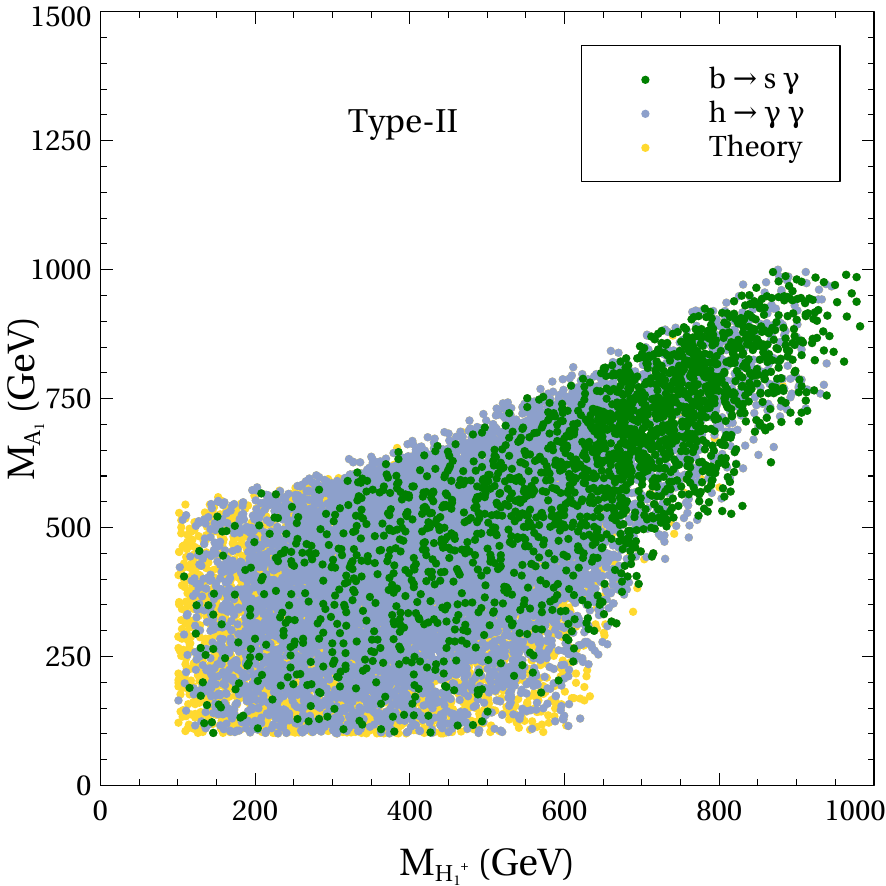}}
\subfigure[]{
\includegraphics[height = 6.5 cm, width = 7 cm]{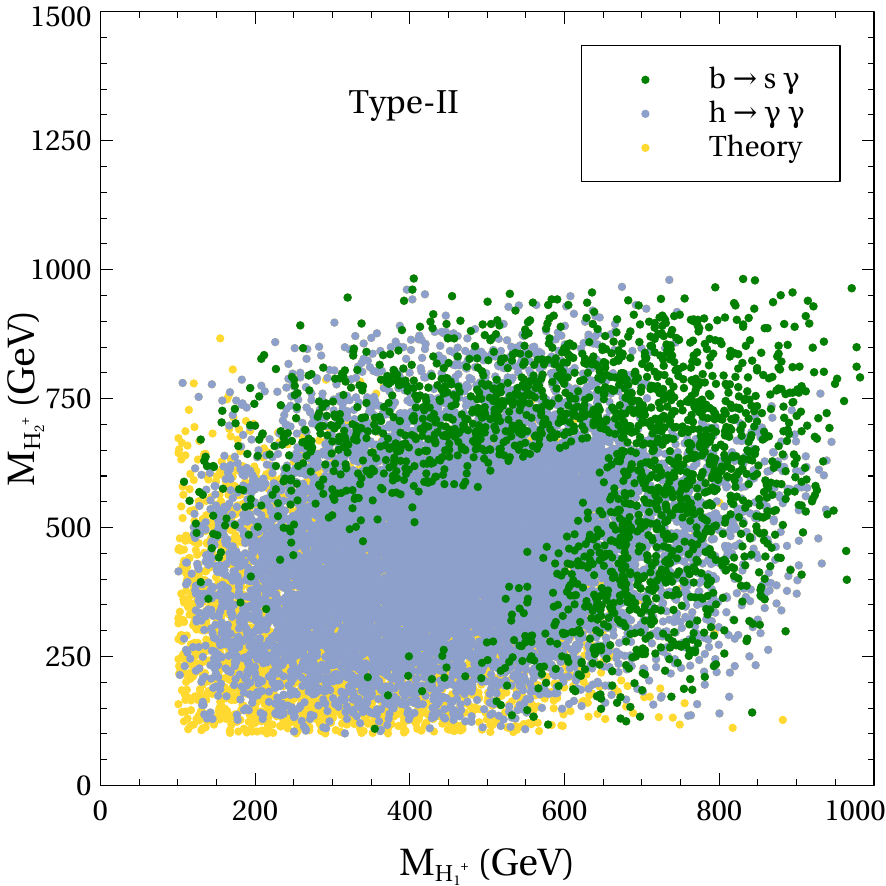}} \\
\subfigure[]{
\includegraphics[height = 6.5 cm, width = 7 cm]{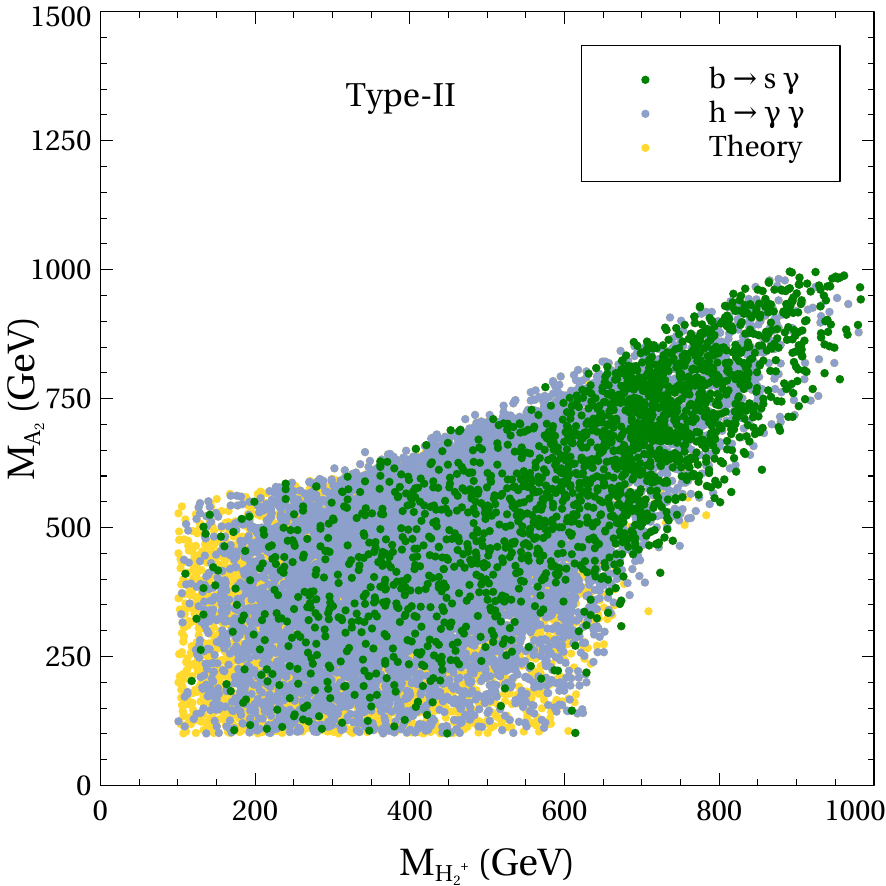}}
}
\caption{Allowed parameter points for the Type-II 3HDM. The color-coding is explained in the legends.}
\label{param-II}
\end{figure}
\begin{figure}[htpb!]{\centering
\subfigure[]{
\includegraphics[height = 6.5 cm, width = 7 cm]{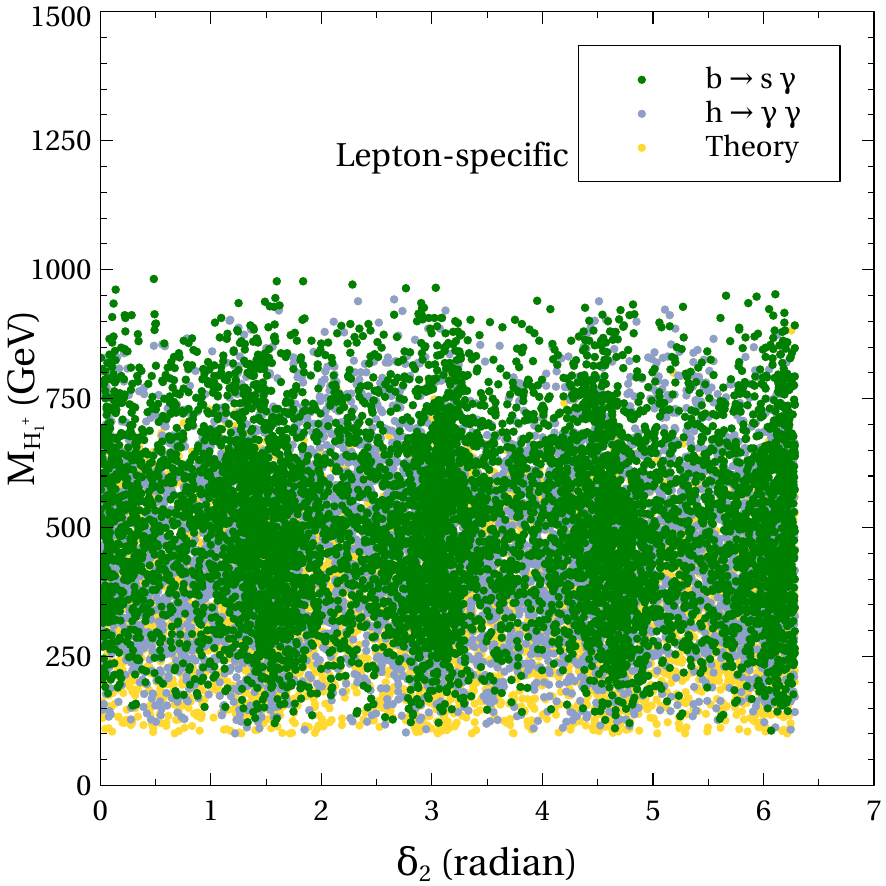}}
\subfigure[]{
\includegraphics[height = 6.5 cm, width = 7 cm]{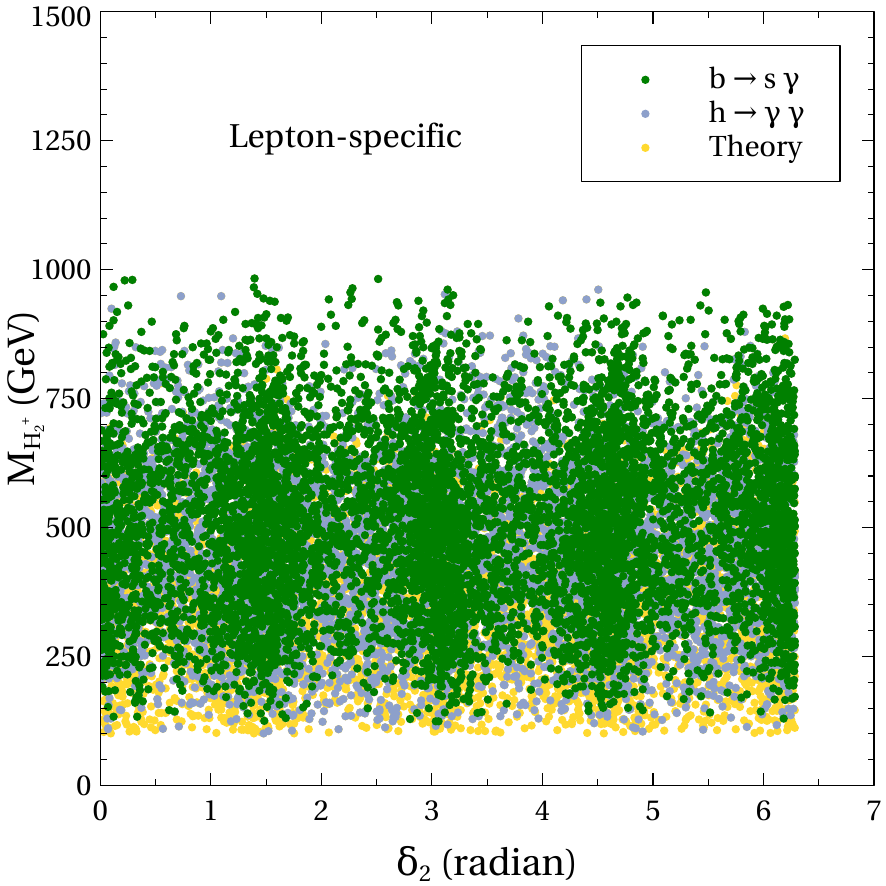}} \\
\subfigure[]{
\includegraphics[height = 6.5 cm, width = 7 cm]{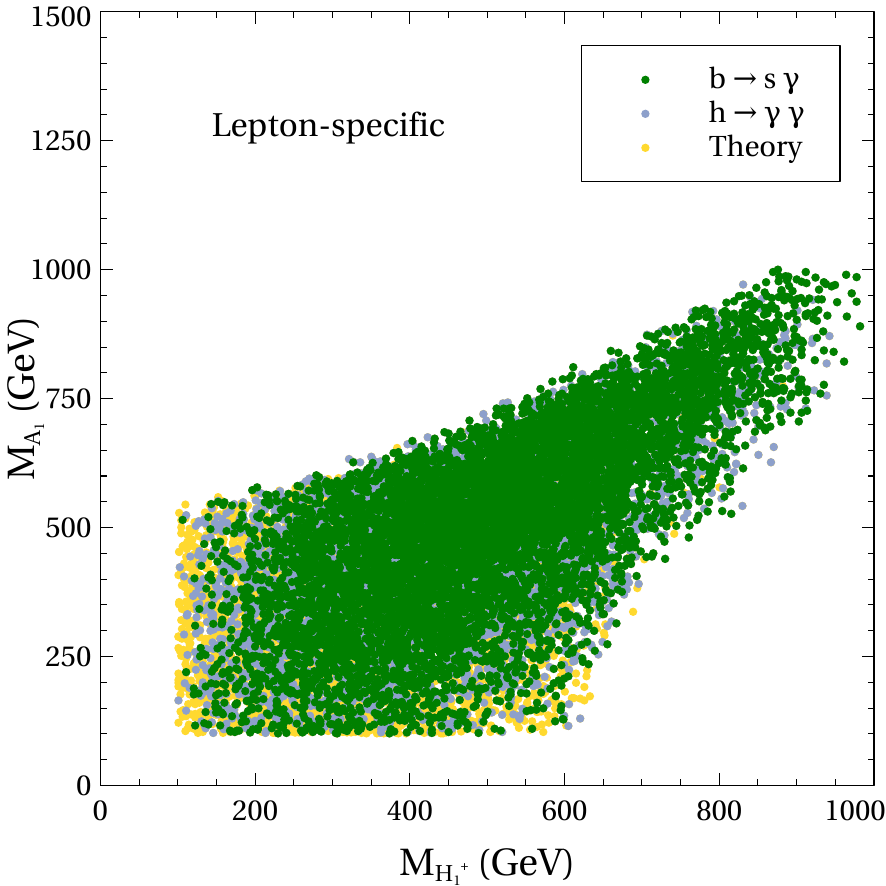}}
\subfigure[]{
\includegraphics[height = 6.5 cm, width = 7 cm]{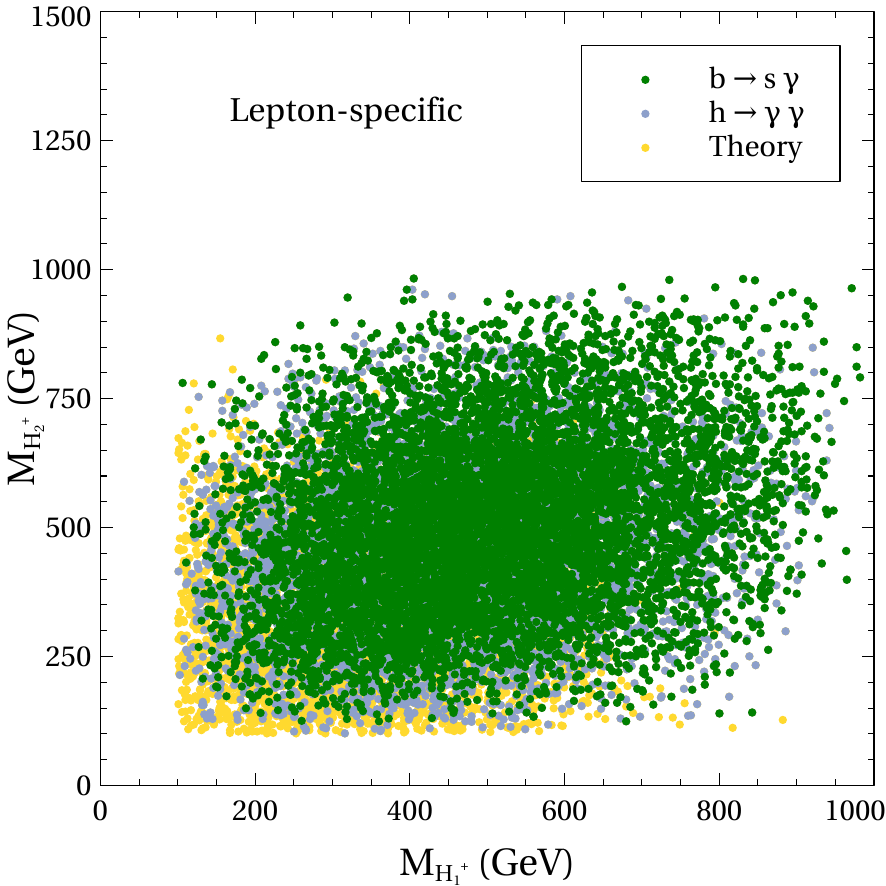}} \\
\subfigure[]{
\includegraphics[height = 6.5 cm, width = 7 cm]{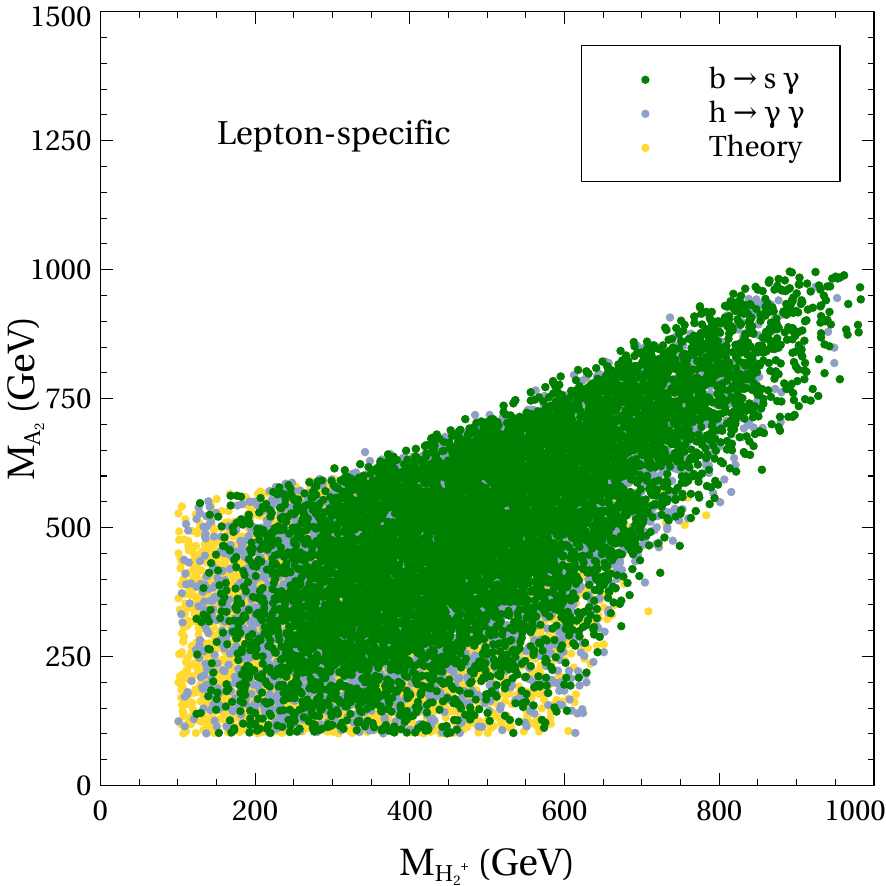}}
}
\caption{Allowed parameter points for the lepton-specific 3HDM. The color-coding is explained in the legends.}
\label{param-LS}
\end{figure}

We adopt the following sequence in accepting parameter points. (i) The first set of points ensure perturbative unitarity and ensure a bounded-from-below potential. (ii) From the first set, we screen the points that satisfy the $h \to \gamma \gamma$ constraint, and finally, (iii) we further select those points that successfully negotiate the bound on the $B \to X_s \gamma$ branching ratio. The parameter points allowed at each of the aforesaid stages are plotted for the Type-II and lepton-specific 3HDMs in Fig.\ref{param-II} and Fig.\ref{param-LS}.

An inspection reveals that the allowed parameter spaces for the two 3HDMs mainly differ in terms of the allowed values of the charged Higgs masses. While the $B \to X_s \gamma$ constraint does not constrain $M_{H_1^+}$ and $M_{H_2^+}$ for the lepton-specific case, it rules out simultaneously low masses for the two charged Higgses.

\begin{figure}[htpb!]{\centering
\subfigure[]{
\includegraphics[height = 5 cm, width = 7 cm]{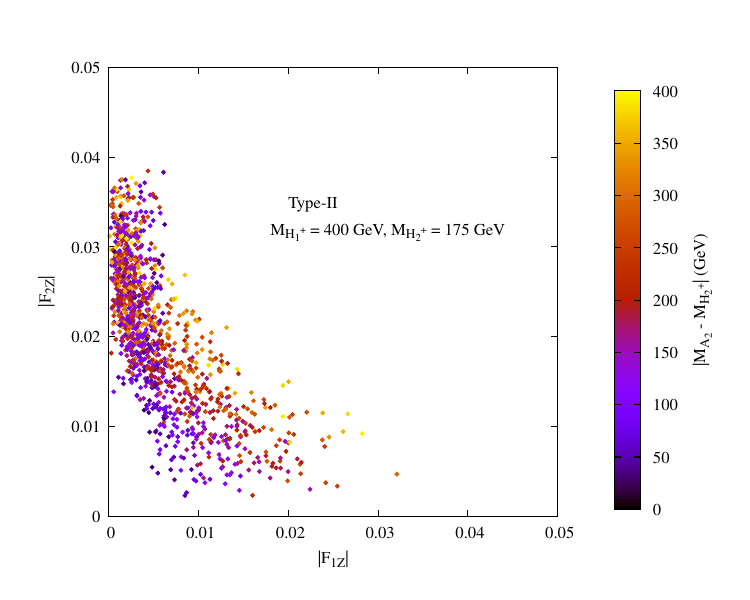}}
\subfigure[]{
\includegraphics[height = 5 cm, width = 7 cm]{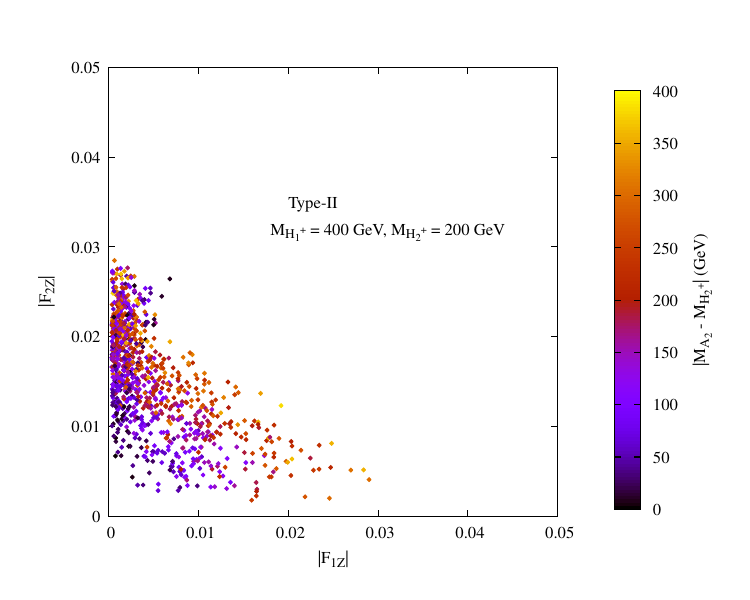}} \\
\subfigure[]{
\includegraphics[height = 5 cm, width = 7 cm]{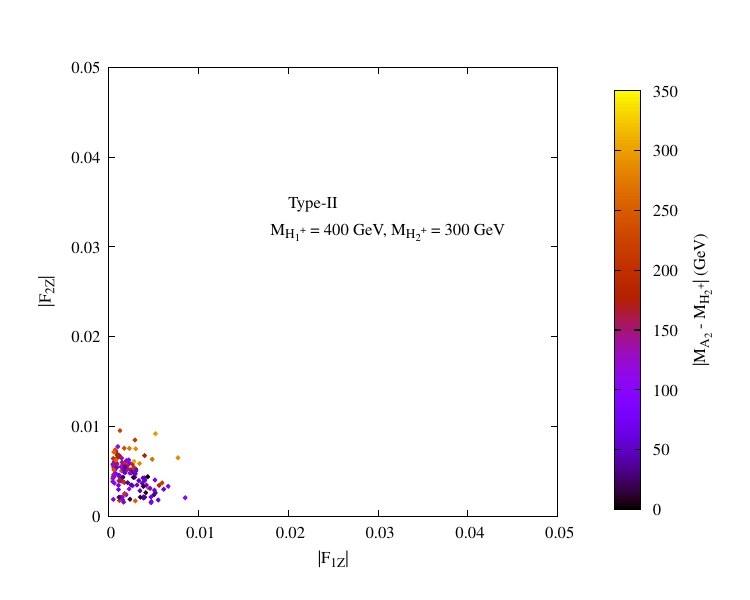}}
\subfigure[]{
\includegraphics[height = 5 cm, width = 7 cm]{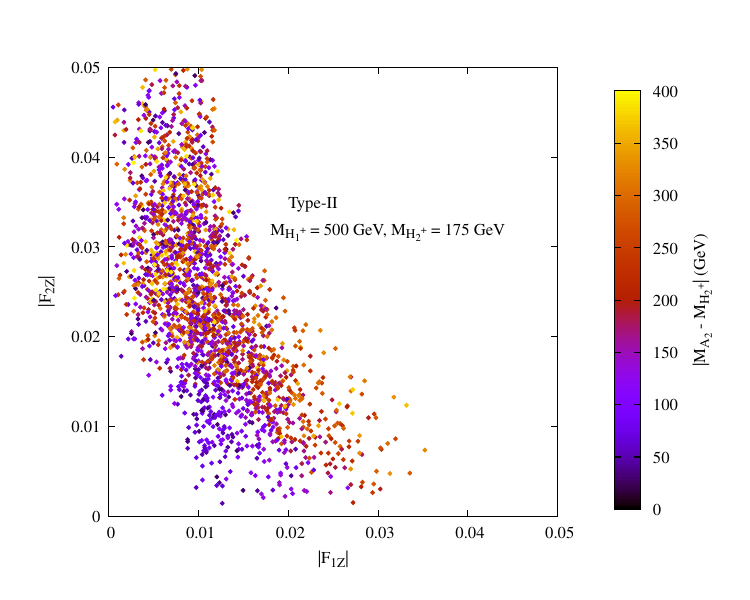}} \\
\subfigure[]{
\includegraphics[height = 5 cm, width = 7 cm]{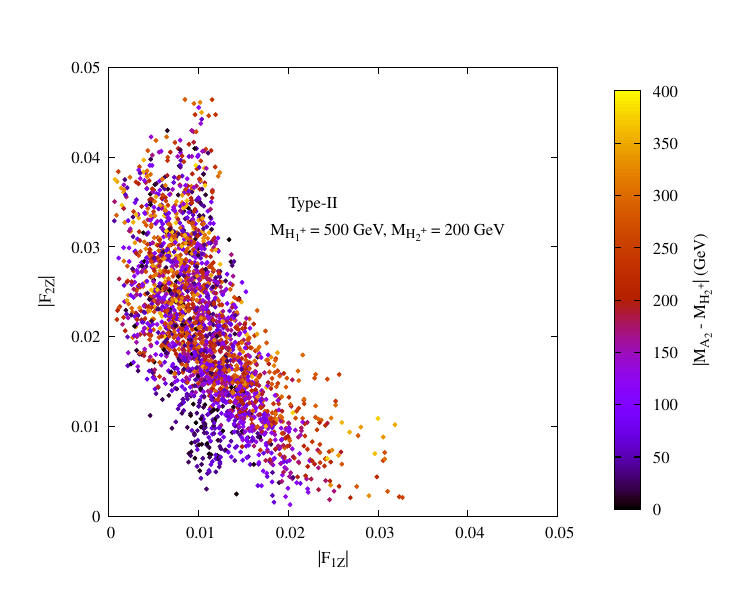}}
\subfigure[]{
\includegraphics[height = 5 cm, width = 7 cm]{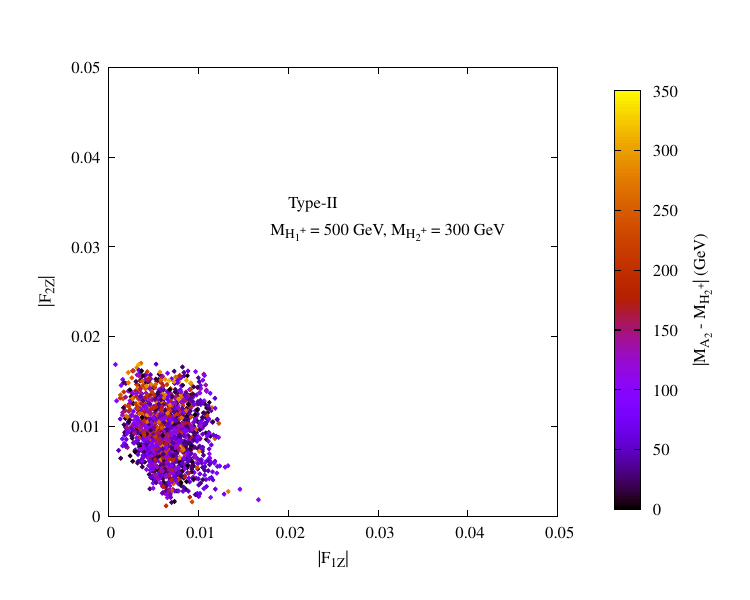}} \\
\subfigure[]{
\includegraphics[height = 5 cm, width = 7 cm]{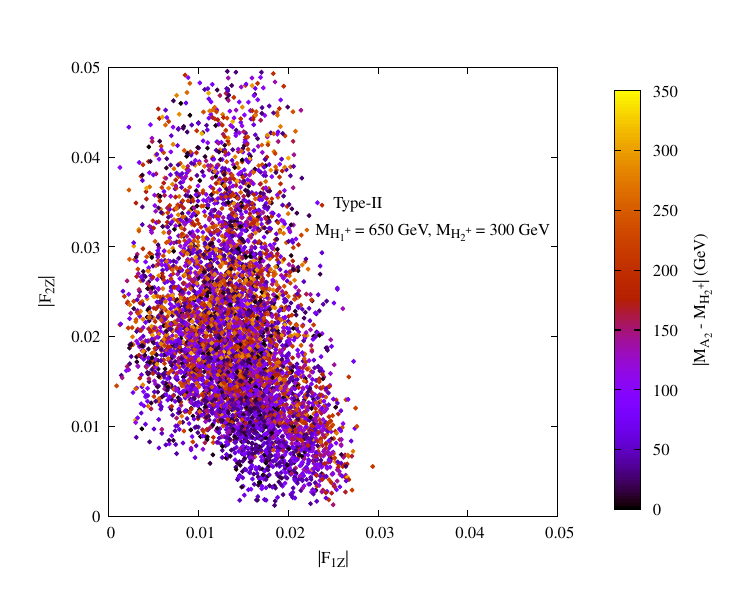}}
\subfigure[]{
\includegraphics[height = 5 cm, width = 7 cm]{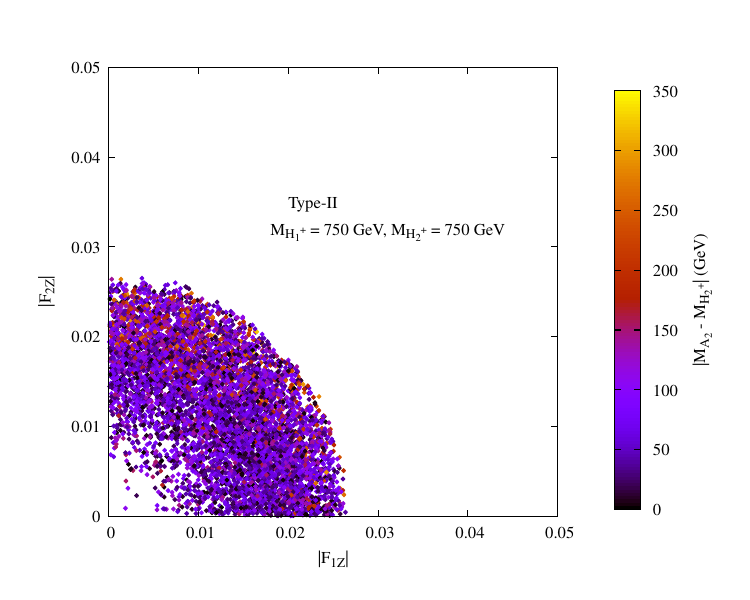}} 
}
\caption{$|F_{2Z}|$ vs. $|F_{1Z}|$ plots for different values of $M_{H_1^+}$ and $M_{H_2^+}$ in Type-II 3HDM.}
\label{FZ-II}
\end{figure}

\begin{figure}[htpb!]{\centering
\subfigure[]{
\includegraphics[height = 5 cm, width = 7 cm]{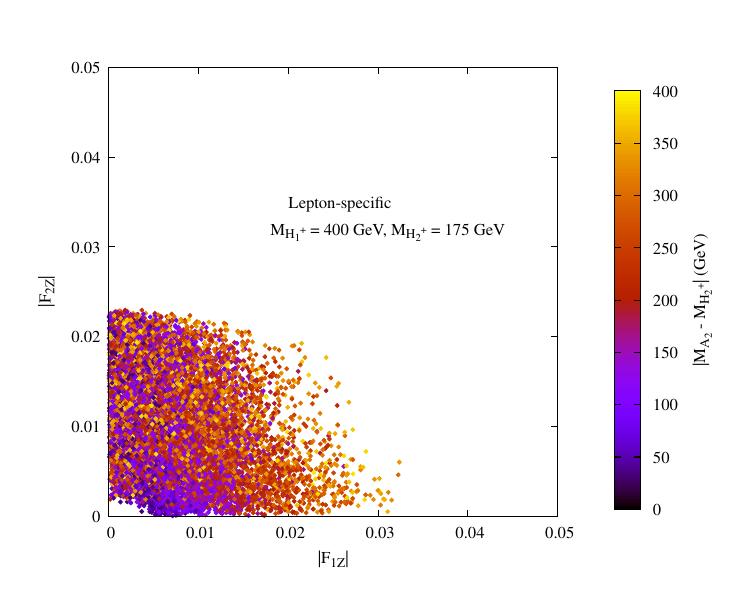}}
\subfigure[]{
\includegraphics[height = 5 cm, width = 7 cm]{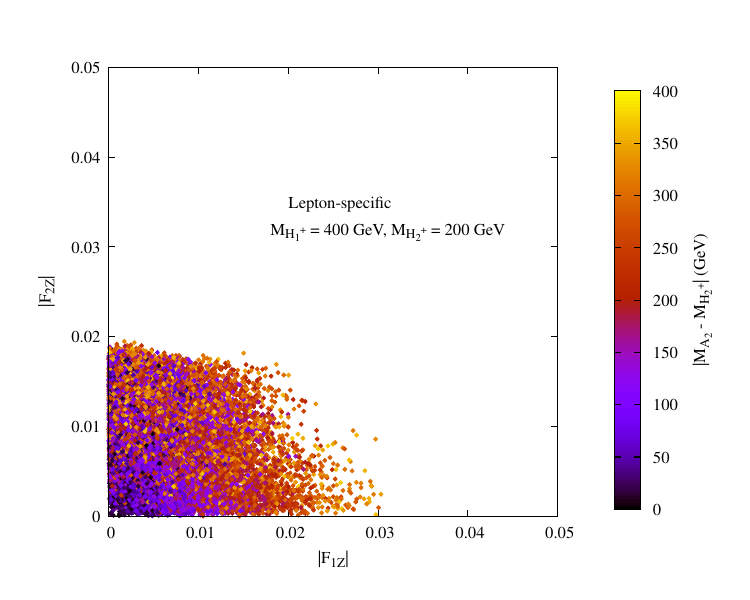}} \\
\subfigure[]{
\includegraphics[height = 5 cm, width = 7 cm]{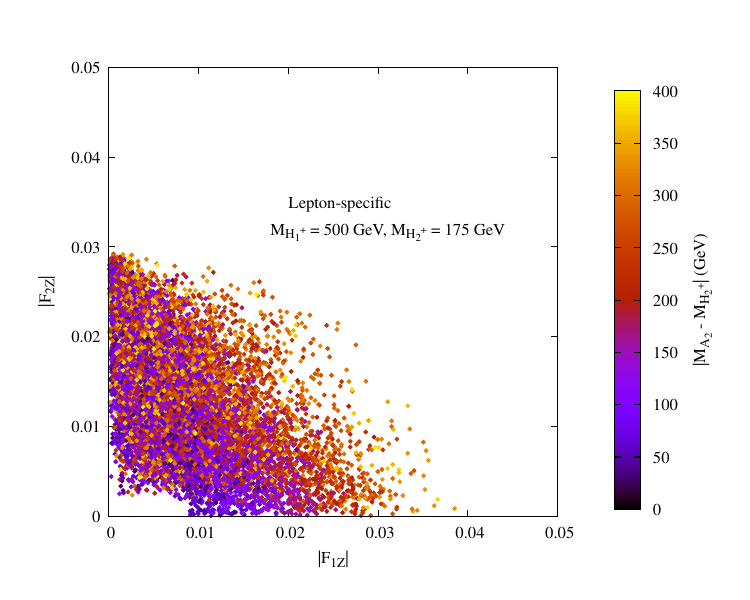}} 
\subfigure[]{
\includegraphics[height = 5 cm, width = 7 cm]{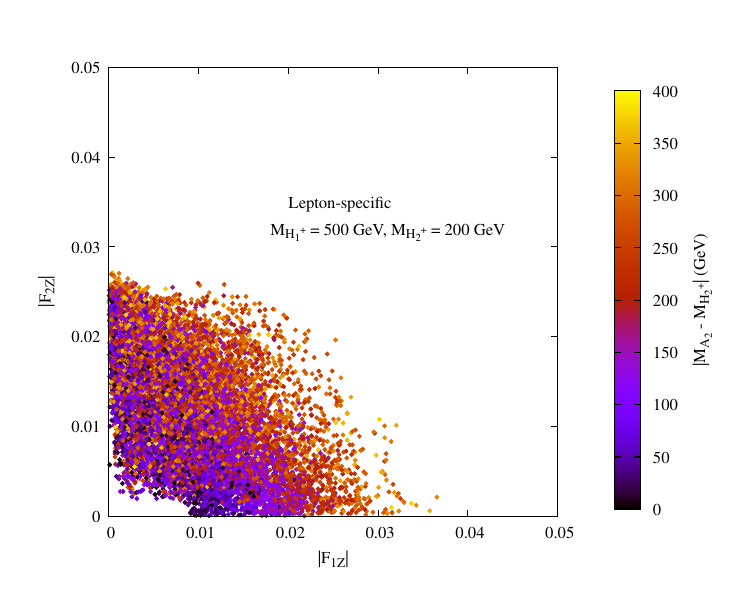}} \\
\subfigure[]{
\includegraphics[height = 5 cm, width = 7 cm]{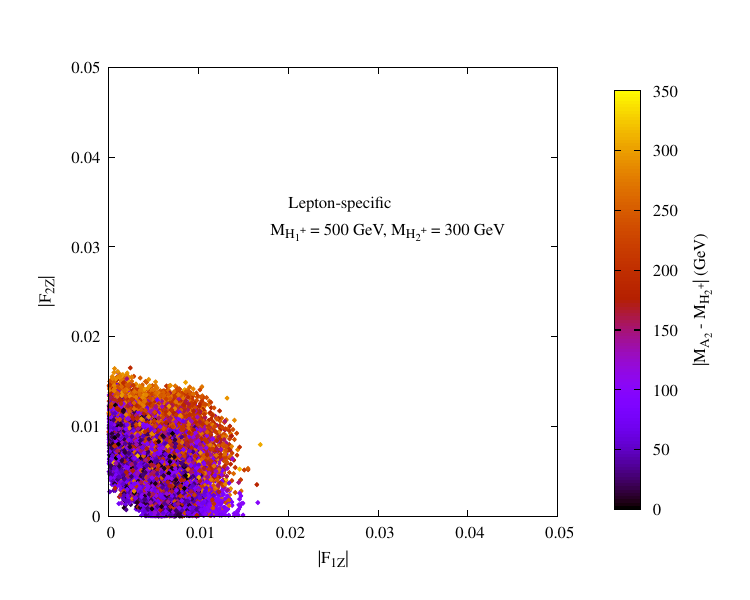}} 
\subfigure[]{
\includegraphics[height = 5 cm, width = 7 cm]{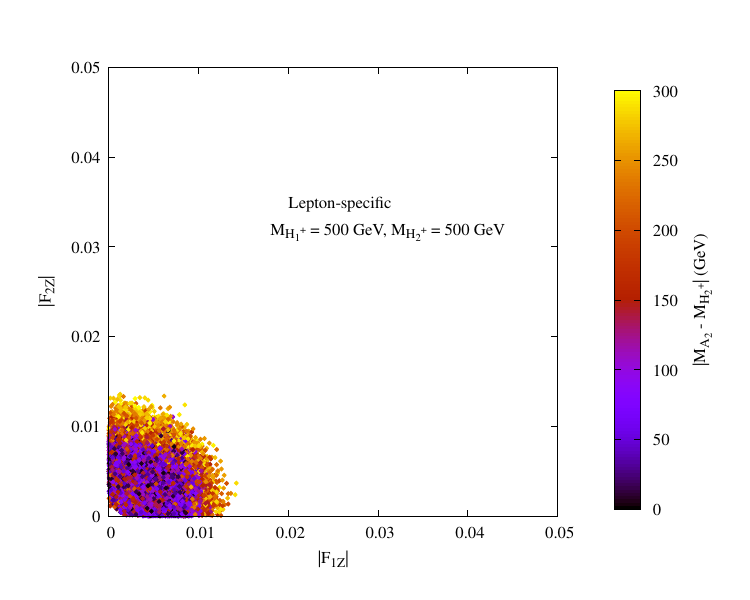}} \\
\subfigure[]{
\includegraphics[height = 5 cm, width = 7 cm]{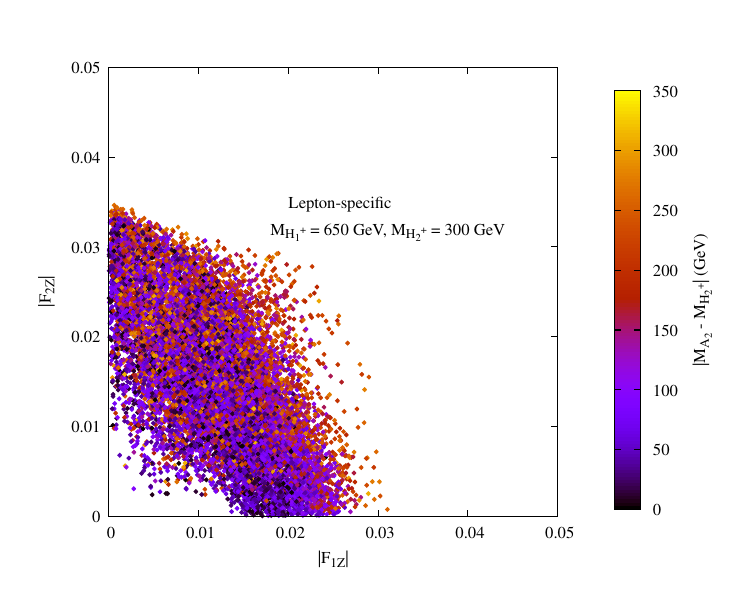}}
\subfigure[]{
\includegraphics[height = 5 cm, width = 7 cm]{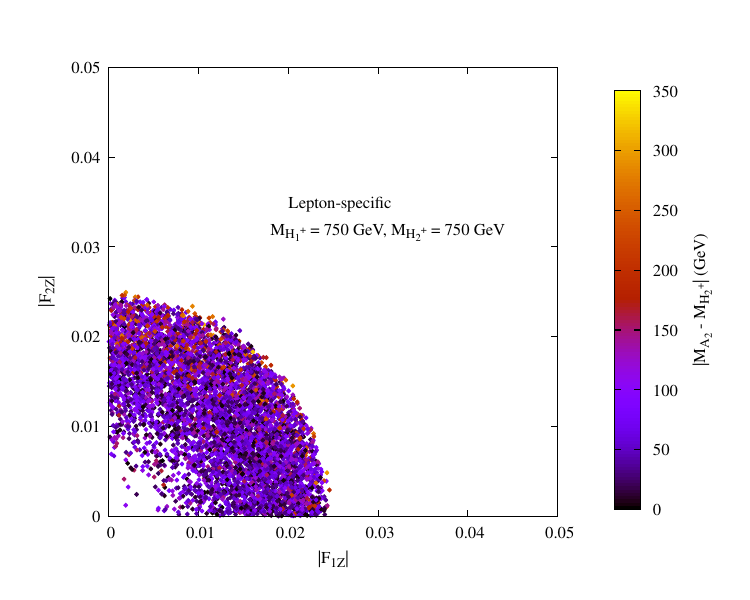}} 
}
\caption{$|F_{2Z}|$ vs. $|F_{1Z}|$ plots for different values of $M_{H_1^+}$ and $M_{H_2^+}$ in lepton-specific 3HDM.}
\label{FZ-LS}
\end{figure}

\begin{figure}[htpb!]{\centering
\subfigure[]{
\includegraphics[height = 5 cm, width = 7 cm]{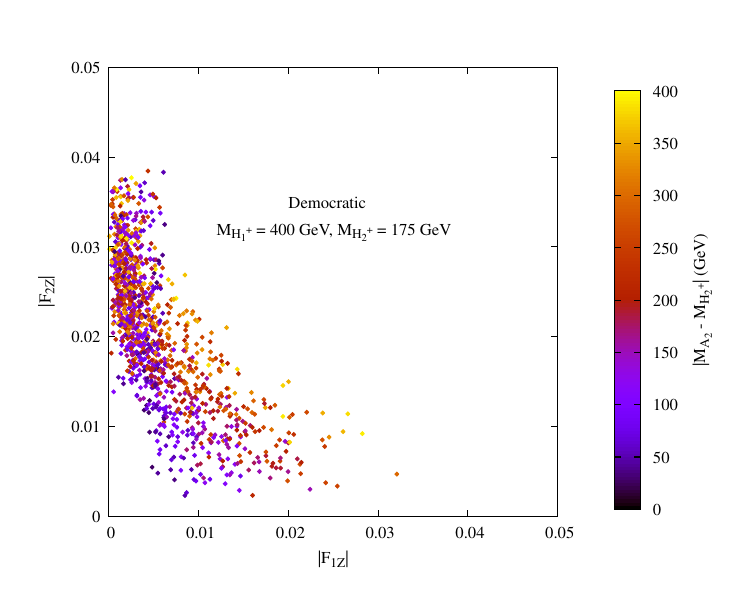}}
\subfigure[]{
\includegraphics[height = 5 cm, width = 7 cm]{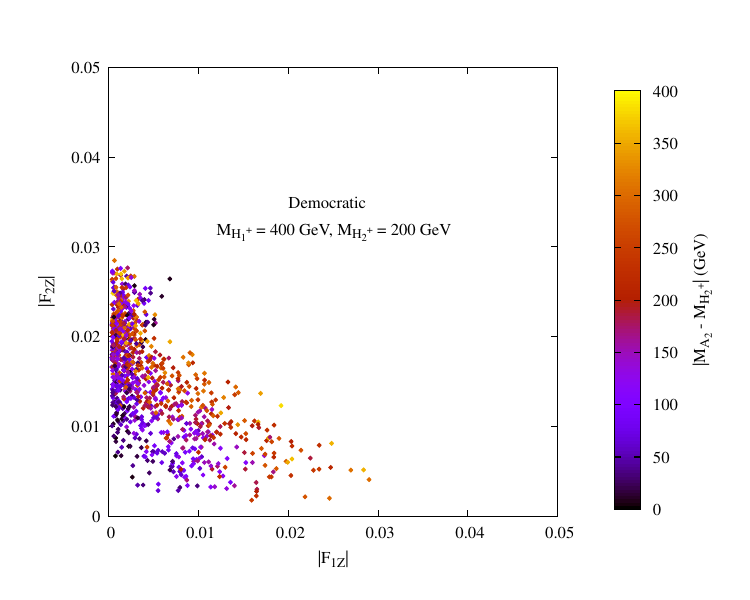}} \\
\subfigure[]{
\includegraphics[height = 5 cm, width = 7 cm]{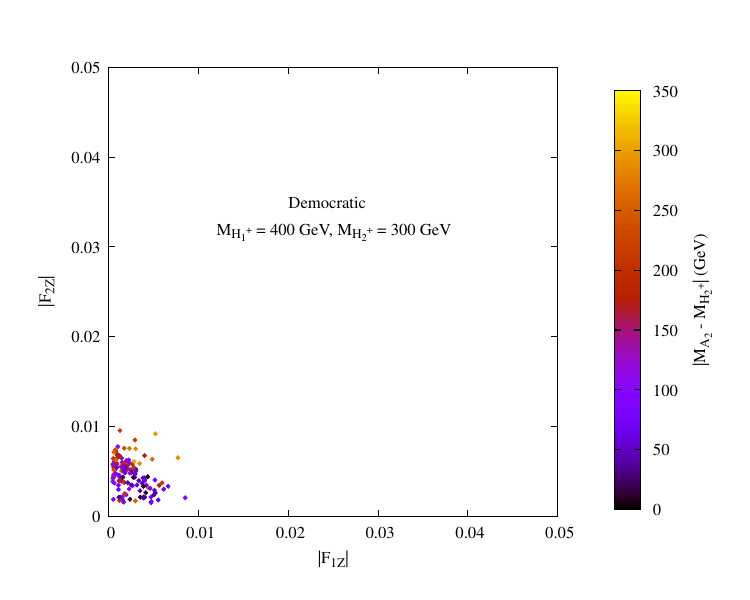}}
\subfigure[]{
\includegraphics[height = 5 cm, width = 7 cm]{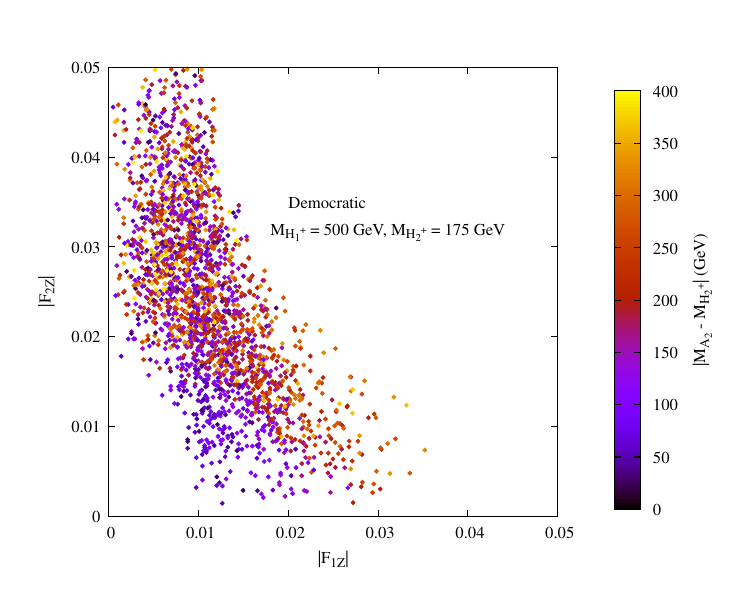}} \\
\subfigure[]{
\includegraphics[height = 5 cm, width = 7 cm]{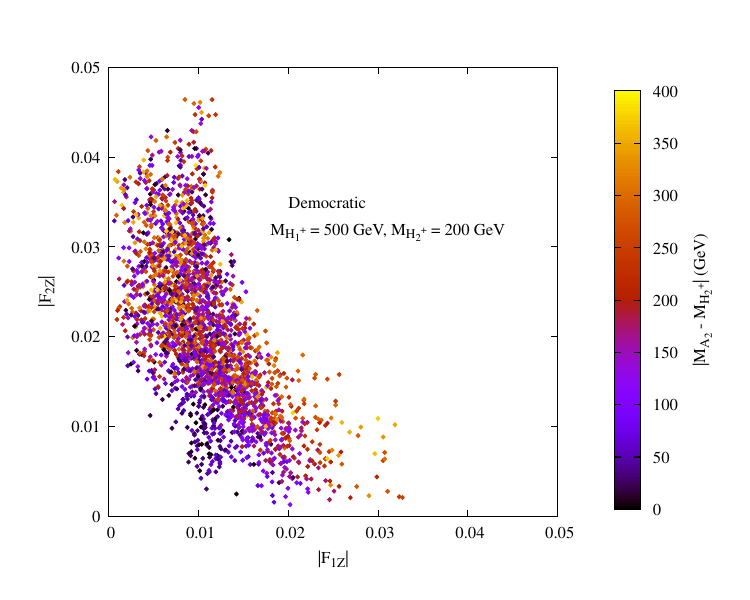}}
\subfigure[]{
\includegraphics[height = 5 cm, width = 7 cm]{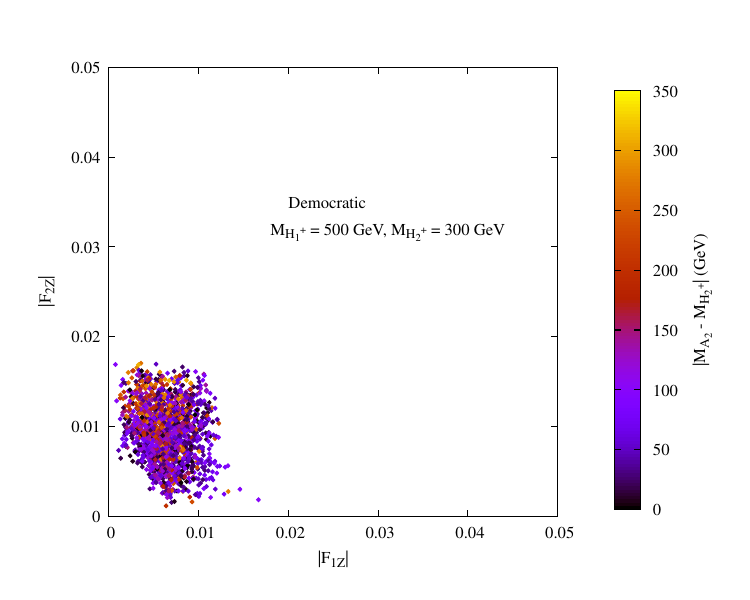}} \\
\subfigure[]{
\includegraphics[height = 5 cm, width = 7 cm]{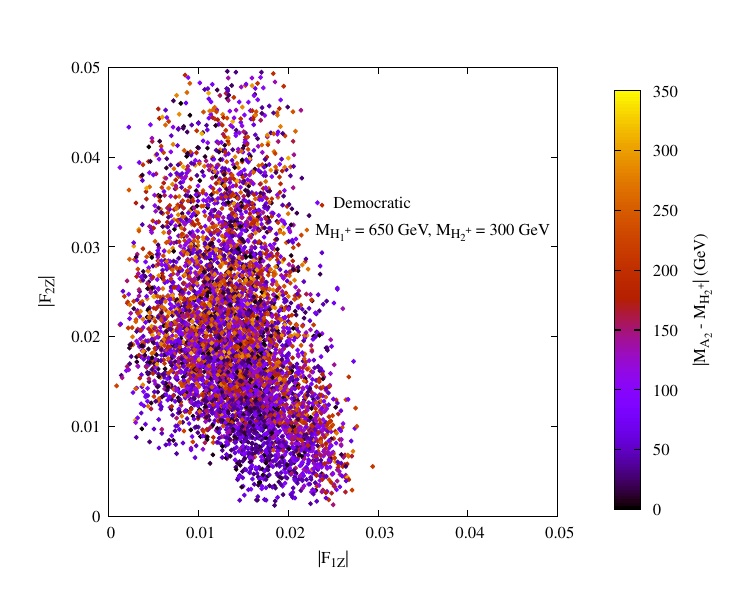}}
\subfigure[]{
\includegraphics[height = 5 cm, width = 7 cm]{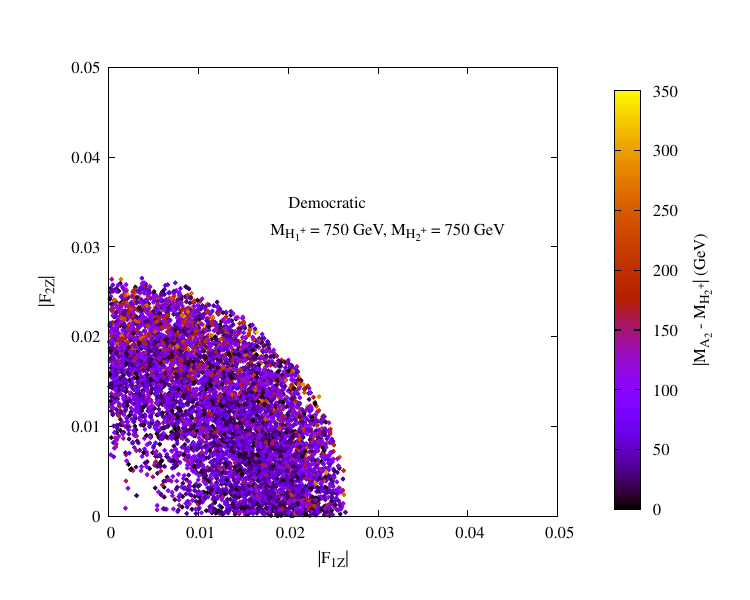}} 
}
\caption{$|F_{2Z}|$ vs. $|F_{1Z}|$ plots for different values of $M_{H_1^+}$ and $M_{H_2^+}$ in Democratic 3HDM.}
\label{FZ-Dem}
\end{figure}

Fig.\ref{FZ-II}-\ref{FZ-Dem} depict 2D color-coded scatter plots in $|F_{1Z}|-|F_{2Z}|$ plane with the colour bar indicating the scalar mass splitting $|M_{A_2}-M_{H_2^+}|$. As can be seen from the plots, the mass splitting $|M_{A_2}-M_{H_2^+}|$ plays a crucial role in determining the strength of one-loop contributions to $H_{1,2}^+ W^- Z$ vertices. The strong sensitivity to the scalar mass splittings is expected due to the custodial symmetry breaking effects. If the splitting is large, the custodial symmetry breaking effects get amplified, producing larger loop contributions. Therefore for the larger mass splittings, $|F_{1(2)Z}|$ are large for all three types of 3HDMs, thereby the yellow points populate the upper parts of the $|F_{1Z}|-|F_{2Z}|$  plane in Fig.\ref{FZ-II}-\ref{FZ-Dem}.

The scatter plots in $|F_{1Z}|-|F_{2Z}|$ plane exhibit a strong positive correlation for all benchmark choices in all three types of 3HDMs. In particular, the parameter points with larger values of |$F_{1Z}|$  systematically correspond to larger values of $F_{2Z}|$. The logic behind this correlation is that, both the contributions to |$F_{1Z}|$ and $F_{2Z}|$ originate from the same diagrams involving same particles. Thus the mass splittings affect both the form factors in a correlated manner, not independently.

With the increase in charged Higgs masses, the strength of form factor decreases and this feature remains same for all variants of 3HDMs. Thus larger values of $M_{H_1^+}$ and $M_{H_2^+}$, the allowed regions in the 
$|F_{1Z}|-|F_{2Z}|$ planes gradually shrink toward the origin. For lighter charged Higgs masses, due to smaller loop suppression, the form factors attain higher values. Thus for the scenarios with $M_{H_{1(2)}^+} \sim 175-200$ GeV, the allowed points cluster towards the upper edges of the scatter plots,  where as increase of the masses to 300 GeV or beyond leads to a noticeable reduction in the attainable magnitudes of form factors.

The basic nature of the 2D colour-coded scatter plots in $|G_{1Z}|-|G_{2Z}|$ plane for the same set of charged Higgs masses and mass splitting $|M_{A_2}-M_{H_2^+}|$ mimics that of the plots in $|F_{1Z}|-|F_{2Z}|$ plane, apart from the magnitude. The magnitude of $G_{1(2)Z}$ is approximately one order of magnitude less than the magnitude of $F_{1(2)Z}$ due to the momentum-suppressed nature of the Lorentz structures.

It is evident from the numerical analysis that the magnitudes of the form factors $|F_{1Z}|$ and $|F_{2Z}|$ strongly depend on the charged Higgs masses $M_{H_1^+}$ and $M_{H_2^+}$ and mildly sensitive on the variants of 3HDMs. For $M_{H_2^+}\sim 175-200$ GeV and $M_{H_2^+} \sim 400-500$ GeV, $|F_{1Z}|$ and $|F_{2Z}|$ span the range: $|F_{1Z}|, |F_{2Z}| \sim 1 \times 10^{-2}$ - $5 \times 10^{-2}$, with the upper end of this range corresponding to scenarios with sizable charged-neutral scalar mass splittings. With the increase of $M_{H_2^+} \sim 300$ GeV, the magnitudes decreases as: $|F_{1Z}|, |F_{2Z}| \sim 2 \times 10^{-2}$. For heavy scalars, {\em i.e.} $M_{H_1^+} \geq 650$ GeV and $M_{H_2^+} \geq 500$ GeV, the magnitudes of $|F_{1Z}|, |F_{2Z}|$ further decrease below $10^{-2}$. For $M_{H_1^+} = M_{H_2^+}$, the plots show a symmetric nature of the allowed parameter space in the $|F_{1Z}|$-$|F_{2Z}|$ plane. This symmetry arises because the loop amplitudes depend only on mass differences and mixing combinations, which become invariant under the interchange of the two charged Higgs states in the degenerate limit. Consequently, no charged scalar is kinematically or dynamically preferred, leading to symmetric parameter distributions.

Although the scalar mass spectrum primarily controls the strength of the one-loop contributions, there is a subdominant but non-negligible effect coming from different variants of 3HDMs. This is by the virtue of different Yukawa couplings between the charged Higgses and fermions, which enter into the diagrams and  contribute to the fermionic counter part of one-loop form factors. Since in type-II and democratic 3HDM, $\Phi_1$ and $\Phi_2$ couple with down-type and up-type quarks respectively and the coupling with lepton is different, the strength of the fermionic counterpart contributing to the form factors appears to be almost same. That is why, the plots in Fig.\ref{FZ-II} and Fig.\ref{FZ-Dem} depict the same nature and magnitude of the form factors for a given value of charged Higgs masses and charged-neutral scalar mass splitting. On the contrary, for lepton-specific 3HDM, where both up-type and down-type quarks couple with $\Phi_2$, there is slight suppression in the maximal attainable values of the form factors compared to the other two aforementioned variants. 

\section{Observability at the LHC}\label{collider}

This section discusses possible signals at the LHC that can be potential probes of the $H_{1,2}^+ W^- Z$ vertex. It is mentioned at the outset that the $g b \to t H^+$ process can have a sizeable production cross section in 2HDMs. And a subsequent $H^+ \to W^+ Z$ decay can shed light on the  $H^+ W^- Z$ coupling. This has been studied before in 2HDMs [Gauhar, Kanemura] and its extensions \cite{Moretti:2015tva,Chakrabarty:2020msl}. It is reported that $\sigma_{gb} \times \text{BR}_{H^+ \to W^+ Z}$ can be $\mathcal{O}(0.1)$ fb for $M_{H^+}< M_t + M_b$ and as large as $\mathcal{O}$(100 fb) for $M_{H^+} > M_t + M_b$. 

Signatures of 3HDMs have been studied at the colliders, a partial list of which is \cite{Moretti:2015bvg, Bandyopadhyay:2016fda, Arhrib:2015hoa, Keus:2015hva, 
Ghosh:2020qbx, Boto:2021gqf, Akeroyd:2017mvg, Chowdhury:2021iag, 
Mondal:2022xpa,Coleppa:2025qst}. In this study, we focus on $H_{1,2}^+$ production using $WZ$ fusion as shown below in Fig.\ref{vbf}. The motivation is to probe the radiative $H_{1,2}^+ W^- Z$  vertices at the production level itself. 
\begin{figure}[htpb!]{\centering
\includegraphics[height = 6 cm, width = 7 cm]{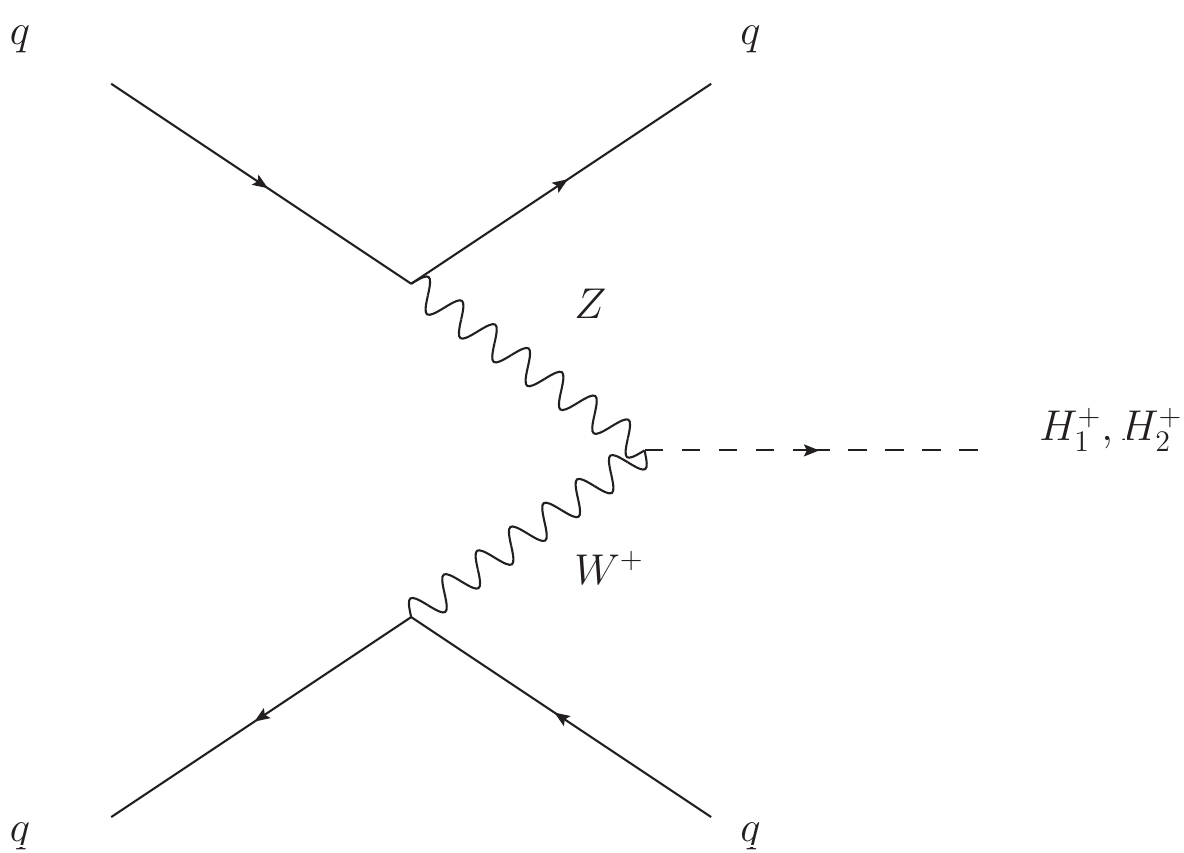}}
\caption{}
\label{vbf}
\end{figure} 
The fact that we have numerical estimates of the one-loop form factors for both $H_1^+$ and $H_2^+$ at this point gives us the scope to compare the corresponding $WZ$ production rates. We implement the model in \texttt{FeynRules}~\cite{Christensen:2008py,Alloul:2013bka} and extract an Universal FeynRules Output~\cite{Degrande:2011ua}. The same is passed on to \texttt{MadGraph5 aMC@NLO}~\cite{Frederix:2018nkq} and $p p \to H_{1,2}^\pm j j$ is subsequently generated. It is added that the \texttt{NN23LO1}
parton distribution function (PDF) is used to compute the cross
sections at the 14 TeV LHC\footnote{An LO PDF such as NN23LO1 must be used for computing LO cross
sections like what we have done here.}. Moreover, the cuts $p_T^j >$ 20 GeV, $|\eta_j| < 5$ and $\Delta R_{jj} > 0.4$ are applied at the time of generation. We look for the subsequent decays $H^+_{1,2} \to t \bar{b},~\tau \bar{\nu},~W^+ Z$. The $\sigma \times \text{BR}$ values are accordingly estimated for 
the lepton specific 3HDM for specific values of the charged Higgs masses. We choose $(M_{H_1^+},M_{H_2^+})$ = (175 GeV, 175 GeV), (200 GeV, 200 GeV), (500 GeV, 500 GeV) for illustration. 
\begin{figure}[htpb!]
{\centering
\includegraphics[height = 7 cm, width = 7 cm]{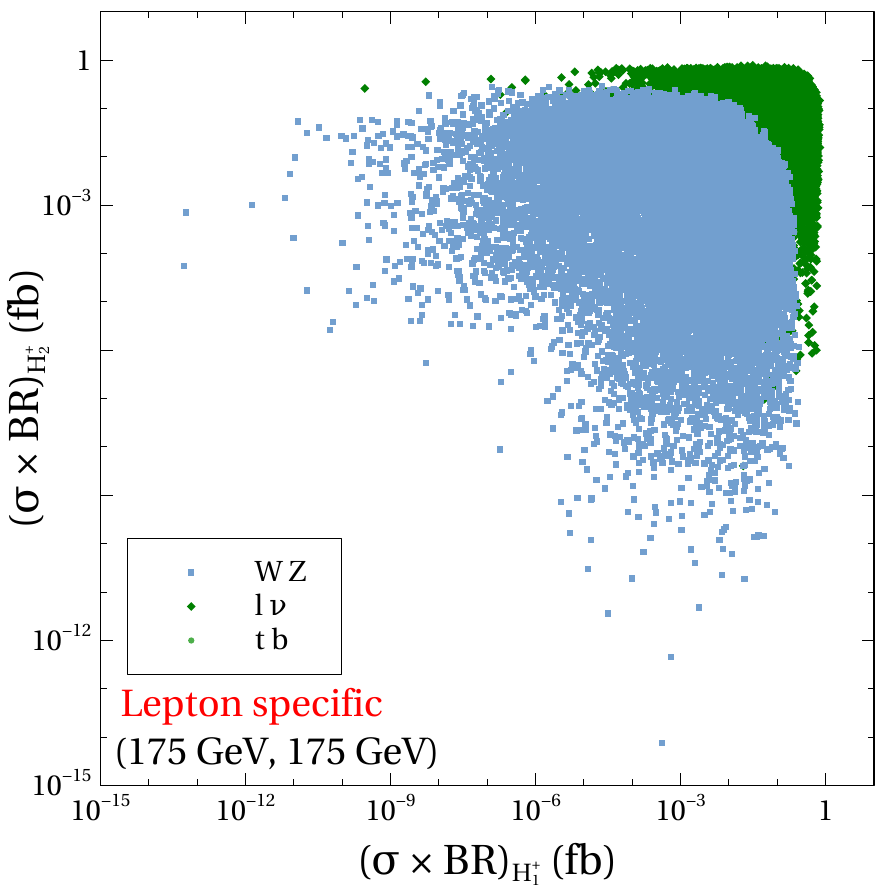}
\includegraphics[height = 7 cm, width = 7 cm]{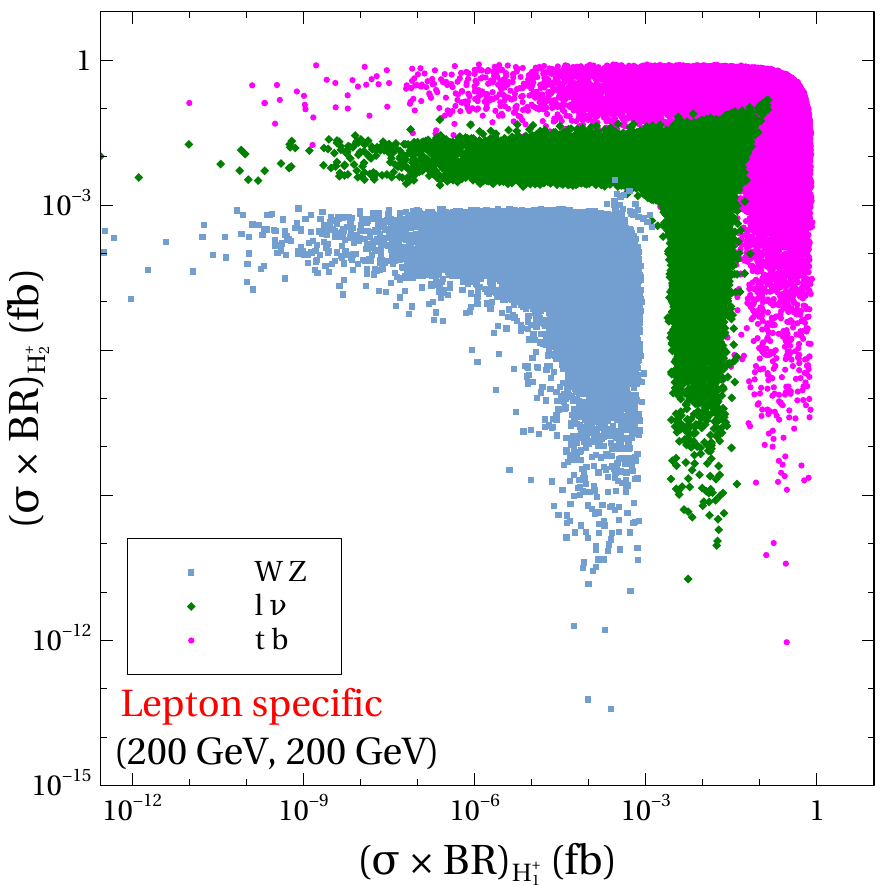} \\
\includegraphics[height = 7 cm, width = 7 cm]{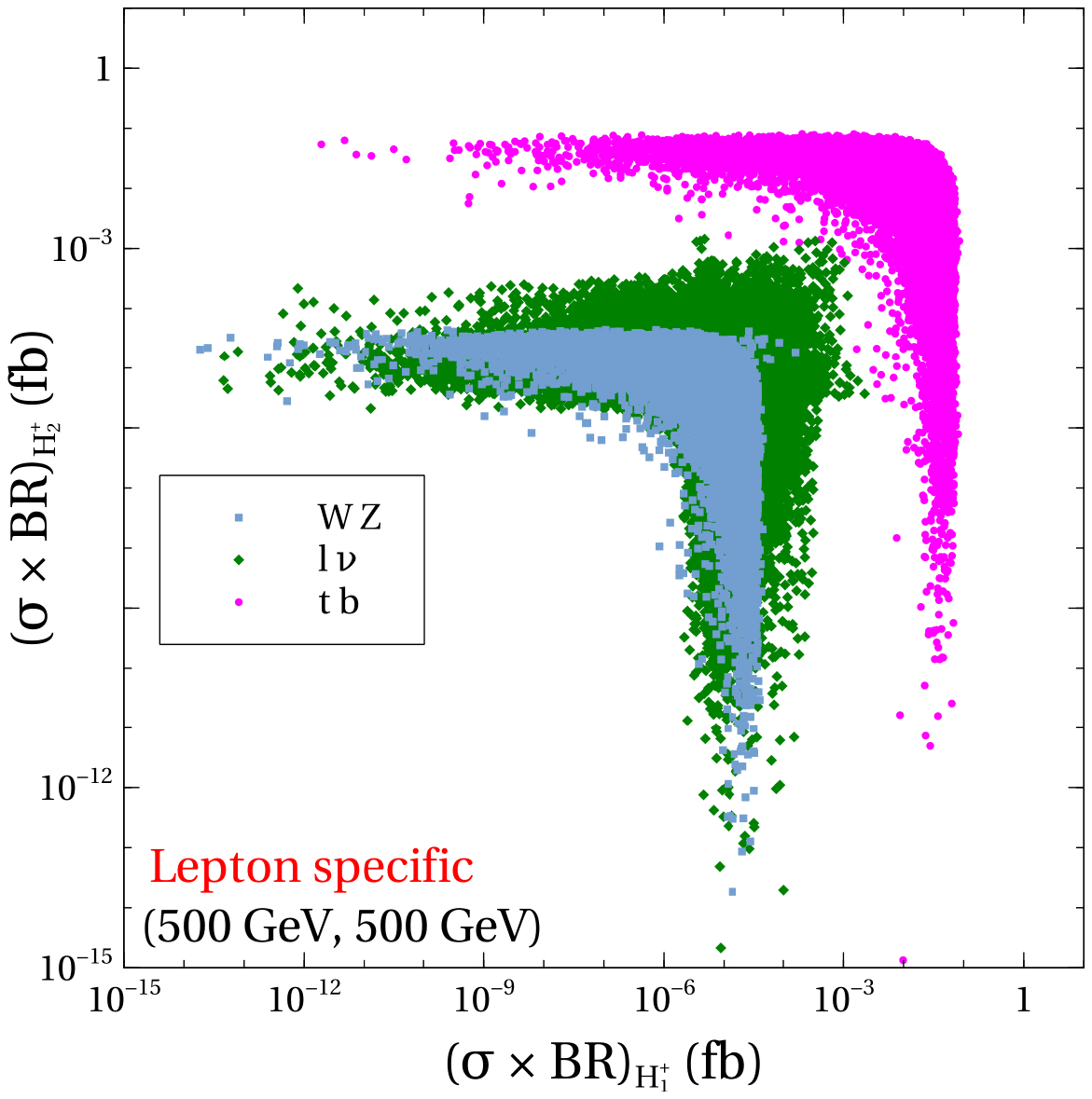}
}
\caption{$\sigma \times$ BR values for $H_1^+$ and $H_2^+$. The color coding is shown in the legends.}
\label{ls_coll1}
\end{figure}
The values of $\sigma \times $ BR are shown in a scatter plot for the allowed parameter points in Fig.\ref{ls_coll1}. The $H_{1,2}^+ \to t \bar{b}$ decay is kinematically blocked for $H^+_{1,2}$ having mass 175 GeV. The $pp \to H_{1,2}^\pm \to \tau \nu$ cross section dominates over 
$pp \to H_{1,2}^\pm \to W^\pm Z$ for most of the parameter space and the former cross section can reach up to $\mathcal{O}$(0.1 fb). Despite the $tb$ mode opening up for $M_{H_1^+}=M_{H_2^+}$ = 200 GeV, it is seen that the $\tau \nu$ mode can still dominate. And this higher leptonic branching fraction is traced back to the enhanced leptonic Yukawas in the lepton-specific case. However, the $tb$ mode takes over for higher charged Higgs masses. This is ascertained in case of $M_{H_1^+}=M_{H_2^+}$ = 500 GeV wherein $pp \to H_{1,2}^\pm \to \tau \nu$ has a cross section not exceeding $\sim 10^{-3}$ fb. The main takeaway here is that charged Higgses with masses $\lesssim$ 200 GeV can give rise to a $jj+\tau_h + \met$ signature. This signature is interesting from the kinematic point of view given a single neutrino in the final state makes the charged Higgs mass potentially recontructible. In addition, the event count at the pre-cut level can also be $gtrsim$ 300 for an integrated luminosity $\mathcal{L}$ = 3000 fb$^{-1}$. 

Next, we study the prospects of $H_1^+ \to H_2^+ H_2,~H_2^+ A_2$ modes. One notes that the leading fermionic decay mode for $H_2$ and $A_2$ is 
$\tau\bar{\tau}$ for the lepton-specific 3HDM whenever the $t\bar{t}$ threshold remains closed. In addition, taking the $H_2^+ \to \tau \nu$ decay implies that $p p \to H_1^\pm \to H_2^\pm H_2 (A_2)$ can give rise to a $jj + 3\tau_h + \met$ final state. 
\begin{figure}[htpb!]
{\centering
\includegraphics[height = 7 cm, width = 7 cm]{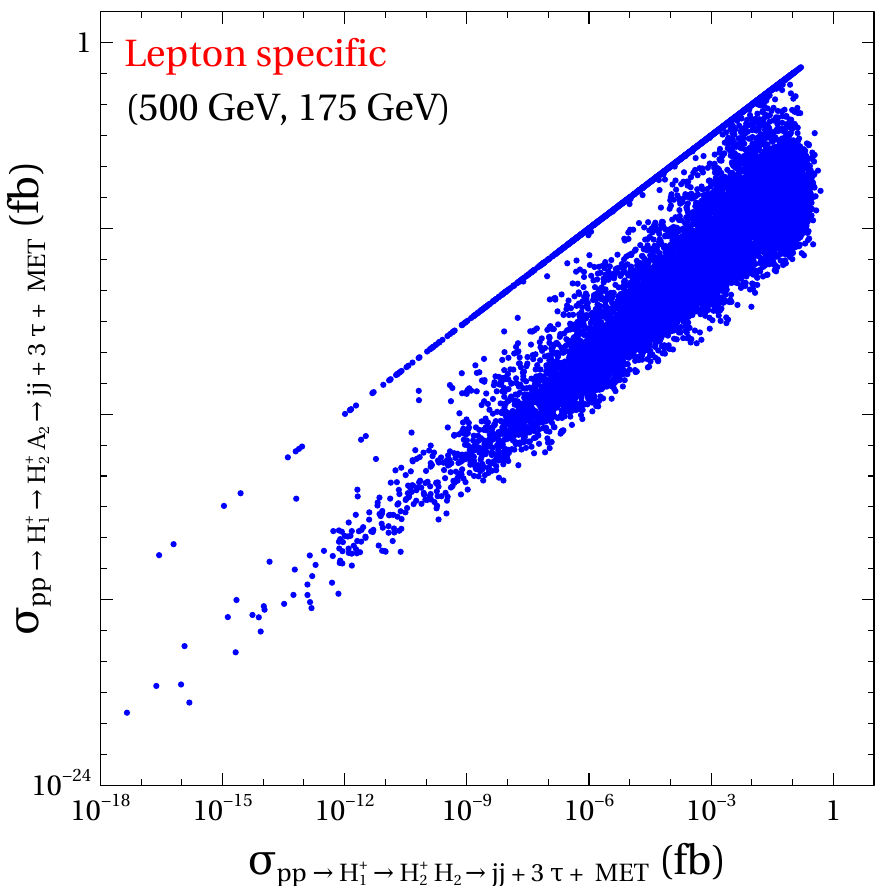}~~~
\includegraphics[height = 7 cm, width = 7 cm]{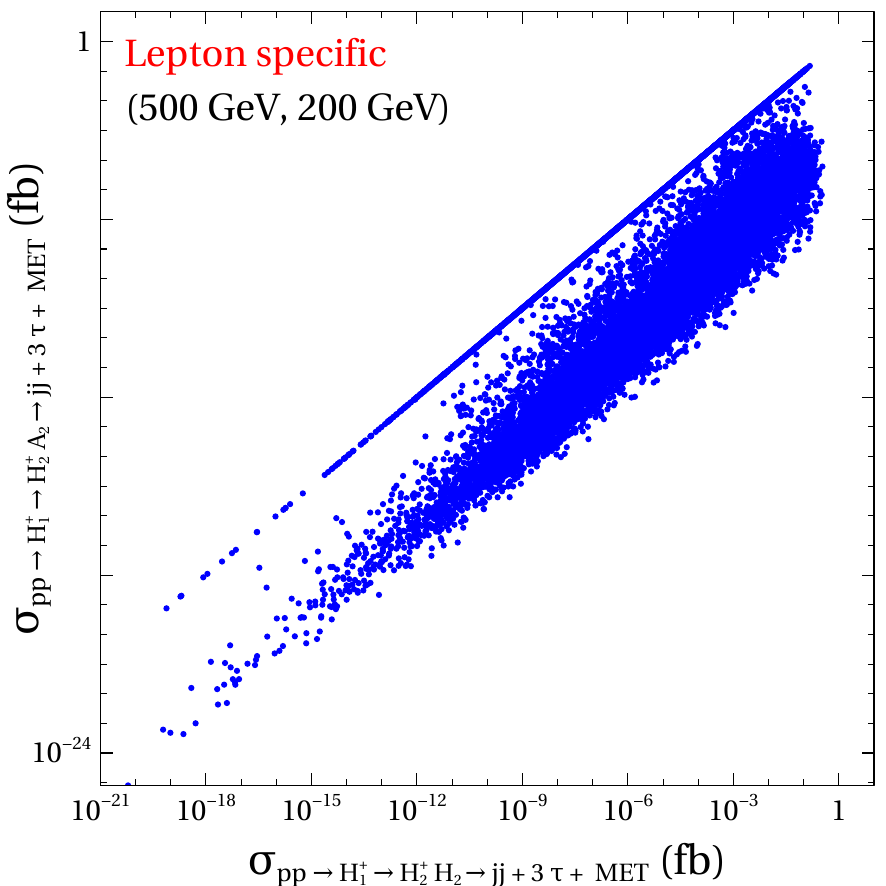}
}
\caption{$\sigma \times$ BR values for $H_1^+$ and $H_2^+$. The color coding is shown in the legends.}
\label{ls_coll2}
\end{figure}
The cross sections for the two aforesaid channels are shown in Fig.\ref{ls_coll2}. A couple of comments are in order. First, $M_{H_2} = M_{H_2^+}$ = 175 GeV, 200 GeV implies that $H_2 \to \tau \bar{\tau}$ is by far the most dominant mode. Secondly, $A_2$ is allowed to be heavier in our analysis and therefore $A_2 \to H_2 Z,~H_2^\pm W^\mp$ also compete with $A_2 \to \tau \bar{\tau}$. Overall, the $jj + 3\tau_h + \met$ originating from $H_2$ is seen to have a cross section $\sim \mathcal{O}$(0.1 fb) despite the multiple cascades involved. The corresponding event count can be $\sim 300$. More importantly, such a signal is kinematically rich and gives the scope to reconstunct both charged Higgs masses thereby being a potential smoking gun of a 3HDM.

\section{Summary and conclusions}
\label{summary}

The weak isospin symmetry of the kinetic terms in multi-Higgs doublet models disallows $H^+ W^- Z$ interactions
at the tree level. However, isospin-breaking effects stemming from the scalar and Yukawa sectors generate 
this vertex at the one-loop level. It is therefore interesting to
estimate the size of this vertex in such models and probe its
observability at the energy frontier. In this study, we have quantified 
the strength of the $H_{1,2}^+ W^- Z$ vertices for flavour conserving 3HDMs. And we have chosen the Type-II, lepton specific and democratic Yukawa structures for illustration. Given the large number of one-loop diagrams, we have arranged them into UV-finite and gauge invariant subsets tagged by the scalar trilinear coupling involved. For a prototype subset, UV-finiteness is explicitly checked for and a numerical validation of Ward identity for the $H_1^+ W^- \gamma$ amplitude is carried out as a check of gauge invariance. 

It is observed that demanding appropriate relations between mixing angles in the scalar sector and scalar masses renders $\Delta T = 0$ thereby making the 3HDM parameter space easier to handle numerically. In addition, we have also pointed out that such relations crucially help segregate the excess contribution to the one-loop form factors over and above the 2HDM. We have folded in constraints coming from perturbative unitarity, Higgs to diphoton decay and $B \to X_s \gamma$ branching ratio in the ensuing scans and computed the one-loop form factors for the surviving parameter points.  
As a highlight of the numerical results, $F_{1Z}$ and $F_{2Z}$ for the lepton specific 3HDM can be $\simeq 0.045$ and $\simeq 0.03$ for $(M_{H_1^+},M_{H_2^+})$ = (500 GeV,200 FeV) respectively, a value sizeably enhanced compared to the $\simeq 0.01$ reported for 2HDMs. 

We have also explored the possibility of probing the $H^+_{1,2}W^- Z$ vertices at the 14 TeV LHC through $WZ$ fusion. The bare $\sigma \times \text{BR}$ values are estimated for $p p \times H^\pm_{1,2}$ with $H^+_{1,2} \to t b,~\tau \nu,~W^+ Z$ for the 3HDM types taken. The lepton specific 3HDM fares particularly well in the $H^+_{1,2} \to \tau \nu$ channel given the enhanced leptonic Yukawas. The $\sigma \times \text{BR}$ value can be $\sim$ 0.1 fb in this case leading to $\mathcal{O}(100)$ events for an integrated luminosity 3000 fb$^{-1}$, modulo kinematical cuts. On the kinematic side, that there is a single neutrino in the signal makes the charged Higgs mass fully reconstructible for this signal. In addition to the above, the lepton specific 3HDM is also shown to accommodate $H_1^+ \to H_2^+ H_2$ with a sizeable branching ratio. With $H_2^+ \to \tau \nu$ and $H_2 \to \tau \tau$, one can obtain a $jj + 3\tau_h + \met$ final state if the hadronic decays of $\tau$ are tagged. This signal therefore gives a scope to reconstruct the masses of both $H_1^+$ and $H_2^+$ up to $\tau_h$ combinatorics thereby being a potential smoking gun signature of a 3HDM. Moreover, a $\sigma \times \text{BR}$ value is $\sim$ 0.1 fb makes it comparable to the $p p \to H_{1,2}^+ j j \to \tau j j + \met$  in terms of event yield.

\section{Acknowledgements}

IC acknowledges support from Department of Science and Technology, Govt. of India, un-
der grant number IFA18-PH214 (INSPIRE Faculty Award). NC acknowledges support from
Department of Science and Technology, Govt. of India, under grant number IFA19-PH237
(INSPIRE Faculty Award). NC also thanks Arani Chakravarti for an useful computational help.

\section{Appendix}

\subsection{Scalar squared mass matrices}

\besub
\bea
(M_{\text{even}}^2)_{11} &=& \frac{\lambda_1 v_1^3 + m_{12}^2 v_2+m_{13}^2 v_3}{v_1}  \nonumber \\
(M_{\text{even}}^2)_{12} &=& v_1 v_2 (\lambda_{12}+\lambda_{12}^{'}+\lambda_{12}^{''})-m_{12}^2 \nonumber \\
(M_{\text{even}}^2)_{13} &=& v_1 v_3 (\lambda_{13}+\lambda_{13}^{'}+\lambda_{13}^{''})-m_{13}^2 \nonumber \\
(M_{\text{even}}^2)_{22} &=& \frac{\lambda_2 v_2^3+m_{12}^2 v_1+m_{23}^2 v_3}{v_2} \nonumber \\
(M_{\text{even}}^2)_{23} &=& v_2 v_3 (\lambda_{23}+\lambda_{23}^{'}+\lambda_{23}^{''})-m_{23}^2 \nonumber \\
(M_{\text{even}}^2)_{33} &=& \frac{\lambda_3 v_3^3+m_{13}^2 v_1+m_{23}^2 v_2}{v_3}
\eea
\eesub

\besub
\bea
(M_{\text{odd}}^2)_{11} &=& \frac{-\lambda_{12}^{''} v_1 v_2^2+v_3 (m_{13}^2-\lambda_{13}^{''} v_1 v_3)+m_{12}^2 v_2}{v_1} \nonumber \\
(M_{\text{odd}}^2)_{12} &=& -m_{12}^2 + \lambda_{12}^{''} v_1 v_2 \nonumber \\
(M_{\text{odd}}^2)_{13} &=& -m_{13}^2 + \lambda_{13}^{''} v_1 v_3 \nonumber \\
(M_{\text{odd}}^2)_{22} &=& \frac{m_{12}^2 v_1 - \lambda_{12}^{''} v_1^2 v_2 + v_3 (m_{23}^2 - \lambda_{23}^{''} v_2 v_3)}{v_2} \nonumber \\
(M_{\text{odd}}^2)_{23} &=& -m_{23}^2 + \lambda_{23}^{''} v_2 v_3 \nonumber \\
(M_{\text{odd}}^2)_{33} &=& \frac{m_{13}^2 v_1 + m_{23}^2 v_2-(\lambda_{13}^{''} v_1^2 + \lambda_{23}^{''} v_2^2)v_3}{v_3} 
\eea
\eesub

\besub
\bea
(M_{\text{ch}}^2)_{11} &=& -\frac{- 2 m_{12}^2 v_2 + (\lambda_{12}^{'} + \lambda_{12}^{''}) v_1 v_2^2 - 2 m_{13}^2 v_3 + (\lambda_{13}^{'} + \lambda_{13}^{''}) v_1 v_3^2}{2 v_1} \nonumber \\
(M_{\text{ch}}^2)_{12} &=&  -m_{12}^2 + \frac{1}{2} (\lambda_{12}^{'} + \lambda_{12}^{''}) v_1 v_2 \nonumber \\
(M_{\text{ch}}^2)_{13} &=& -m_{13}^2 + \frac{1}{2} (\lambda_{13}^{'} + \lambda_{13}^{''}) v_1 v_3 \nonumber \\
(M_{\text{ch}}^2)_{22} &=& -\frac{- 2 m_{12}^2 v_1 + (\lambda_{12}^{'} + \lambda_{12}^{''}) v_1 v_2^2 - 2 m_{23}^2 v_3 + (\lambda_{23}^{'} + \lambda_{23}^{''}) v_2 v_3^2}{2 v_2} \nonumber \\
(M_{\text{ch}}^2)_{23} &=& -m_{23}^2 + \frac{1}{2} (\lambda_{23}^{'} + \lambda_{23}^{''}) v_2 v_3 \nonumber \\
(M_{\text{ch}}^2)_{33} &=& \frac{2 m_{13}^2 v_1 - (\lambda_{13}^{'} + \lambda_{13}^{''}) v_1^2 v_3 + 2 m_{23}^2 v_2 - (\lambda_{23}^{'} + \lambda_{23}^{''}) v_2^2 v_3}{2 v_3} 
\eea
\eesub

\subsection{One-loop diagrams}

\subsubsection{E-type amplitudes}

\begin{figure}[htpb!]{\centering
\subfigure[$E_1$]{
\includegraphics[height = 3 cm, width = 6 cm]{E1-eps-converted-to.pdf}} 
\subfigure[$E_2$]{
\includegraphics[height = 3 cm, width = 6 cm]{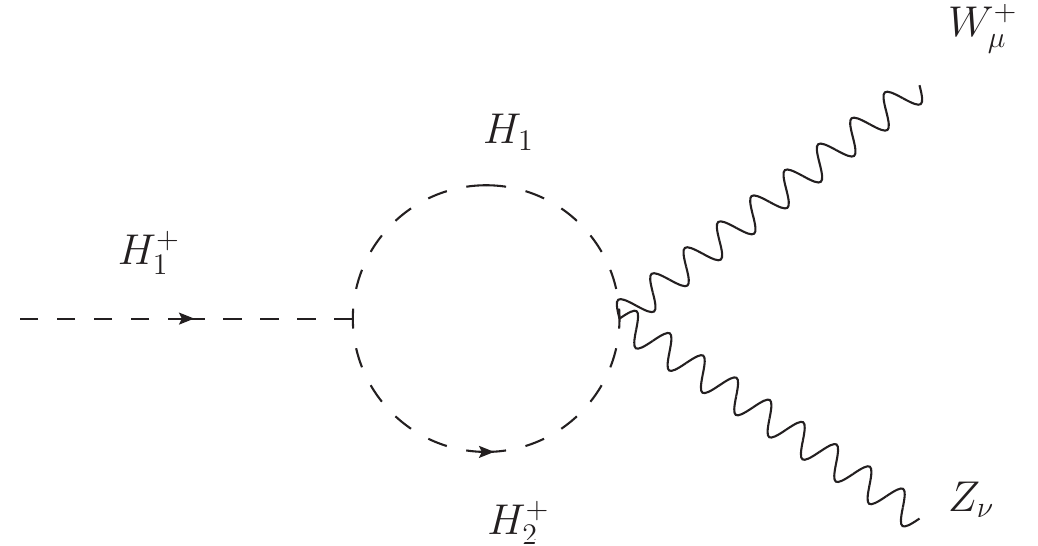}}\\
\subfigure[$E_3$]{
\includegraphics[height = 3 cm, width = 6 cm]{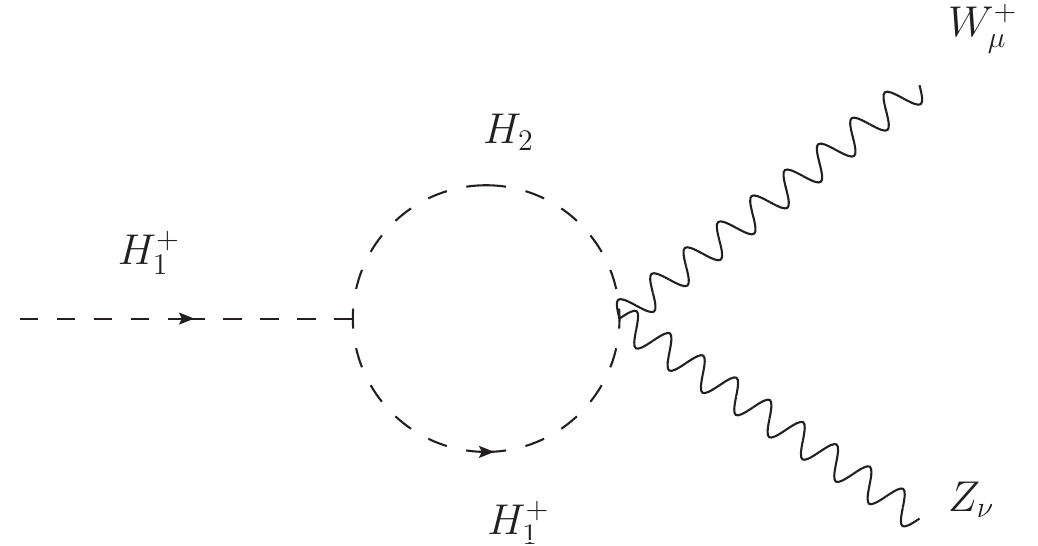}} 
\subfigure[$E_4$]{
\includegraphics[height = 3 cm, width = 6 cm]{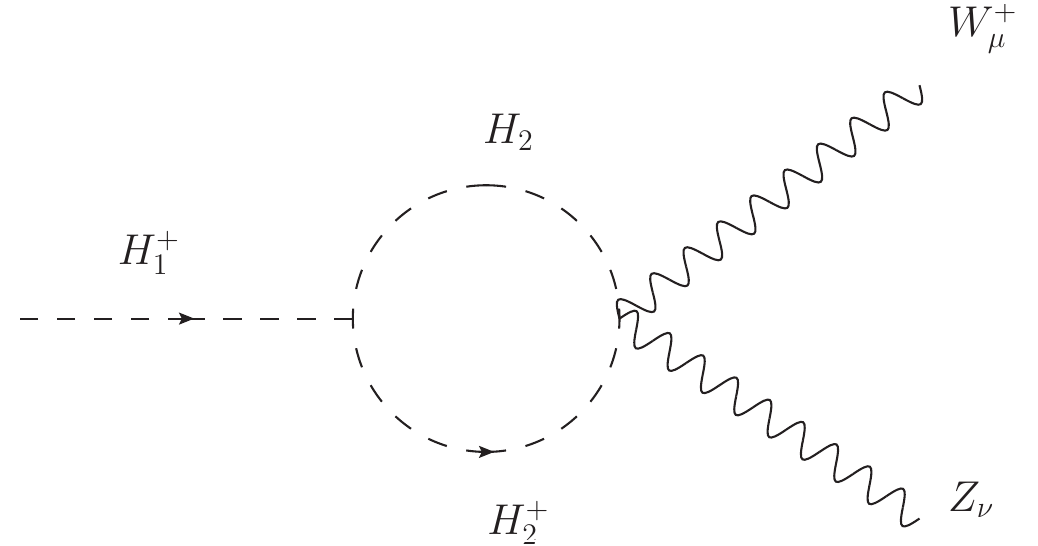}}
\subfigure[$E_5$]{
\includegraphics[height = 3 cm, width = 6 cm]{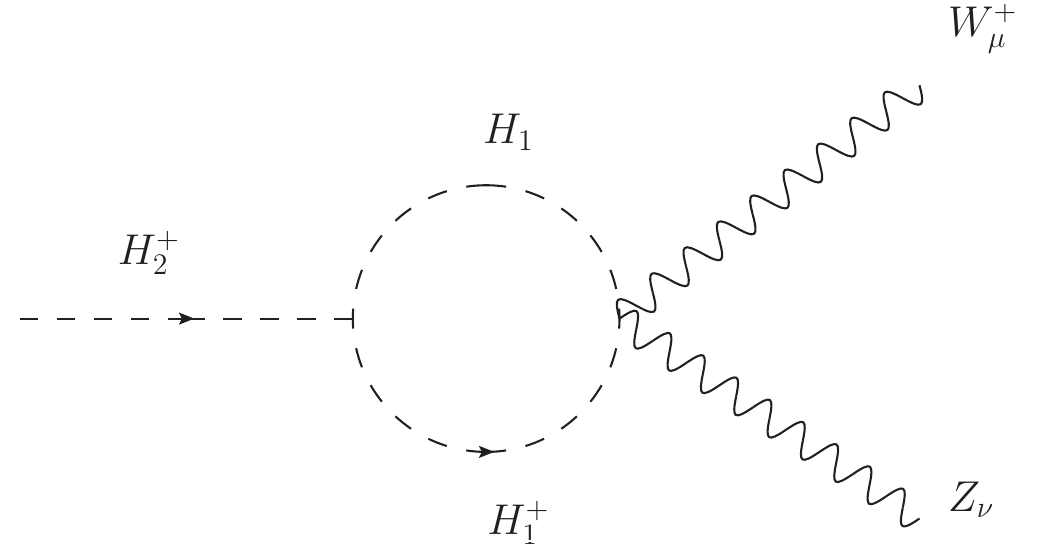}}
\subfigure[$E_6$]{
\includegraphics[height = 3 cm, width = 6 cm]{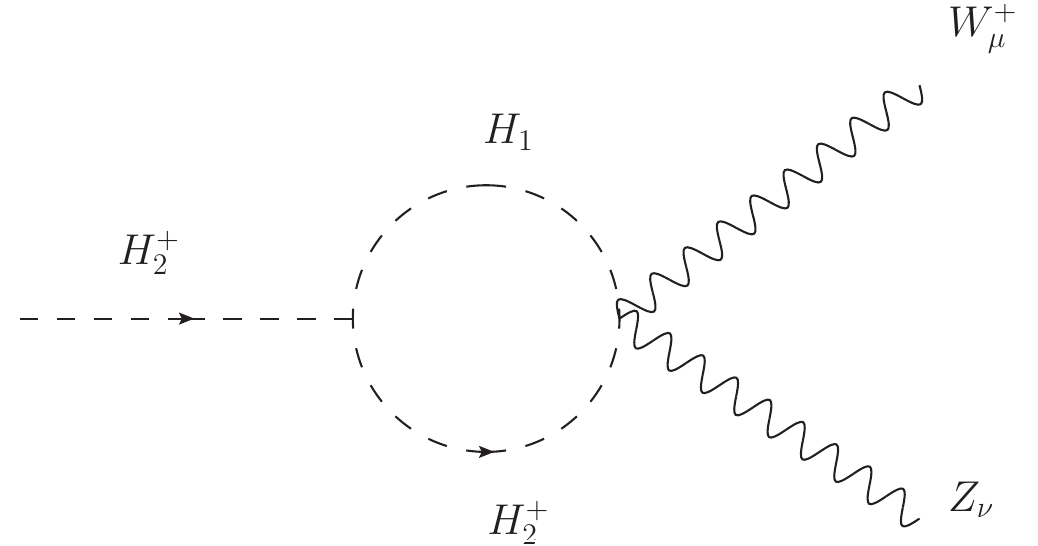}} \\
\subfigure[$E_7$]{
\includegraphics[height = 3 cm, width = 6 cm]{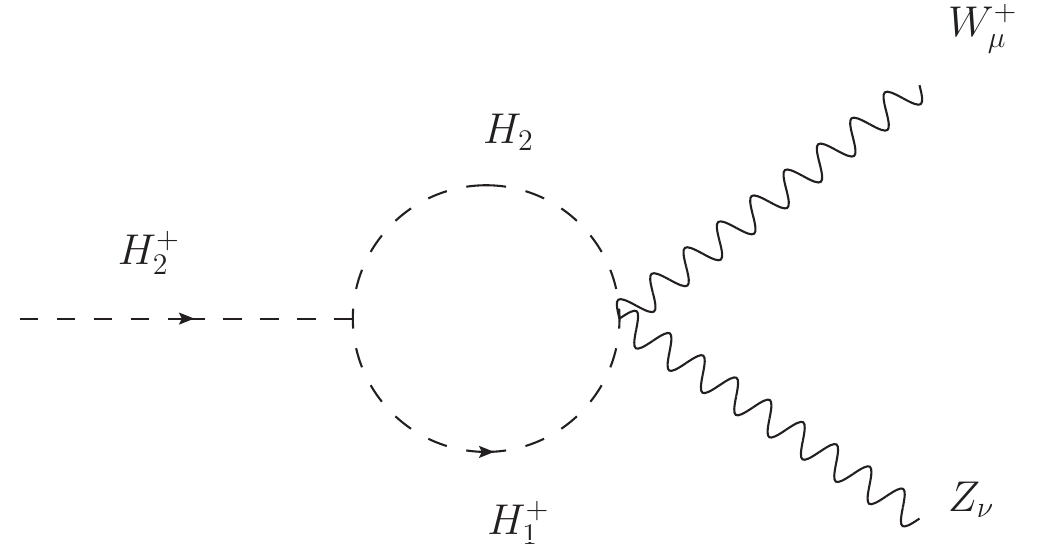}}
\subfigure[$E_8$]{
\includegraphics[height = 3 cm, width = 6 cm]{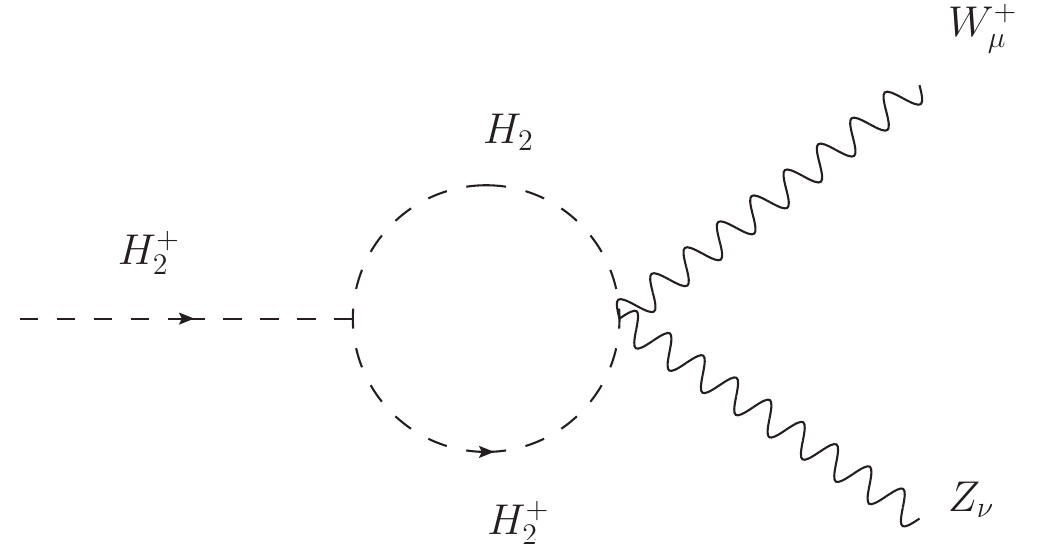}}
} \\
\caption{}
\label{fig-1}
\end{figure}

\begin{center}
\begin{table}[htb!]
\begin{tabular}{|c|c|}\hline
~~~~Diagram number ~~~~~& ~~~~$ F_Z$~~~~ \\ \hline \hline
$E_1$ & ~~~~$\frac{1}{16 \pi^2 v}~\frac{s_W^2}{c_W}~ c_{\alpha_3 + \delta_2}~ \lambda_{H_1^+ H_1^- H_1}~B_0(q^2, H_1^+, H_1)$~~~~ \\ 
$E_2$ & $\frac{1}{16 \pi^2 v}~\frac{s_W^2}{c_W}~ s_{\alpha_3 + \delta_2} ~ \lambda_{H_1^+ H_2^- H_1}~B_0(q^2, H_2^+, H_1)$ \\ 
$E_3$ & -$\frac{1}{16 \pi^2 v}~\frac{s_W^2}{c_W}~ s_{\alpha_3 + \delta_2} ~ \lambda_{H_1^+ H_1^- H_2}~B_0(q^2, H_1^+, H_2)$ \\  
 $E_4$ & $\frac{1}{16 \pi^2 v}~\frac{s_W^2}{c_W}~ c_{\alpha_3 + \delta_2} ~ \lambda_{H_1^+ H_2^- H_2}~B_0(q^2, H_2^+, H_2)$ \\ \hline \hline 
$E_5$ & $\frac{1}{16 \pi^2 v}~\frac{s_W^2}{c_W}~ c_{\alpha_3 + \delta_2} ~ \lambda_{H_2^+ H_1^- H_1}~B_0(q^2, H_1^+, H_1)$ \\  
$E_6$ & $\frac{1}{16 \pi^2 v}~\frac{s_W^2}{c_W}~ s_{\alpha_3 + \delta_2} ~ \lambda_{H_2^+ H_2^- H_1}~B_0(q^2, H_2^+, H_1)$ \\ 
$E_7$ & -$\frac{1}{16 \pi^2 v}~\frac{s_W^2}{c_W}~ s_{\alpha_3 + \delta_2} ~ \lambda_{H_2^+ H_1^- H_2}~B_0(q^2, H_1^+, H_2)$ \\
$E_8$ & $\frac{1}{16 \pi^2 v}~\frac{s_W^2}{c_W}~ c_{\alpha_3 + \delta_2} ~ \lambda_{H_2^+ H_2^- H_2}~B_0(q^2, H_2^+, H_2)$ \\ \hline \hline
 \end{tabular}
\caption{}
\label{tab-1}
\end{table}
\end{center}

\subsubsection{L-, M-, N-type amplitudes}

\begin{figure}[htpb!]{\centering
\subfigure[$L_1$]{
\includegraphics[height = 2 cm, width = 5.2 cm]{L1-eps-converted-to.pdf}} 
\subfigure[$M_1$]{
\includegraphics[height = 2 cm, width = 5.2 cm]{M1-eps-converted-to.pdf}} 
\subfigure[$N_1$]{
\includegraphics[height = 2 cm, width = 5.2 cm]{N1-eps-converted-to.pdf}} \\
\subfigure[$L_2$]{
\includegraphics[height = 2 cm, width = 5.2 cm]{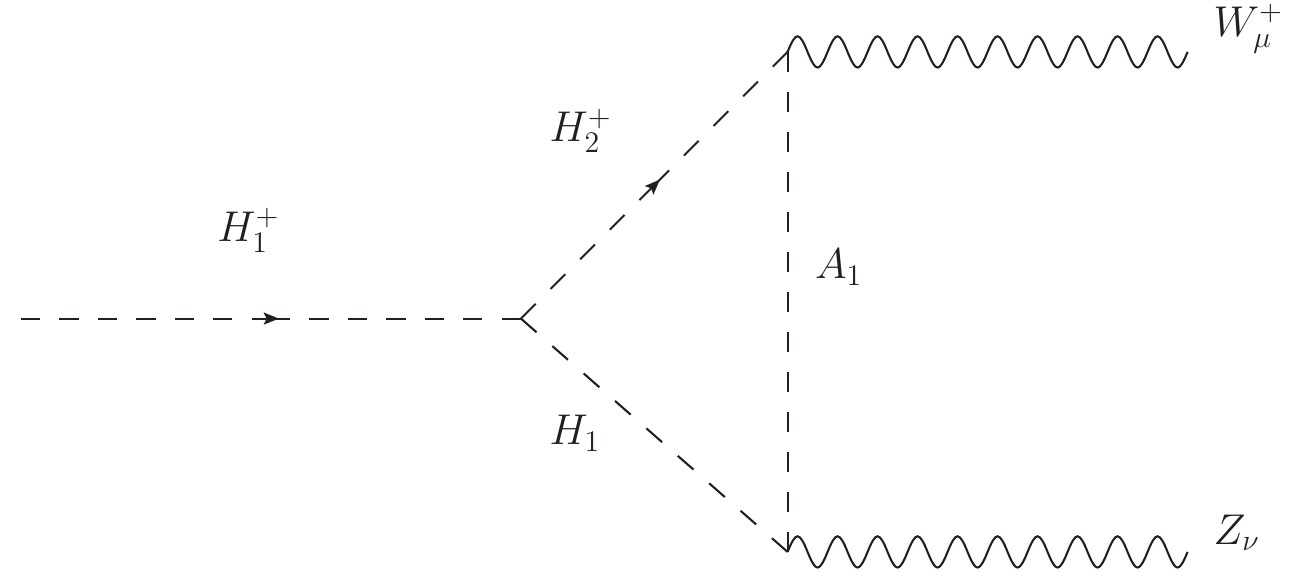}} 
\subfigure[$M_2$]{
\includegraphics[height = 2 cm, width = 5.2 cm]{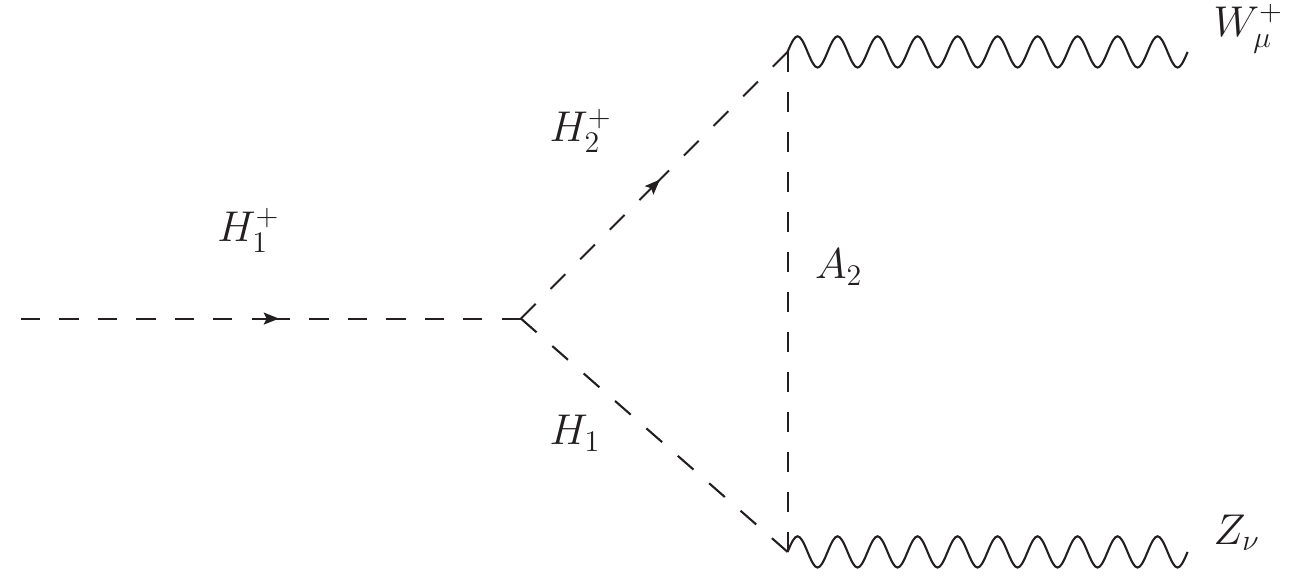}} 
\subfigure[$N_2$]{
\includegraphics[height = 2 cm, width = 5.2 cm]{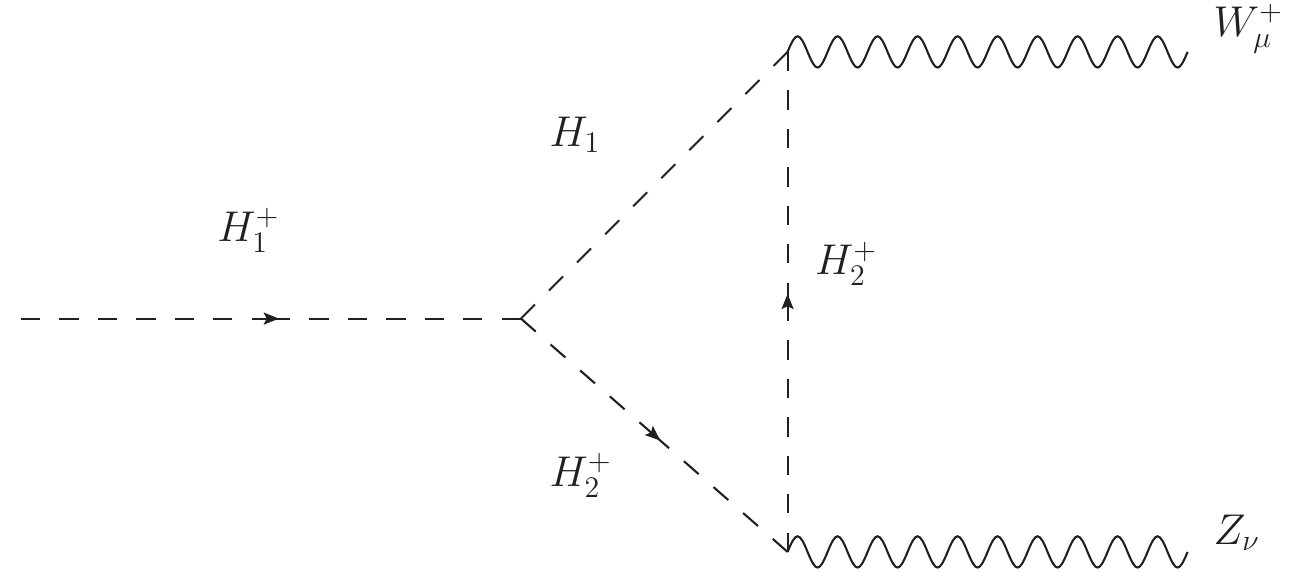}} \\
\subfigure[$L_3$]{
\includegraphics[height = 2 cm, width = 5.2 cm]{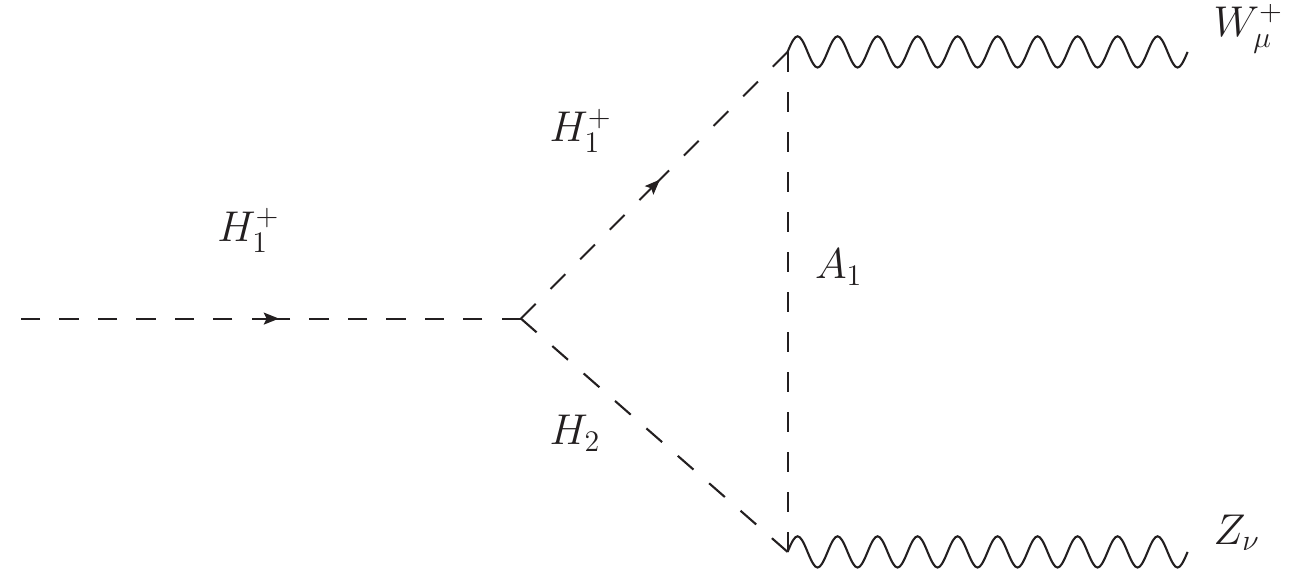}} 
\subfigure[$M_3$]{
\includegraphics[height = 2 cm, width = 5.2 cm]{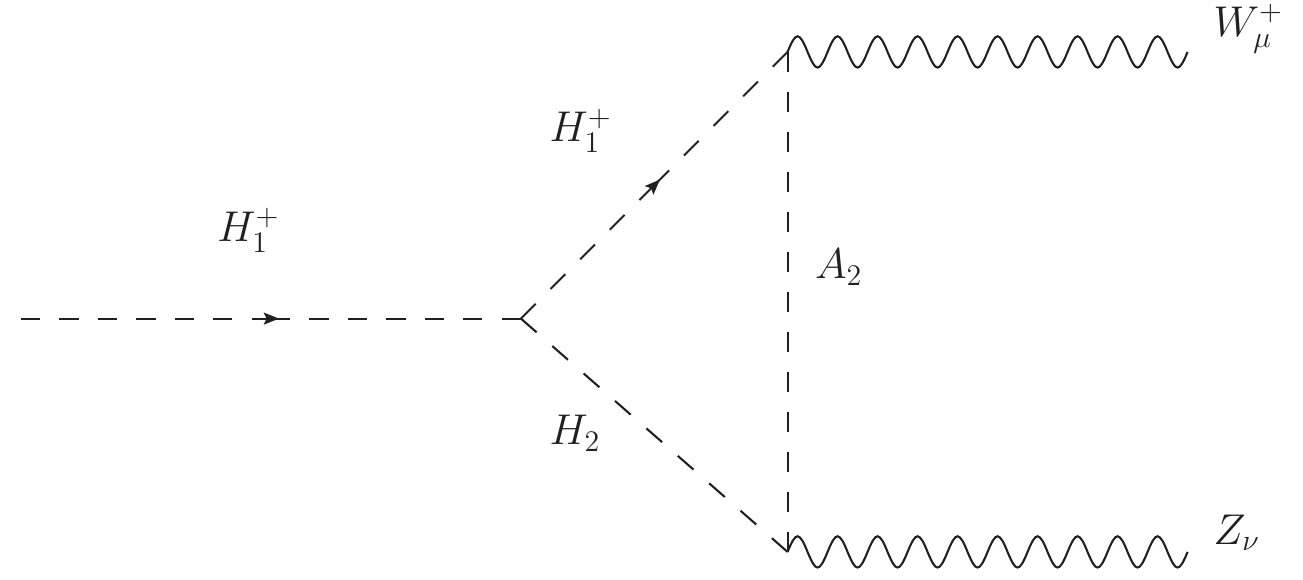}} 
\subfigure[$N_3$]{
\includegraphics[height = 2 cm, width = 5.2 cm]{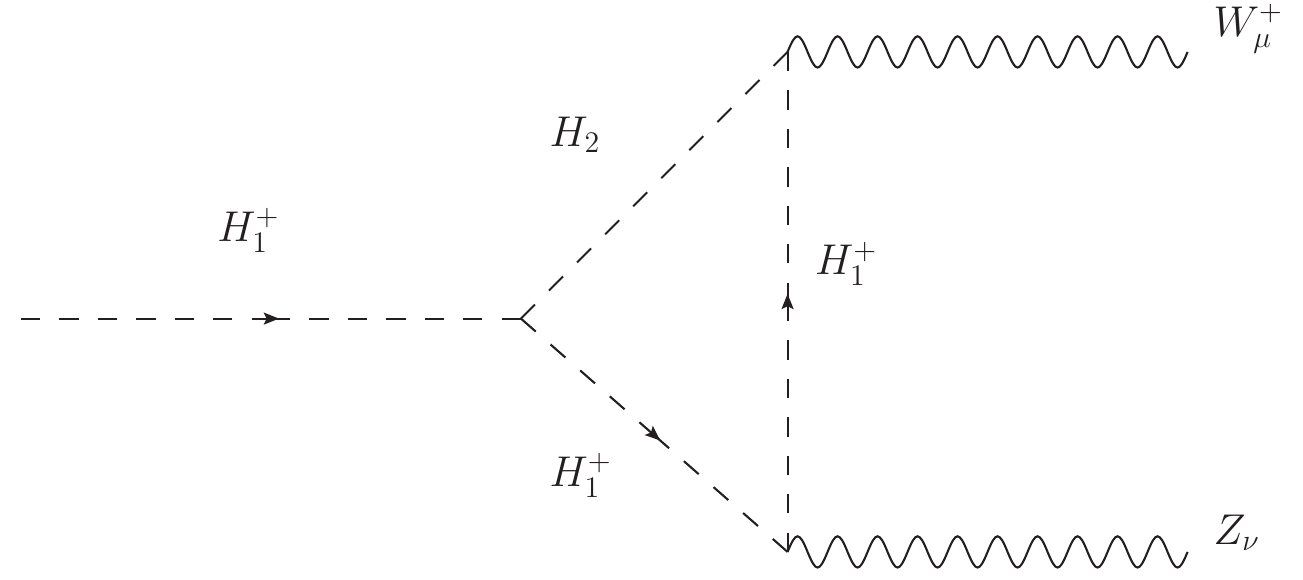}} \\
\subfigure[$L_4$]{
d\includegraphics[height = 2 cm, width = 5.2 cm]{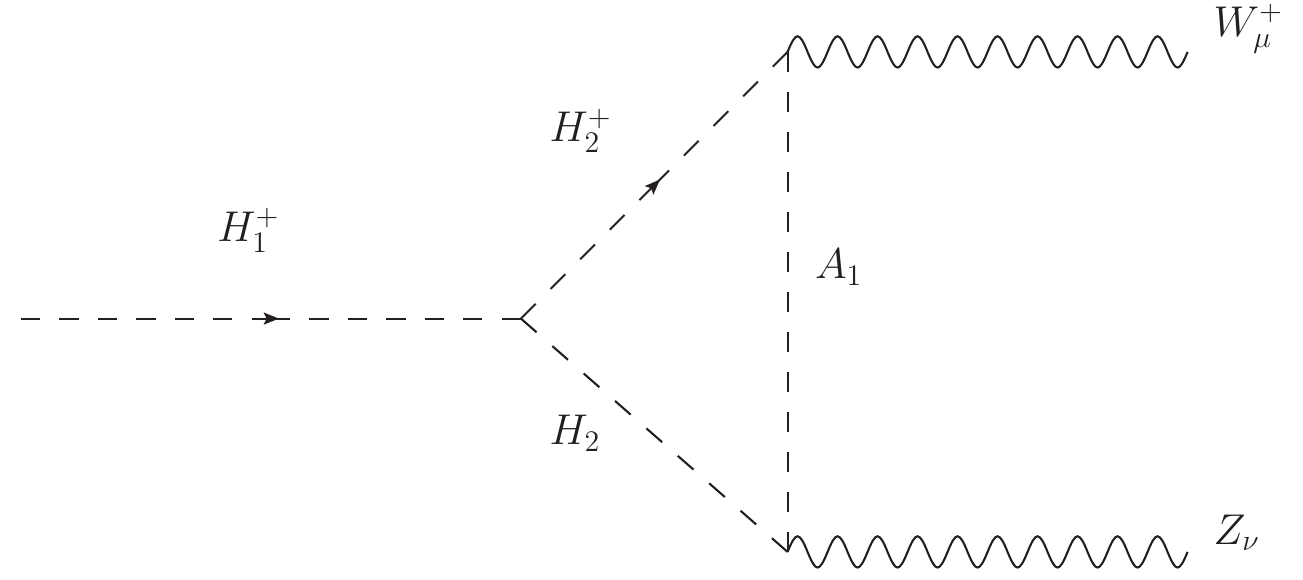}} 
\subfigure[$M_4$]{
\includegraphics[height = 2 cm, width = 5.2 cm]{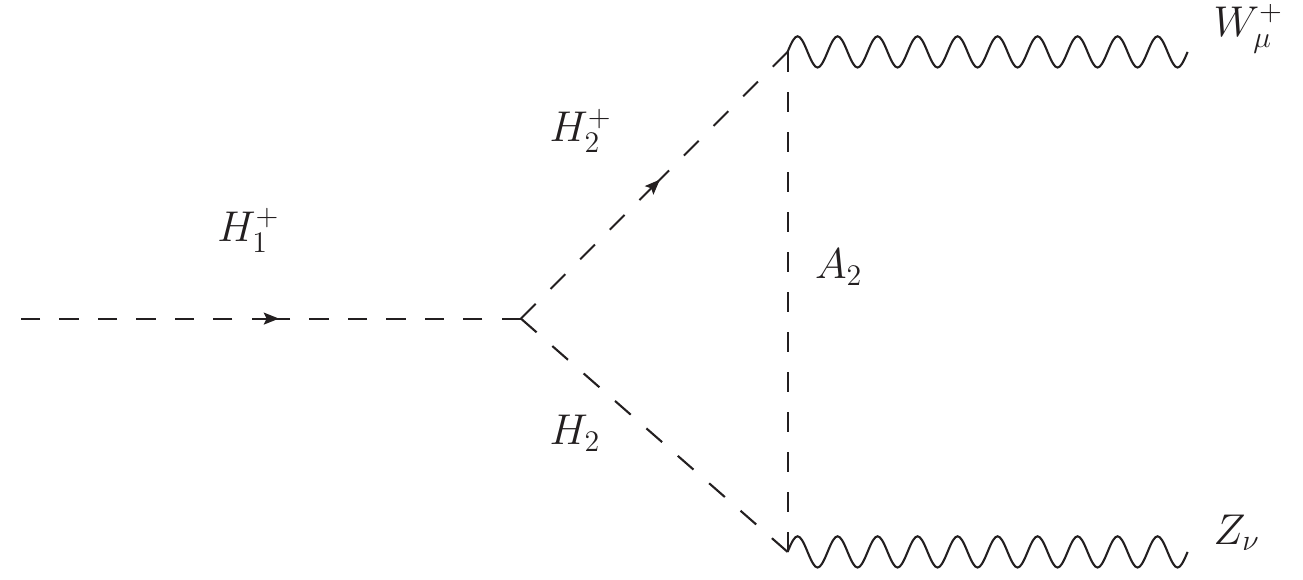}} 
\subfigure[$N_4$]{
\includegraphics[height = 2 cm, width = 5.2 cm]{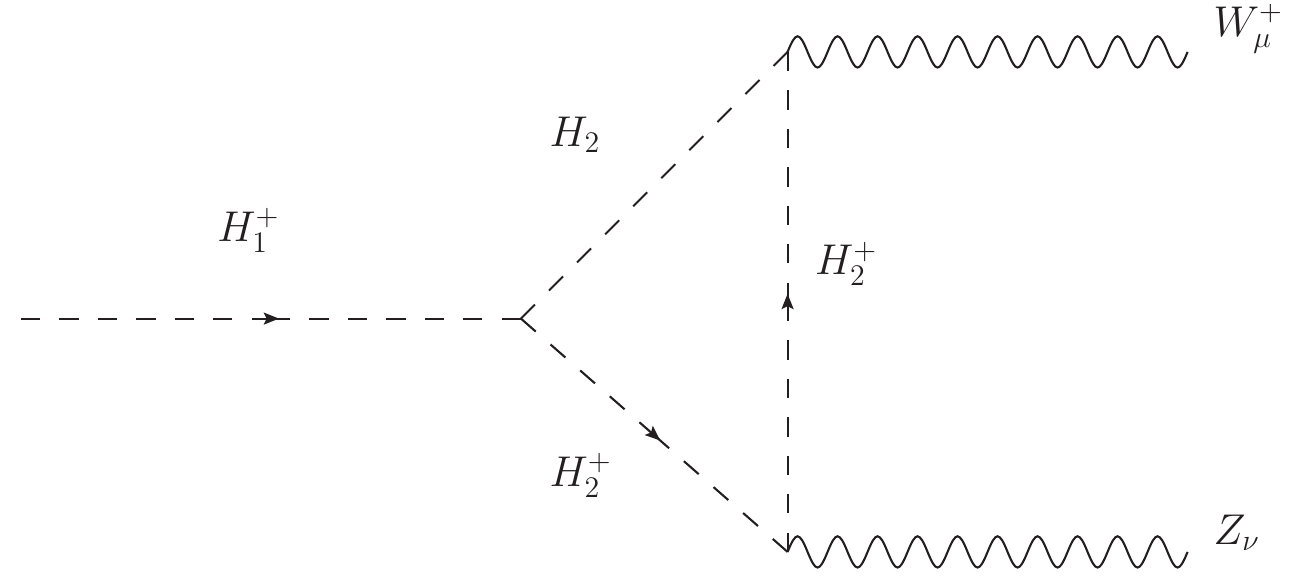}} 
\subfigure[$L_5$]{
\includegraphics[height = 2 cm, width = 5.2 cm]{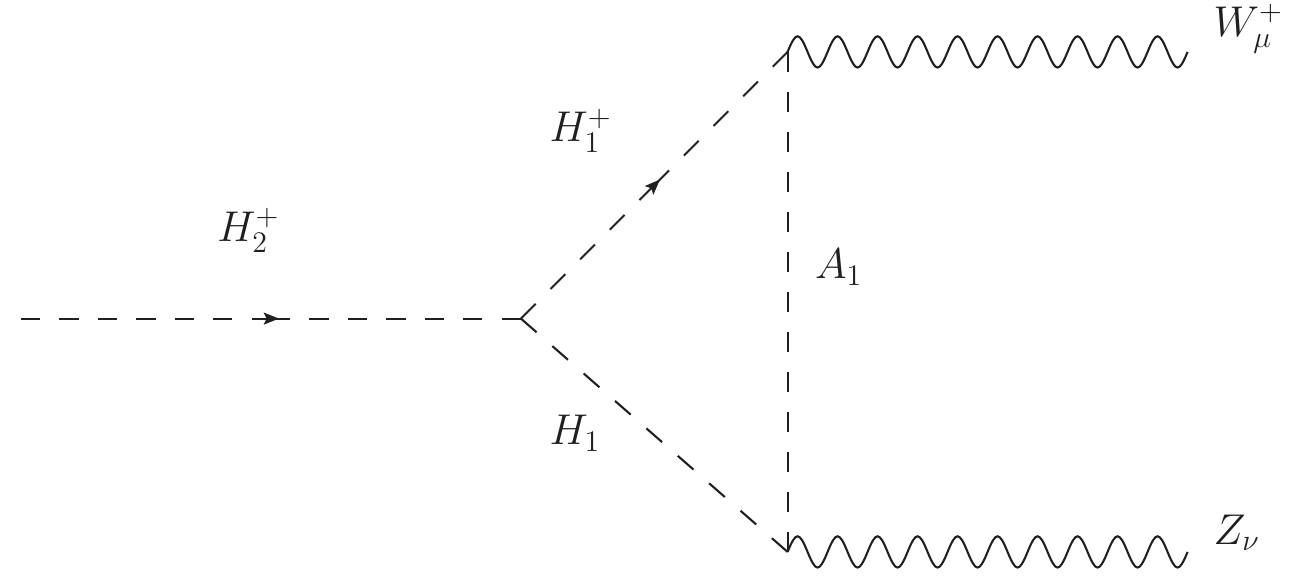}} 
\subfigure[$M_5$]{
\includegraphics[height = 2 cm, width = 5.2 cm]{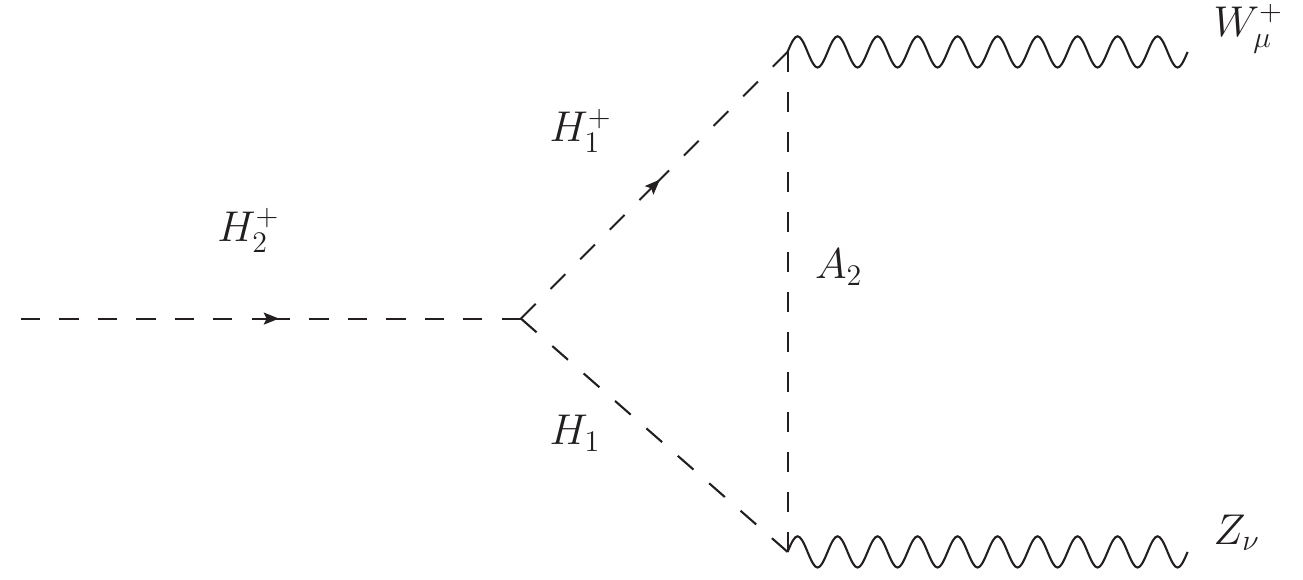}} 
\subfigure[$N_5$]{
\includegraphics[height = 2 cm, width = 5.2 cm]{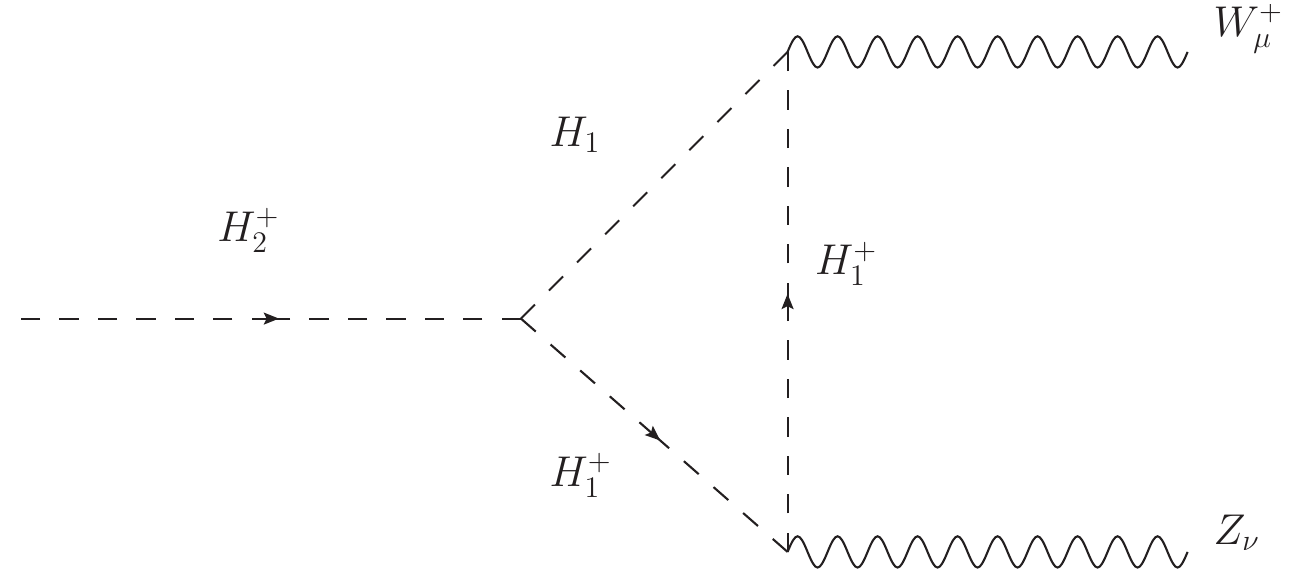}} \\
\subfigure[$L_6$]{
\includegraphics[height = 2 cm, width = 5.2 cm]{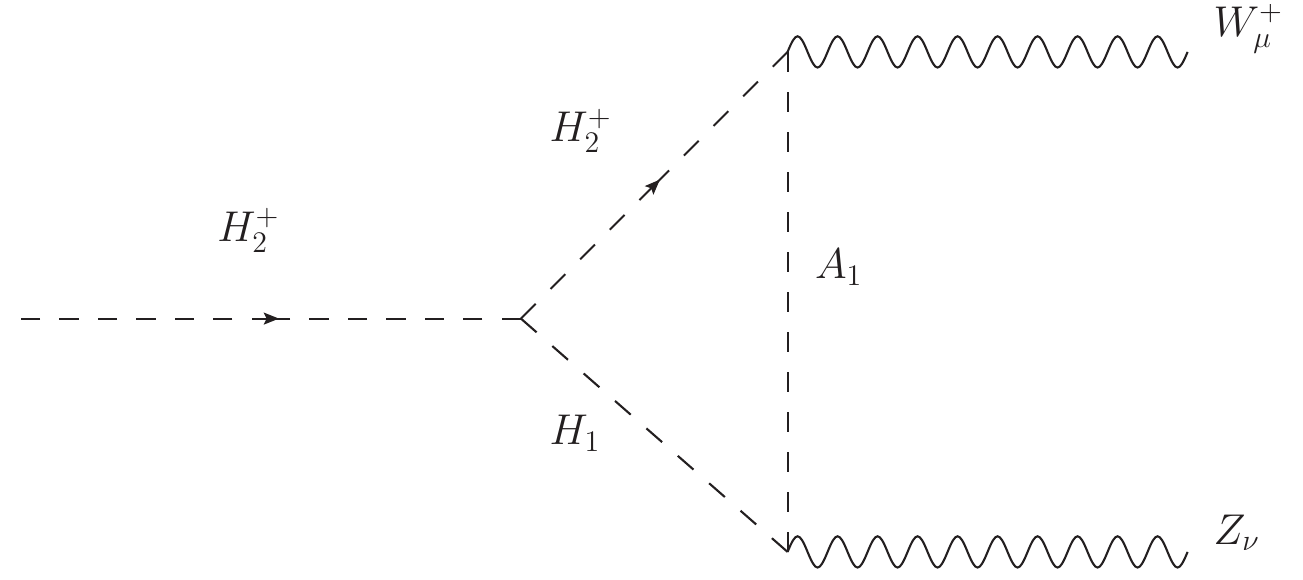}} 
\subfigure[$M_6$]{
\includegraphics[height = 2 cm, width = 5.2 cm]{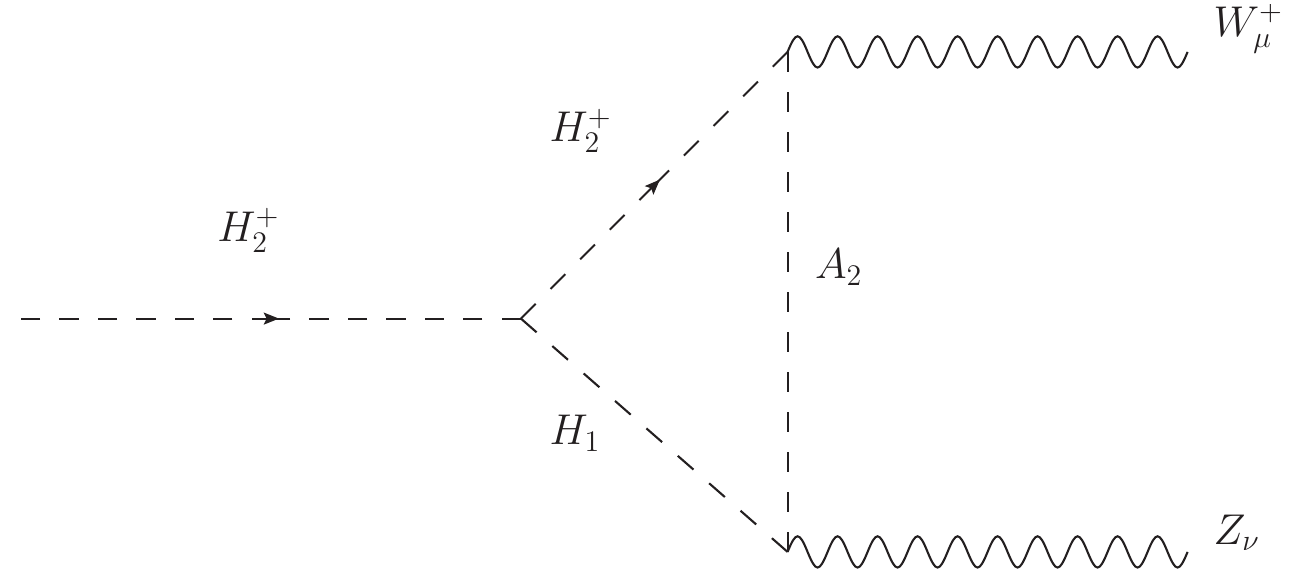}} 
\subfigure[$N_6$]{
\includegraphics[height = 2 cm, width = 5.2 cm]{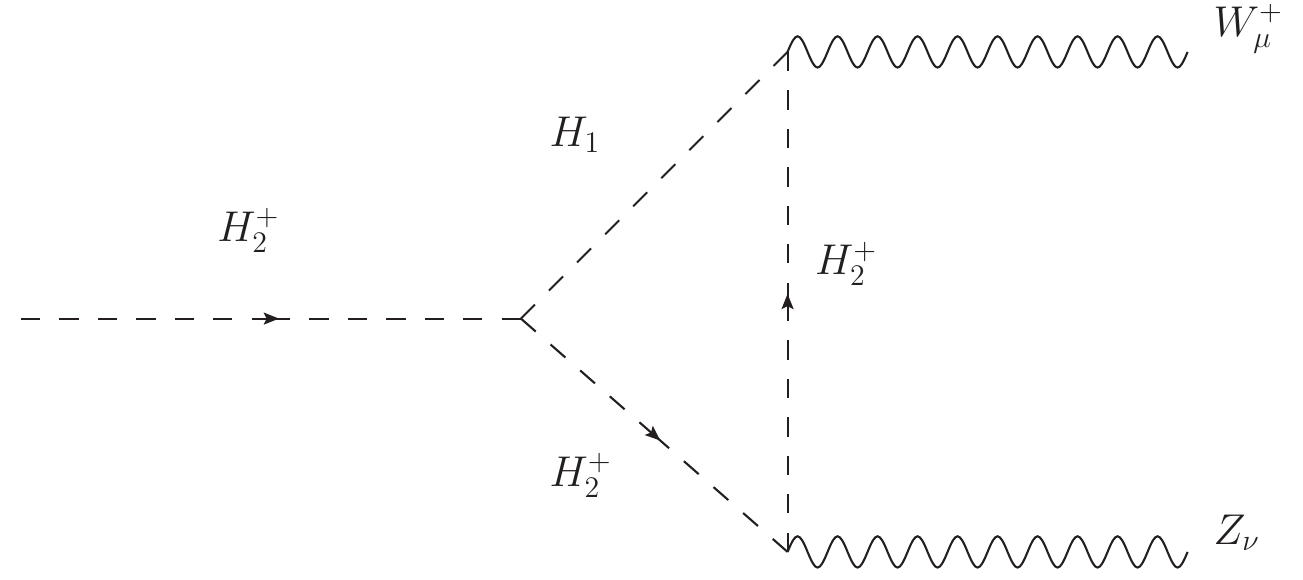}} \\
\subfigure[$L_7$]{
\includegraphics[height = 2 cm, width = 5.2 cm]{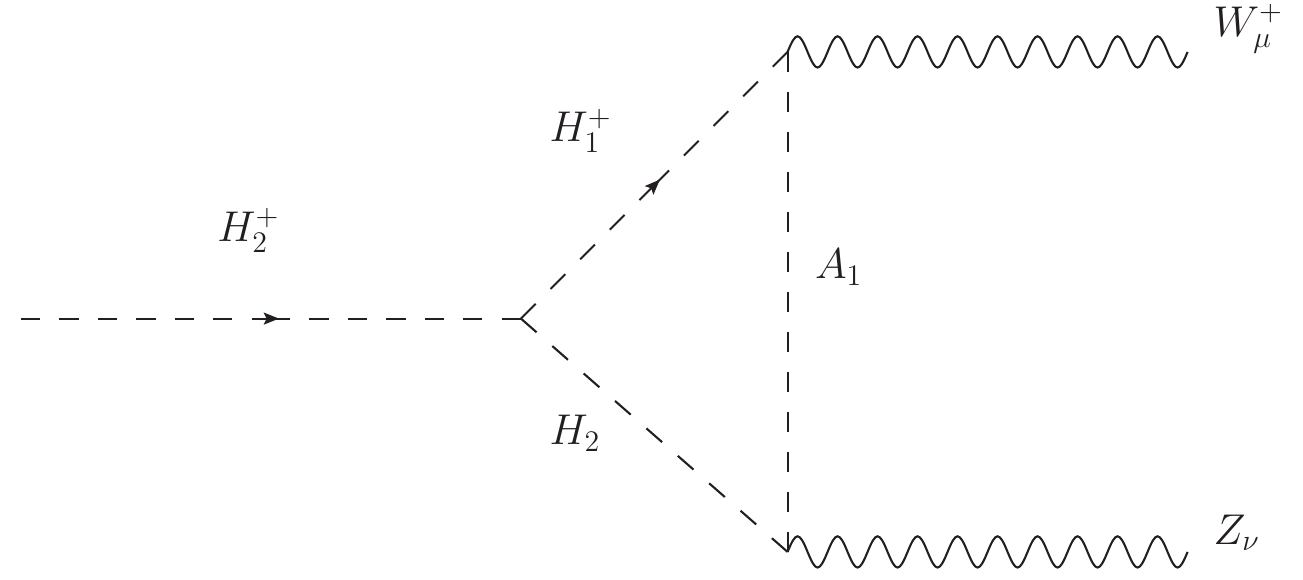}} 
\subfigure[$M_7$]{
\includegraphics[height = 2 cm, width = 5.2 cm]{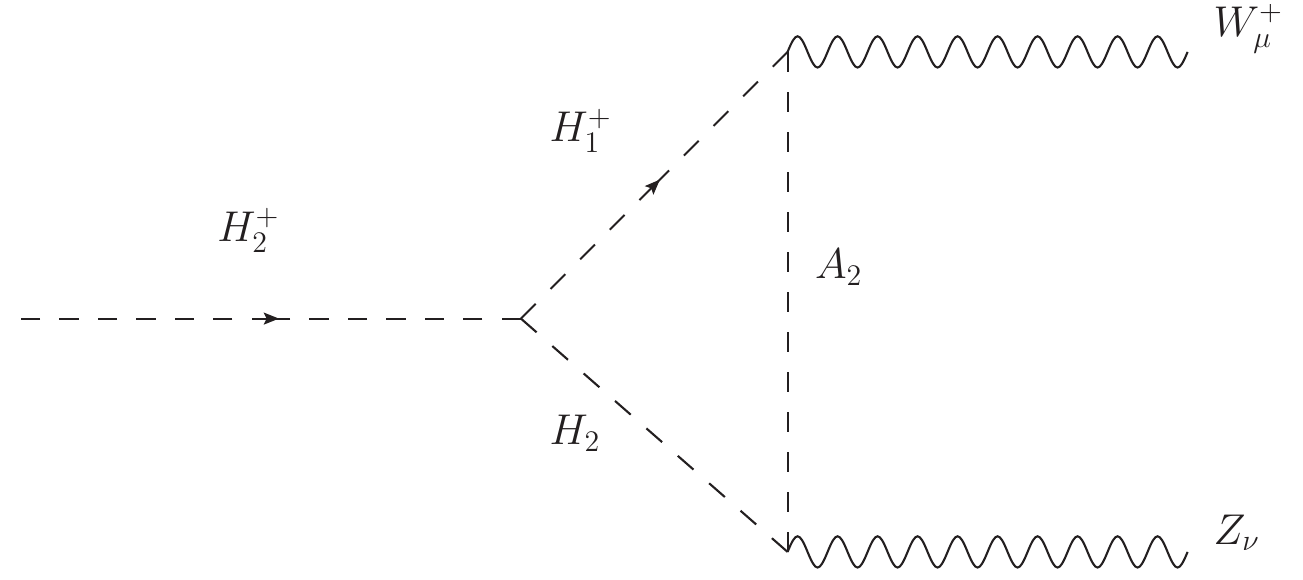}} 
\subfigure[$N_7$]{
\includegraphics[height = 2 cm, width = 5.2 cm]{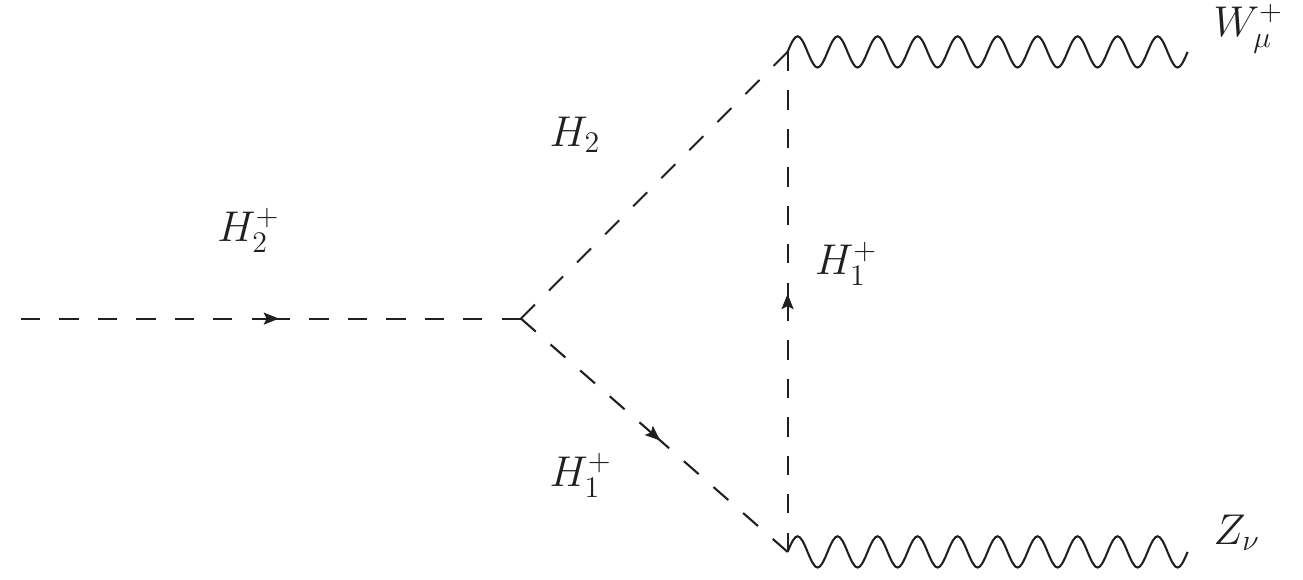}} 
} \\
\subfigure[$L_8$]{
\includegraphics[height = 2 cm, width = 5.2 cm]{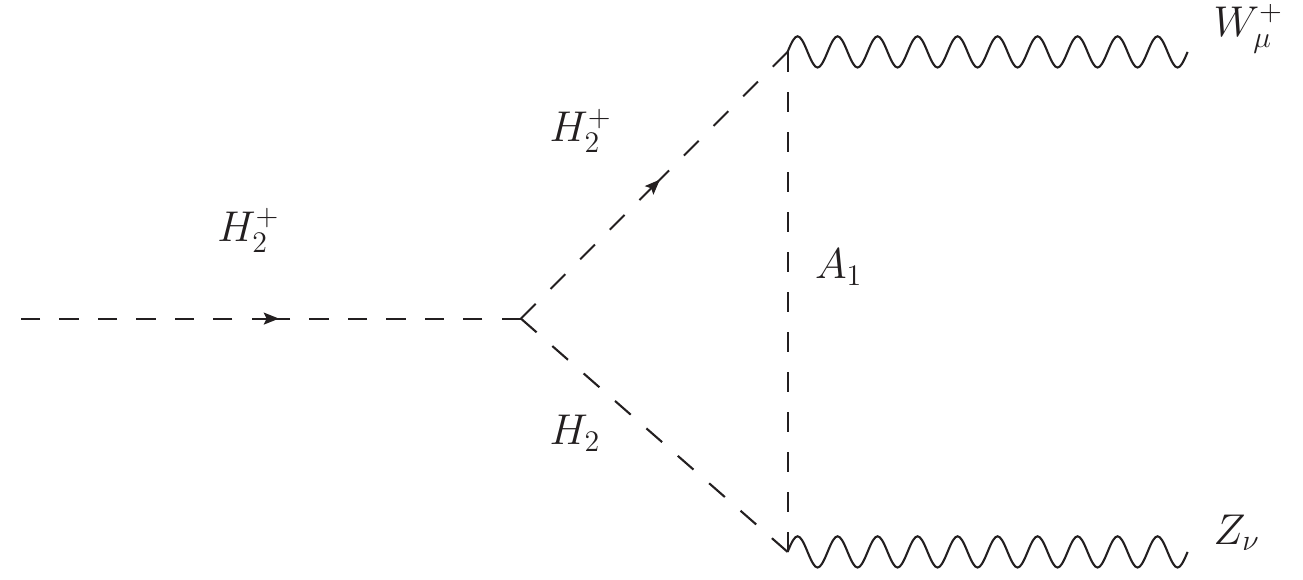}} 
\subfigure[$M_8$]{
\includegraphics[height = 2 cm, width = 5.2 cm]{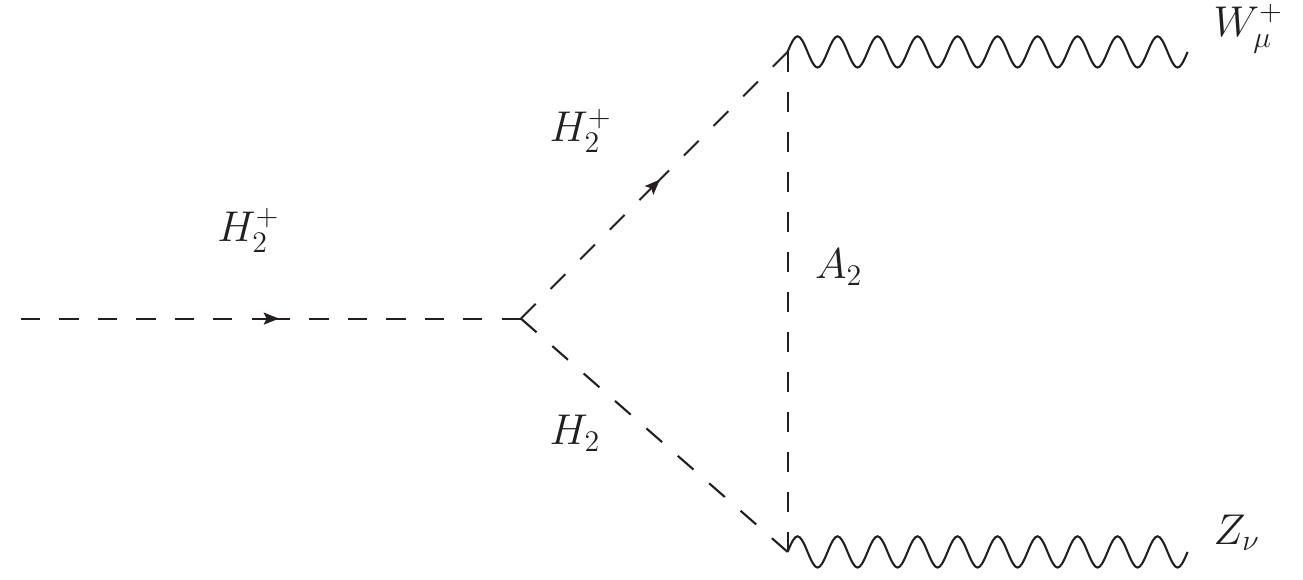}} 
\subfigure[$N_8$]{
\includegraphics[height = 2 cm, width = 5.2 cm]{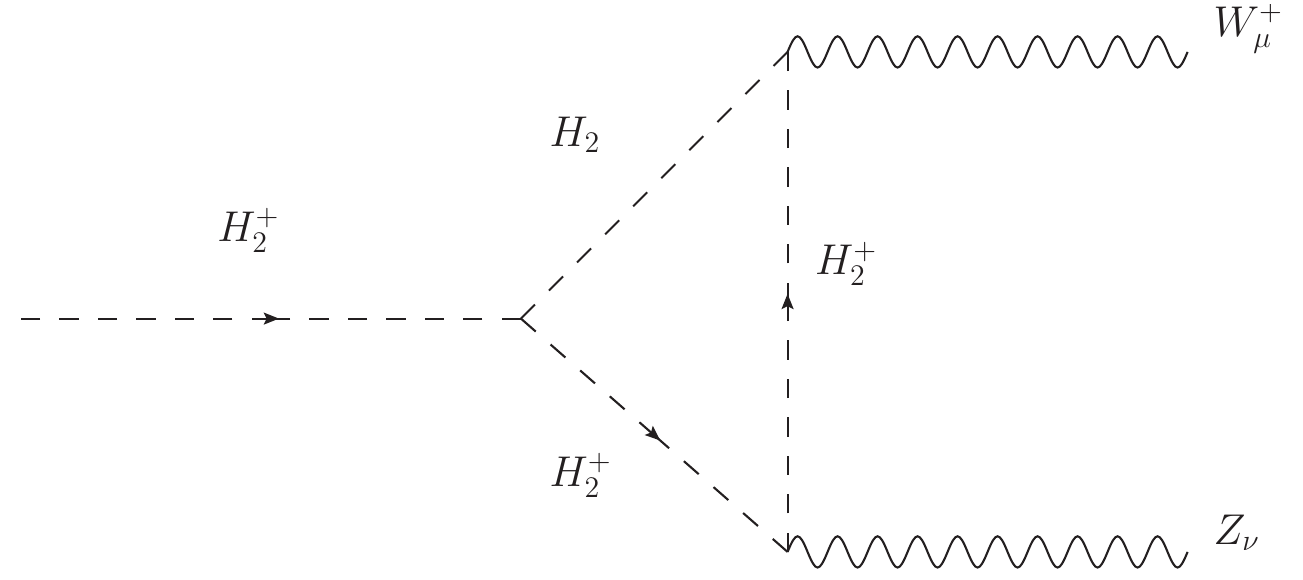}}
\caption{}
\label{fig-3}
\end{figure}

\begin{center}
\begin{table}[htb!]
\resizebox{17cm}{!}{
\begin{tabular}{|c|c|c|}\hline
Diagram no. & $ F_Z$ & $G_Z$ \\ \hline
$L_1$ & $ \left(\frac{-2 \lambda_{H_1^+ H_1^- H_1}}{16 \pi^2 v ~c_W}\right) (c_{\delta_1-\delta_2}~ c_{\alpha_3+\delta_1}) C_{24}(H_1^+,A_1,H_1)$ & $ \left(\frac{-2 \lambda_{H_1^+ H_1^- H_1}}{16 \pi^2 v ~c_W}\right) M_W^2 (c_{\delta_1-\delta_2}~ c_{\alpha_3+\delta_1}) C_{1223}(H_1^+,A_1,H_1)$ \\ 
$M_1$ &  $ \left(\frac{-2 \lambda_{H_1^+ H_1^- H_1}}{16 \pi^2 v ~c_W}\right) (s_{\delta_1-\delta_2}~ s_{\alpha_3+\delta_1}) C_{24}(H_1^+,A_2,H_1)$ & $ \left(\frac{-2 \lambda_{H_1^+ H_1^- H_1}}{16 \pi^2 v ~c_W}\right) M_W^2 (s_{\delta_1-\delta_2}~ s_{\alpha_3+\delta_1}) C_{1223}(H_1^+,A_2,H_1)$\\  
$N_1$ & $ \left(\frac{2 c_{2W}}{c_W}\right) \frac{\lambda_{H_1^+ H_1^- H_1}}{16 \pi^2 v} (c_{\alpha_3+\delta_2}) C_{24}(H_1, H_1^+, H_1^+)$ & $ \left(\frac{2 c_{2W}}{c_W}\right) \frac{\lambda_{H_1^+ H_1^- H_1}}{16 \pi^2 v} M_W^2 (c_{\alpha_3+\delta_2}) C_{1223}(H_1, H_1^+, H_1^+)$ \\  \hline 
$L_2$ & $ \left(\frac{-2 \lambda_{H_1^+ H_2^- H_1}}{16 \pi^2 v ~c_W}\right) (-s_{\delta_1-\delta_2}~ c_{\alpha_3+\delta_1}) C_{24}(H_2^+,A_1,H_1)$ & $ \left(\frac{-2 \lambda_{H_1^+ H_2^- H_1}}{16 \pi^2 v ~c_W}\right) M_W^2 (-s_{\delta_1-\delta_2}~ c_{\alpha_3+\delta_1}) C_{1223}(H_2^+,A_1,H_1)$\\ 
$M_2$ & $ \left(\frac{-2 \lambda_{H_1^+ H_2^- H_1}}{16 \pi^2 v ~c_W}\right) (c_{\delta_1-\delta_2}~ s_{\alpha_3+\delta_1}) C_{24}(H_2^+,A_2,H_1)$ &  $ \left(\frac{-2 \lambda_{H_1^+ H_2^- H_1}}{16 \pi^2 v ~c_W}\right) M_W^2 (c_{\delta_1-\delta_2}~ s_{\alpha_3+\delta_1}) C_{1223}(H_2^+,A_2,H_1)$\\  
$N_2$ & $ \left(\frac{2 c_{2W}}{c_W}\right) \frac{\lambda_{H_1^+ H_2^- H_1}}{16 \pi^2 v} (s_{\alpha_3+\delta_2}) C_{24}(H_1, H_2^+, H_2^+)$ & $ \left(\frac{2 c_{2W}}{c_W}\right) \frac{\lambda_{H_1^+ H_2^- H_1}}{16 \pi^2 v} M_W^2 (s_{\alpha_3+\delta_2}) C_{1223}(H_1, H_2^+, H_2^+)$ \\  \hline
$L_3$ & $ \left(\frac{-2 \lambda_{H_1^+ H_1^- H_2}}{16 \pi^2 v ~c_W}\right) (-c_{\delta_1-\delta_2}~ s_{\alpha_3+\delta_1}) C_{24}(H_1^+,A_1,H_2)$ & $ \left(\frac{-2 \lambda_{H_1^+ H_1^- H_2}}{16 \pi^2 v ~c_W}\right) M_W^2 (-c_{\delta_1-\delta_2}~ s_{\alpha_3+\delta_1}) C_{1223}(H_1^+,A_1,H_2)$\\ 
$M_3$ &  $ \left(\frac{-2 \lambda_{H_1^+ H_1^- H_2}}{16 \pi^2 v ~c_W}\right) (s_{\delta_1-\delta_2}~ c_{\alpha_3+\delta_1}) C_{24}(H_1^+,A_2,H_2)$ & $ \left(\frac{-2 \lambda_{H_1^+ H_1^- H_2}}{16 \pi^2 v ~c_W}\right) M_W^2 (s_{\delta_1-\delta_2}~ c_{\alpha_3+\delta_1}) C_{1223}(H_1^+,A_2,H_2)$\\  
$N_3$ &  $ \left(\frac{2 c_{2W}}{c_W}\right) \frac{\lambda_{H_1^+ H_1^- H_2}}{16 \pi^2 v} (-s_{\alpha_3+\delta_2}) C_{24}(H_2, H_1^+, H_1^+)$ & $ \left(\frac{2 c_{2W}}{c_W}\right) \frac{\lambda_{H_1^+ H_1^- H_2}}{16 \pi^2 v} M_W^2 (-s_{\alpha_3+\delta_2}) C_{1223}(H_2, H_1^+, H_1^+)$ \\ \hline 
$L_4$ & $ \left(\frac{-2 \lambda_{H_1^+ H_2^- H_2}}{16 \pi^2 v ~c_W}\right) (s_{\delta_1-\delta_2}~ s_{\alpha_3+\delta_1}) C_{24}(H_2^+,A_1,H_2)$ & $ \left(\frac{-2 \lambda_{H_1^+ H_2^- H_2}}{16 \pi^2 v ~c_W}\right) M_W^2 (s_{\delta_1-\delta_2}~ s_{\alpha_3+\delta_1}) C_{1223}(H_2^+,A_1,H_2)$\\ 
$M_4$ & $ \left(\frac{-2 \lambda_{H_1^+ H_2^- H_2}}{16 \pi^2 v ~c_W}\right) (c_{\delta_1-\delta_2}~ c_{\alpha_3+\delta_1}) C_{24}(H_2^+,A_2,H_2)$ & $ \left(\frac{-2 \lambda_{H_1^+ H_2^- H_2}}{16 \pi^2 v ~c_W}\right)M_W^2 (c_{\delta_1-\delta_2}~ c_{\alpha_3+\delta_1}) C_{1223}(H_2^+,A_2,H_2)$\\ 
$N_4$ & $ \left(\frac{2 c_{2W}}{c_W}\right) \frac{\lambda_{H_1^+ H_2^- H_2}}{16 \pi^2 v} (c_{\alpha_3+\delta_2}) C_{24}(H_2, H_2^+, H_2^+)$ & $ \left(\frac{2 c_{2W}}{c_W}\right) \frac{\lambda_{H_1^+ H_2^- H_2}}{16 \pi^2 v} M_W^2 (c_{\alpha_3+\delta_2}) C_{1223}(H_2, H_2^+, H_2^+)$\\ \hline \hline
$L_5$ & $ \left(\frac{-2 \lambda_{H_2^+ H_1^- H_1}}{16 \pi^2 v ~c_W}\right) (c_{\delta_1-\delta_2}~ c_{\alpha_3+\delta_1}) C_{24}(H_1^+,A_1,H_1)$ & $ \left(\frac{-2 \lambda_{H_2^+ H_1^- H_1}}{16 \pi^2 v ~c_W}\right)M_W^2 (c_{\delta_1-\delta_2}~ c_{\alpha_3+\delta_1}) C_{1223}(H_1^+,A_1,H_1)$\\
$M_5$ & $ \left(\frac{-2 \lambda_{H_2^+ H_1^- H_1}}{16 \pi^2 v ~c_W}\right) (s_{\delta_1-\delta_2}~ s_{\alpha_3+\delta_1}) C_{24}(H_1^+,A_2,H_1)$ & $ \left(\frac{-2 \lambda_{H_2^+ H_1^- H_1}}{16 \pi^2 v ~c_W}\right)M_W^2 (s_{\delta_1-\delta_2}~ s_{\alpha_3+\delta_1}) C_{1223}(H_1^+,A_2,H_1)$\\
$N_5$ & $ \left(\frac{2 c_{2W}}{c_W}\right) \frac{\lambda_{H_2^+ H_1^- H_1}}{16 \pi^2 v} (c_{\alpha_3+\delta_2}) C_{24}(H_1, H_1^+, H_1^+)$ & $ \left(\frac{2 c_{2W}}{c_W}\right) \frac{\lambda_{H_2^+ H_1^- H_1}}{16 \pi^2 v} M_W^2 (c_{\alpha_3+\delta_2}) C_{1223}(H_1, H_1^+, H_1^+)$ \\ \hline
$L_6$ & $ \left(\frac{-2 \lambda_{H_2^+ H_2^- H_1}}{16 \pi^2 v ~c_W}\right) (-s_{\delta_1-\delta_2}~ c_{\alpha_3+\delta_1}) C_{24}(H_2^+,A_1,H_1)$ & $ \left(\frac{-2 \lambda_{H_2^+ H_2^- H_1}}{16 \pi^2 v ~c_W}\right) M_W^2 (-s_{\delta_1-\delta_2}~ c_{\alpha_3+\delta_1}) C_{1223}(H_2^+,A_1,H_1)$\\
$M_6$ & $ \left(\frac{-2 \lambda_{H_2^+ H_2^- H_1}}{16 \pi^2 v ~c_W}\right) (c_{\delta_1-\delta_2}~ s_{\alpha_3+\delta_1}) C_{24}(H_2^+,A_2,H_1)$ & $ \left(\frac{-2 \lambda_{H_2^+ H_2^- H_1}}{16 \pi^2 v ~c_W}\right)M_W^2 (c_{\delta_1-\delta_2}~ s_{\alpha_3+\delta_1}) C_{1223}(H_2^+,A_2,H_1)$\\
$N_6$ & $ \left(\frac{2 c_{2W}}{c_W}\right) \frac{\lambda_{H_2^+ H_2^- H_1}}{16 \pi^2 v} (s_{\alpha_3+\delta_2}) C_{24}(H_1, H_2^+, H_2^+)$ & $ \left(\frac{2 c_{2W}}{c_W}\right) \frac{\lambda_{H_2^+ H_2^- H_1}}{16 \pi^2 v} M_W^2 (s_{\alpha_3+\delta_2}) C_{1223}(H_1, H_2^+, H_2^+)$ \\ \hline
$L_7$ & $ \left(\frac{-2 \lambda_{H_2^+ H_1^- H_2}}{16 \pi^2 v ~c_W}\right) (-c_{\delta_1-\delta_2}~ s_{\alpha_3+\delta_1}) C_{24}(H_1^+,A_1,H_2)$ & $ \left(\frac{-2 \lambda_{H_2^+ H_1^- H_2}}{16 \pi^2 v ~c_W}\right) M_W^2 (-c_{\delta_1-\delta_2}~ s_{\alpha_3+\delta_1}) C_{1223}(H_1^+,A_1,H_2)$\\
$M_7$ & $ \left(\frac{-2 \lambda_{H_2^+ H_1^- H_2}}{16 \pi^2 v ~c_W}\right) (s_{\delta_1-\delta_2}~ c_{\alpha_3+\delta_1}) C_{24}(H_1^+,A_2,H_2)$ & $ \left(\frac{-2 \lambda_{H_2^+ H_1^- H_2}}{16 \pi^2 v ~c_W}\right) M_W^2 (s_{\delta_1-\delta_2}~ c_{\alpha_3+\delta_1}) C_{1223}(H_1^+,A_2,H_2)$\\
$N_7$ & $ \left(\frac{2 c_{2W}}{c_W}\right) \frac{\lambda_{H_2^+ H_1^- H_2}}{16 \pi^2 v} (-s_{\alpha_3+\delta_2}) C_{24}(H_2, H_1^+, H_1^+)$ & $ \left(\frac{2 c_{2W}}{c_W}\right) \frac{\lambda_{H_2^+ H_1^- H_2}}{16 \pi^2 v} M_W^2 (-s_{\alpha_3+\delta_2}) C_{1223}(H_2, H_1^+, H_1^+)$ \\
\hline
$L_8$ & $ \left(\frac{-2 \lambda_{H_2^+ H_2^- H_2}}{16 \pi^2 v ~c_W}\right) (s_{\delta_1-\delta_2}~ s_{\alpha_3+\delta_1}) C_{24}(H_2^+,A_1,H_2)$ & $ \left(\frac{-2 \lambda_{H_2^+ H_2^- H_2}}{16 \pi^2 v ~c_W}\right) M_W^2 (s_{\delta_1-\delta_2}~ s_{\alpha_3+\delta_1}) C_{1223}(H_2^+,A_1,H_2)$ \\
$M_8$ &  $ \left(\frac{-2 \lambda_{H_2^+ H_2^- H_2}}{16 \pi^2 v ~c_W}\right) (c_{\delta_1-\delta_2}~ c_{\alpha_3+\delta_1}) C_{24}(H_2^+,A_2,H_2)$ & $ \left(\frac{-2 \lambda_{H_2^+ H_2^- H_2}}{16 \pi^2 v ~c_W}\right)M_W^2 (c_{\delta_1-\delta_2}~ c_{\alpha_3+\delta_1}) C_{1223}(H_2^+,A_2,H_2)$\\ 
$N_8$ & $ \left(\frac{2 c_{2W}}{c_W}\right) \frac{\lambda_{H_2^+ H_2^- H_2}}{16 \pi^2 v} (c_{\alpha_3+\delta_2}) C_{24}(H_2, H_2^+, H_2^+)$ & $ \left(\frac{2 c_{2W}}{c_W}\right) \frac{\lambda_{H_2^+ H_2^- H_2}}{16 \pi^2 v} M_W^2 (c_{\alpha_3+\delta_2}) C_{1223}(H_2, H_2^+, H_2^+)$\\ \hline \hline
 \end{tabular}}
\caption{}
\label{tab-3}
\end{table}
\end{center}

\subsubsection{$E^\prime$-type amplitudes}

\begin{figure}[htpb!]{\centering
\subfigure[$E_1^{'}$]{
\includegraphics[height = 3 cm, width = 6 cm]{E1p-eps-converted-to.pdf}} 
\subfigure[$E_2^{'}$]{
\includegraphics[height = 3 cm, width = 6 cm]{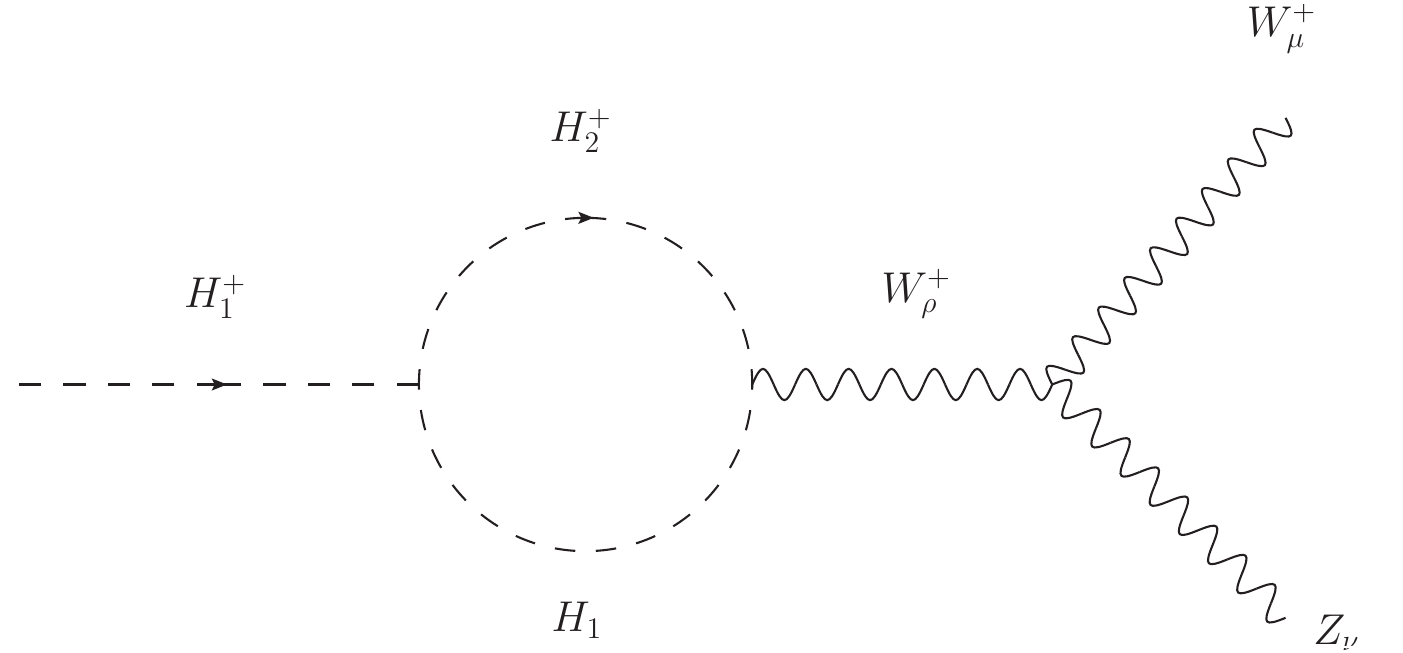}}\\
\subfigure[$E_3^{'}$]{
\includegraphics[height = 3 cm, width = 6 cm]{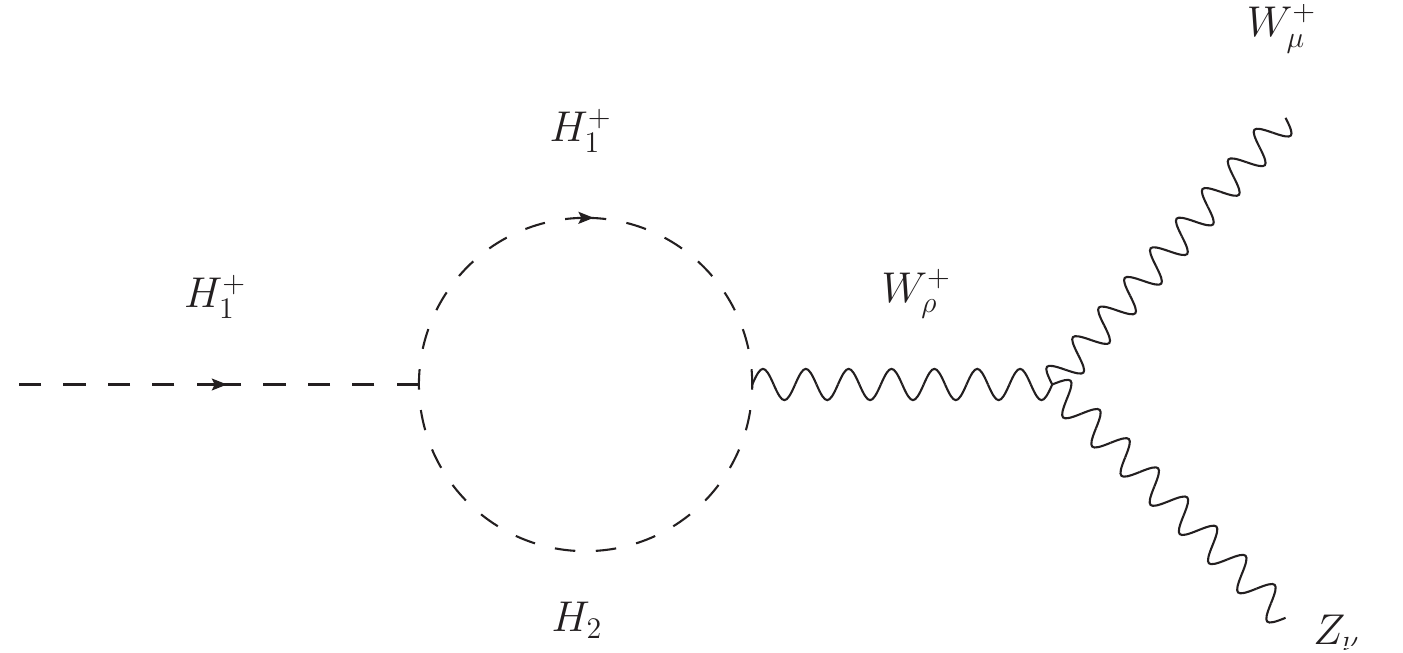}} 
\subfigure[$E_4^{'}$]{
\includegraphics[height = 3 cm, width = 6 cm]{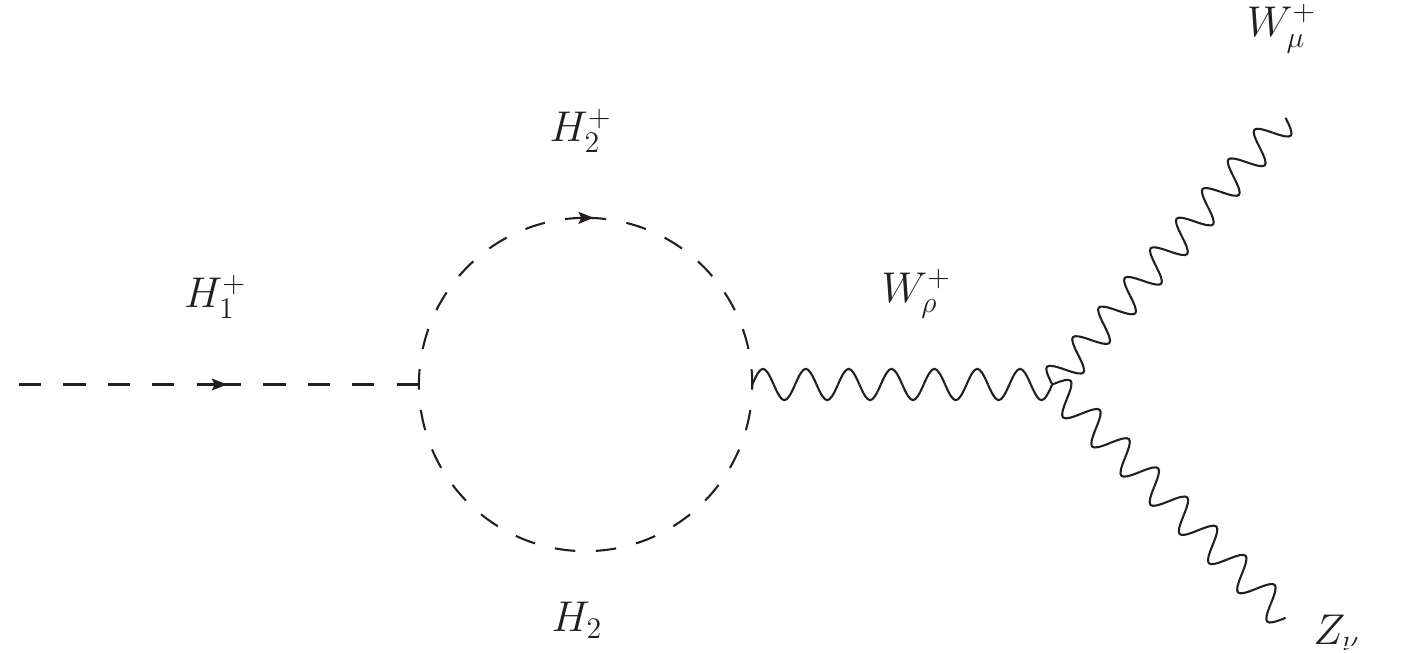}}
\subfigure[$E_5^{'}$]{
\includegraphics[height = 3 cm, width = 6 cm]{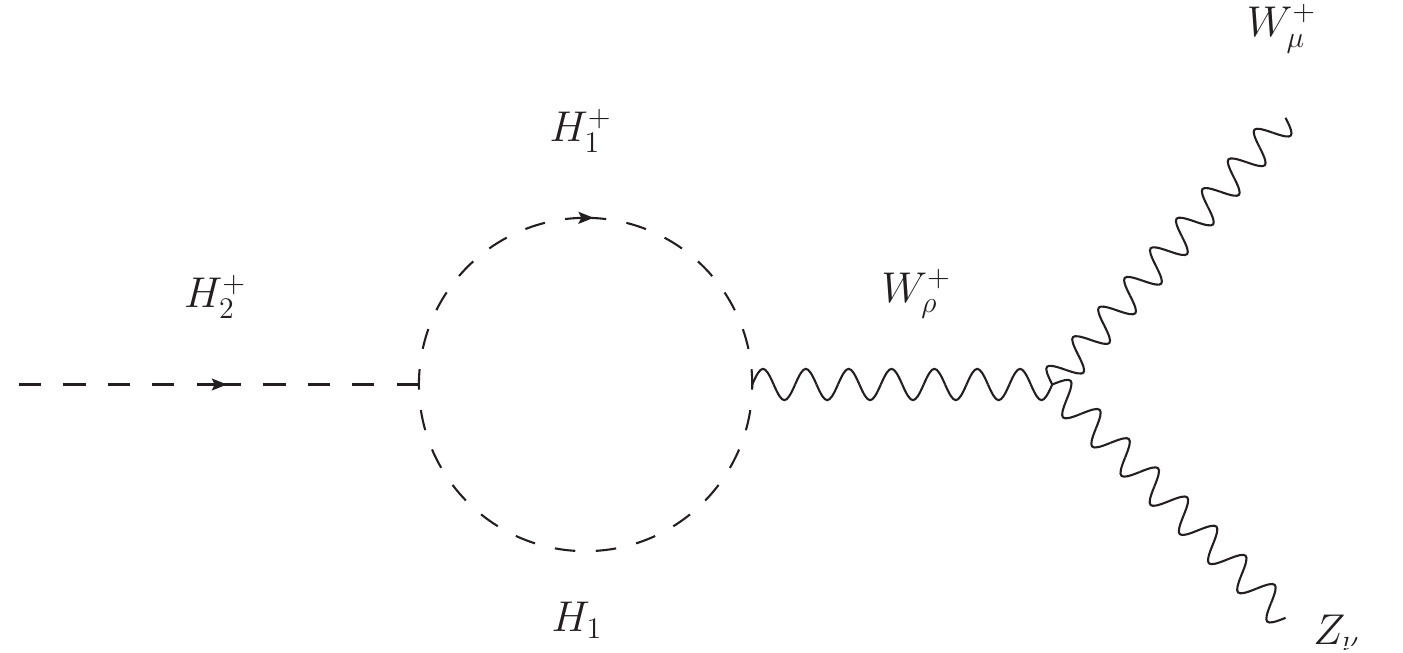}}
\subfigure[$E_6^{'}$]{
\includegraphics[height = 3 cm, width = 6 cm]{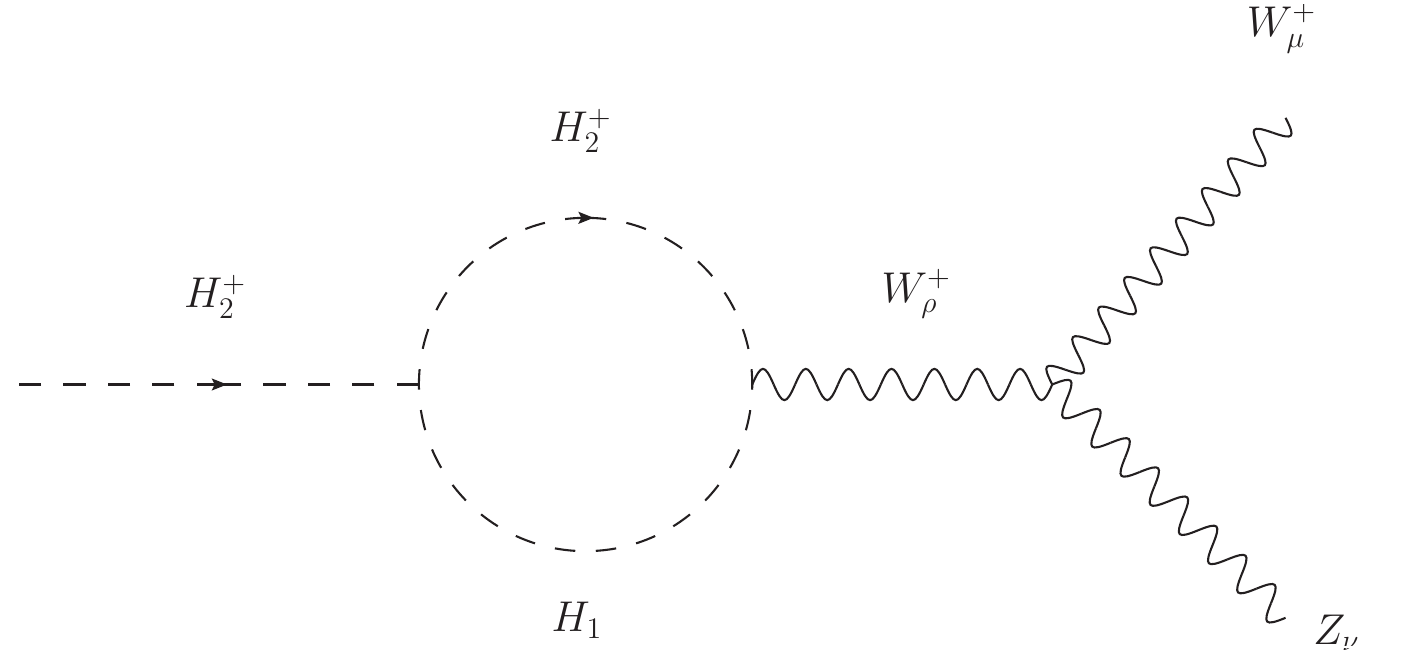}} \\
\subfigure[$E_7^{'}$]{
\includegraphics[height = 3 cm, width = 6 cm]{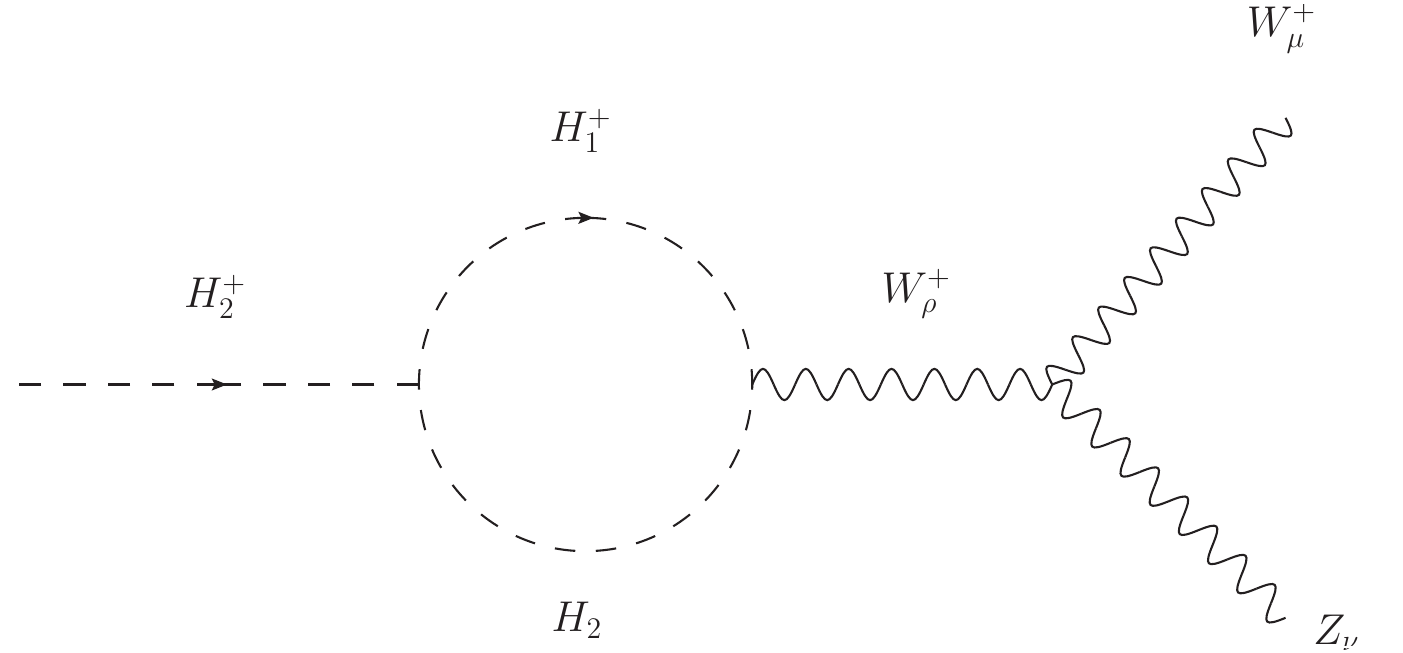}}
\subfigure[$E_8^{'}$]{
\includegraphics[height = 3 cm, width = 6 cm]{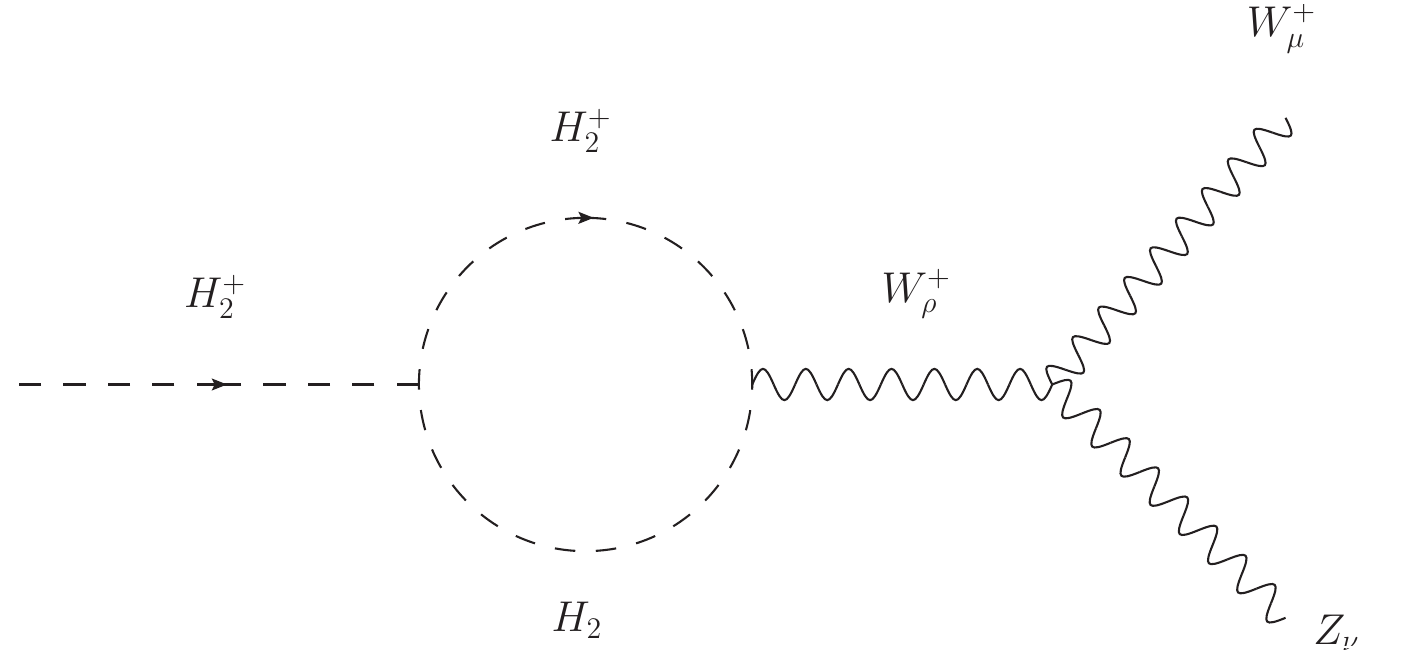}}
} \\
\caption{}
\label{fig-5}
\end{figure}

\subsubsection{Fermionic diagrams}

\begin{figure}[htpb!]{\centering
\includegraphics[height = 3 cm, width = 5.5 cm]{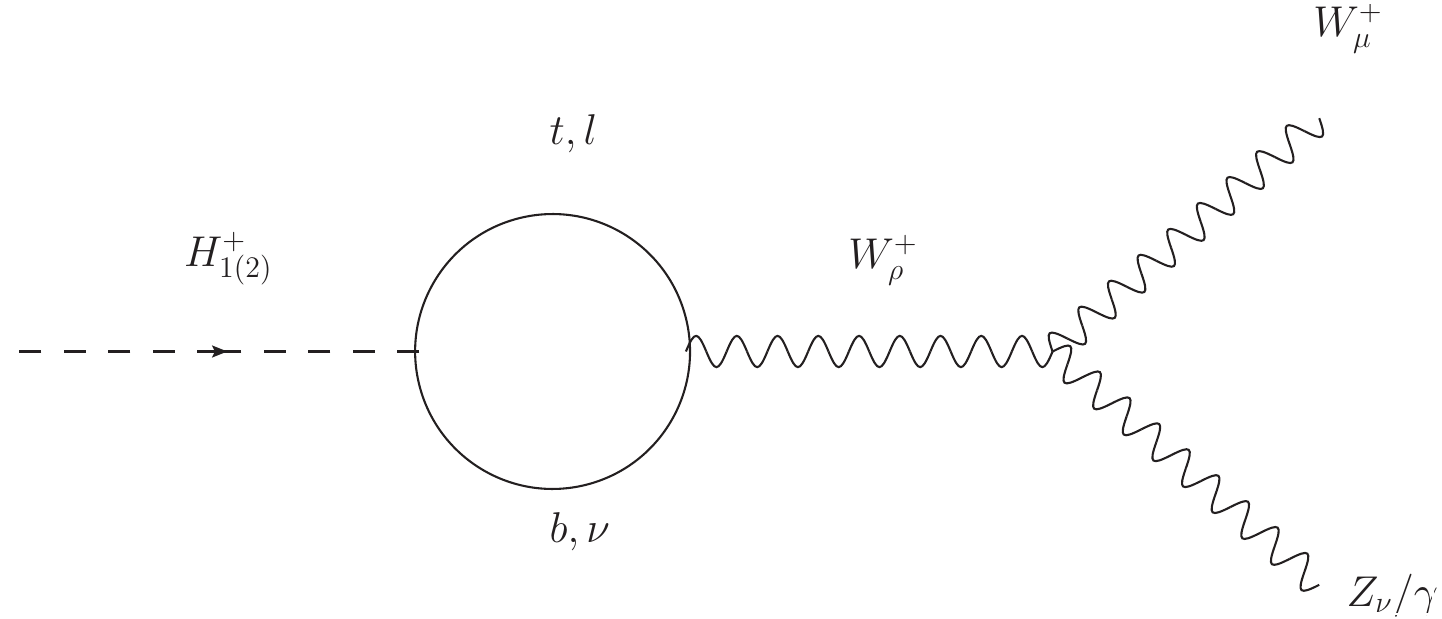}~~
\includegraphics[height = 3 cm, width = 5.5 cm]{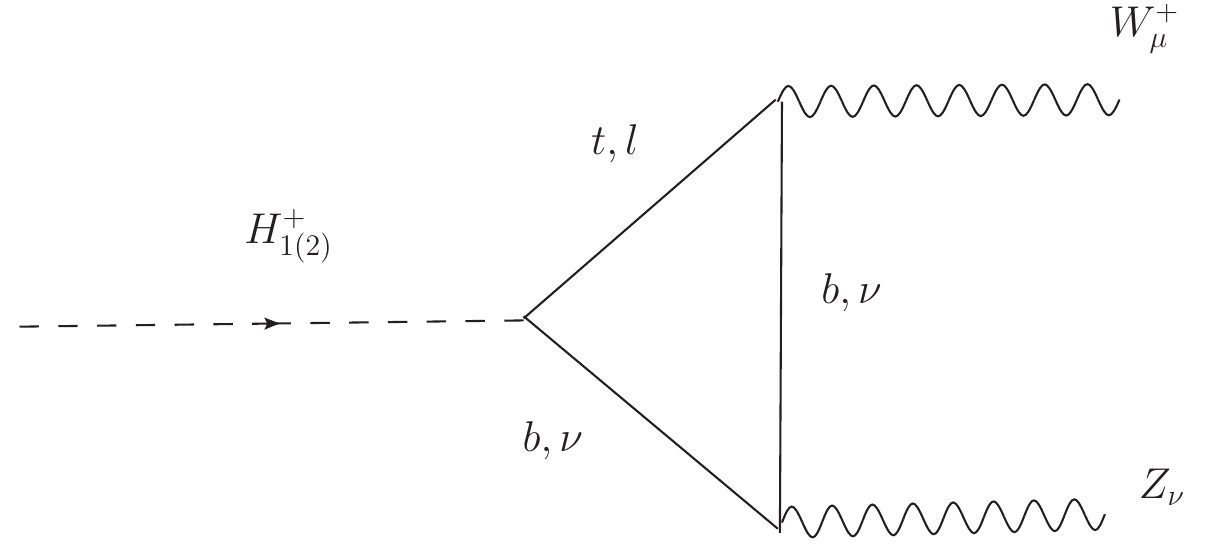}~~
\includegraphics[height = 3 cm, width = 5.5 cm]{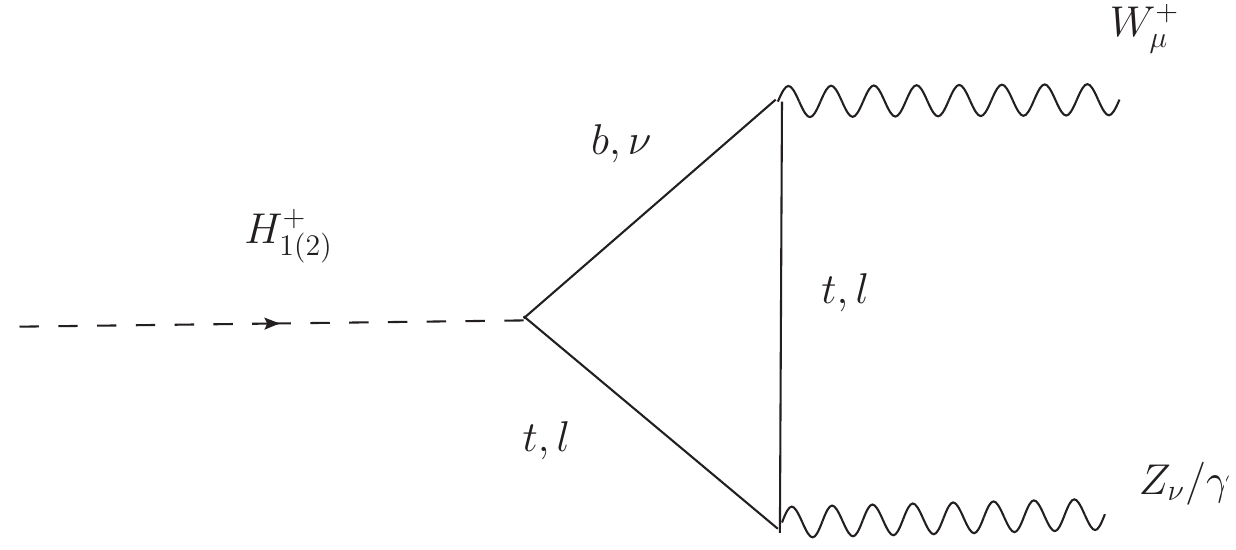}
} \\
\caption{}
\label{fig-ferm}
\end{figure}

\begin{center}
\begin{table}[htb!]
\begin{tabular}{|c|c|}\hline
~~~~Diagram number ~~~~~& ~~~~$ F_Z$~~~~ \\ \hline \hline
$E_1^{'}$ & ~~~~$\frac{1}{16 \pi^2 v}~\frac{s_W^2}{c_W}~ c_{\alpha_3 + \delta_2}~ \lambda_{H_1^+ H_1^- H_1}~[B_0(q^2,H_1^+, H_1)+2 B_1(q^2, H_1^+,H_1)]$~~~~ \\
$E_2^{'}$ & $\frac{1}{16 \pi^2 v}~\frac{s_W^2}{c_W}~ s_{\alpha_3 + \delta_2} ~ \lambda_{H_1^+ H_2^- H_1}~[B_0(q^2, H_2^+, H_1)+ 2 B_1(q^2, H_2^+, H_1)]$ \\  
$E_3^{'}$ & -$\frac{1}{16 \pi^2 v}~\frac{s_W^2}{c_W}~ s_{\alpha_3 + \delta_2} ~ \lambda_{H_1^+ H_1^- H_2}~[B_0(q^2,H_1^+, H_2)+2 B_1(q^2, H_1^+, H_2)]$ \\   
$E_4^{'}$ & $\frac{1}{16 \pi^2 v}~\frac{s_W^2}{c_W}~ c_{\alpha_3 + \delta_2} ~ \lambda_{H_1^+ H_2^- H_2}~[B_0(q^2,H_2^+, H_2) + 2 B_1(q^2,H_2^+, H_2)]$ \\ \hline \hline
$E_5^{'}$ & $\frac{1}{16 \pi^2 v}~\frac{s_W^2}{c_W}~ c_{\alpha_3 + \delta_2} ~ \lambda_{H_2^+ H_1^- H_1}~[B_0(q^2,H_1^+, H_1) + 2 B_1(q^2, H_1^+, H_1)] $ \\ 
$E_6^{'}$ & $\frac{1}{16 \pi^2 v}~\frac{s_W^2}{c_W}~ s_{\alpha_3 + \delta_2} ~ \lambda_{H_2^+ H_2^- H_1}~[B_0(q^2,H_2^+, H_1)+ 2 B_1(q^2,H_2^+, H_1)]$ \\ 
$E_7^{'}$ & -$\frac{1}{16 \pi^2 v}~\frac{s_W^2}{c_W}~ s_{\alpha_3 + \delta_2} ~ \lambda_{H_2^+ H_1^- H_2}~[B_0(q^2,H_1^+, H_2) + 2 B_1(q^2,H_1^+, H_2)$] \\
$E_8^{'}$ & $\frac{1}{16 \pi^2 v}~\frac{s_W^2}{c_W}~ c_{\alpha_3 + \delta_2} ~ \lambda_{H_2^+ H_2^- H_2}~[B_0(q^2, H_2^+, H_2)+2 B_1(q^2,  H_2^+, H_2)]$ \\ 
 \hline \hline
 \end{tabular}
\caption{}
\label{tab-4}
\end{table}
\end{center}

\subsubsection{$O$-type amplitudes}

\begin{figure}[htpb!]{\centering
\subfigure[$O_1$]{
\includegraphics[height = 3 cm, width = 6 cm]{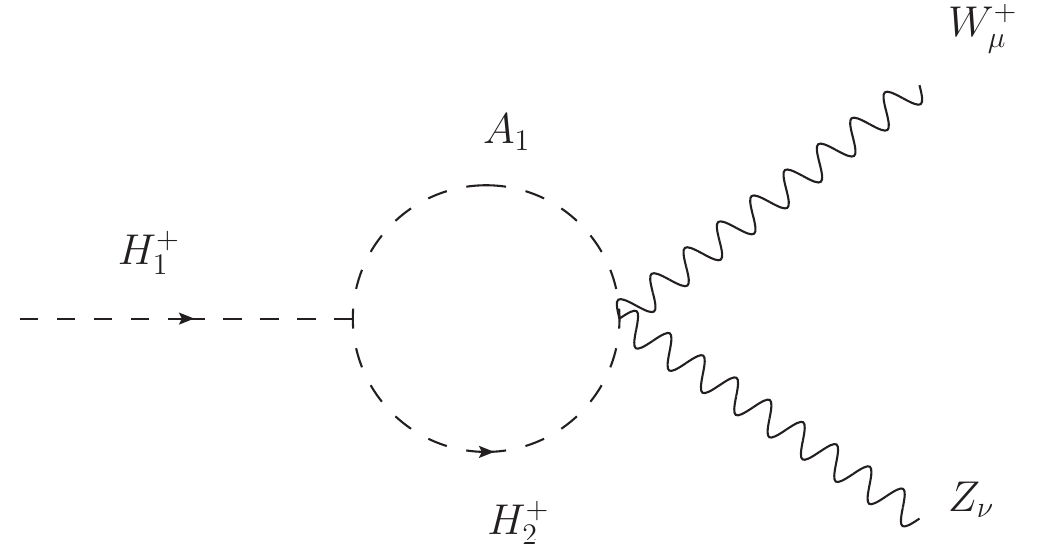}} 
\subfigure[$O_2$]{
\includegraphics[height = 3 cm, width = 6 cm]{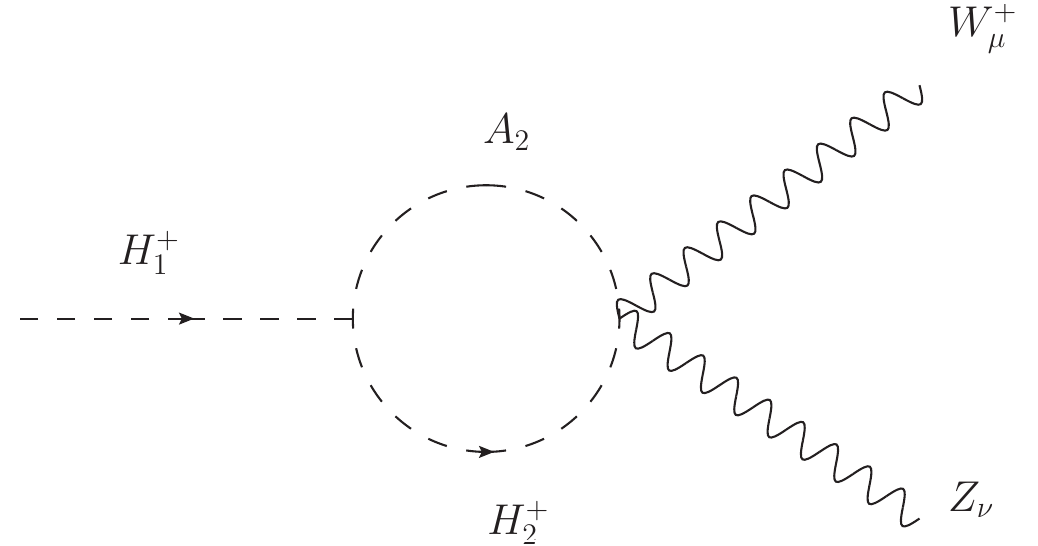}} \\
\subfigure[$O_3$]{
\includegraphics[height = 3 cm, width = 6 cm]{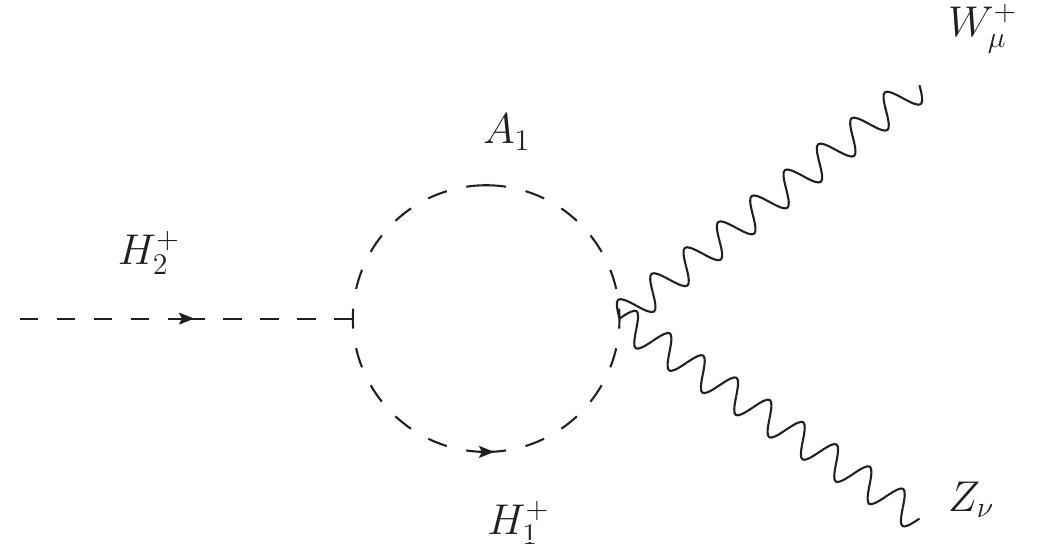}}
\subfigure[$O_4$]{
\includegraphics[height = 3 cm, width = 6 cm]{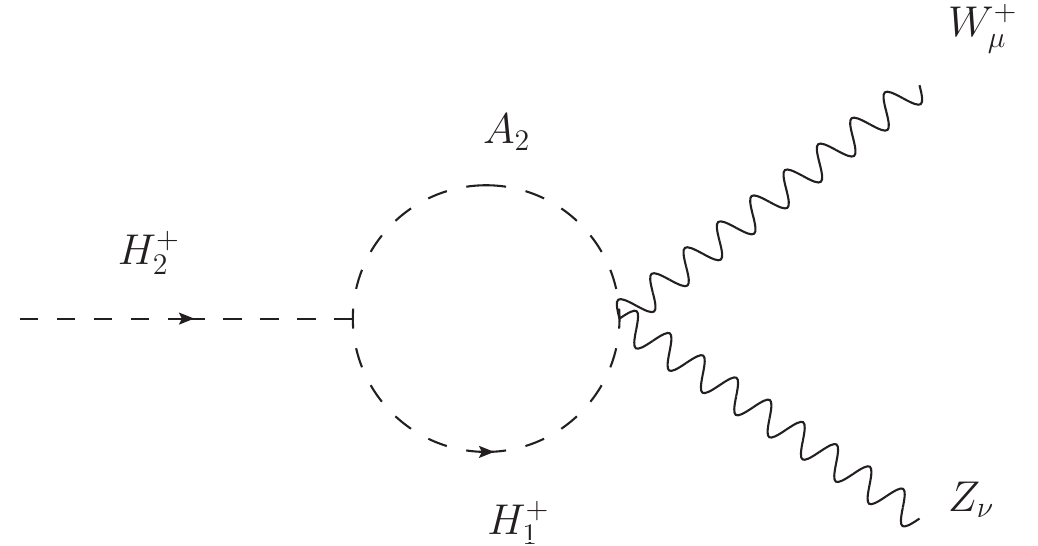}} 
} \\
\caption{}
\label{fig-2}
\end{figure}

\begin{center}
\begin{table}[htb!]
\begin{tabular}{|c|c|}\hline
~~~~Diagram number ~~~~~& ~~~~$ F_Z$~~~~ \\ \hline \hline  
$O_1$ & $-\frac{1}{16 \pi^2 v}~\frac{s_W^2}{c_W}~ s_{\delta_1 - \delta_2} ~ \lambda_{H_1^+ H_2^- A_1}~B_0(q^2,  H_2^+, A_1)$ \\  
$O_2$ & $\frac{1}{16 \pi^2 v}~\frac{s_W^2}{c_W}~ c_{\delta_1 - \delta_2} ~ \lambda_{H_1^+ H_2^- A_2}~B_0(q^2, H_2^+, A_2)$ \\ \hline \hline  
$O_3$ & $\frac{1}{16 \pi^2 v}~\frac{s_W^2}{c_W}~ c_{\delta_1 - \delta_2} ~ \lambda_{H_2^+ H_1^- A_1}~B_0(q^2, H_1^+, A_1)$ \\ 
$O_4$ & $\frac{1}{16 \pi^2 v}~\frac{s_W^2}{c_W}~ s_{\delta_1 - \delta_2} ~ \lambda_{H_2^+ H_1^- A_2}~B_0(q^2, H_1^+, A_2)$ \\ \hline \hline
 \end{tabular}
\caption{Corrected table}
\label{tab-2}
\end{table}
\end{center}

\subsubsection{$X$-,$Y$-,$Z$-type amplitudes}

\begin{figure}[htpb!]{\centering
\subfigure[$X_1$]{
\includegraphics[height = 2 cm, width = 5.2 cm]{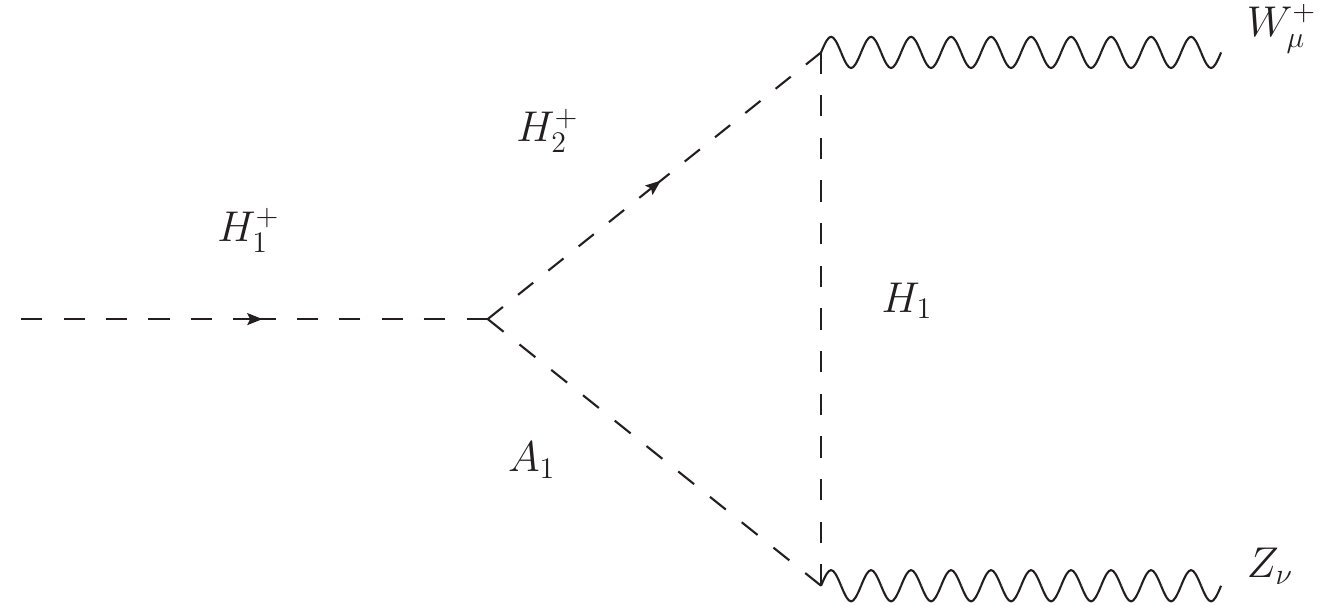}} 
\subfigure[$Y_1$]{
\includegraphics[height = 2 cm, width = 5.2 cm]{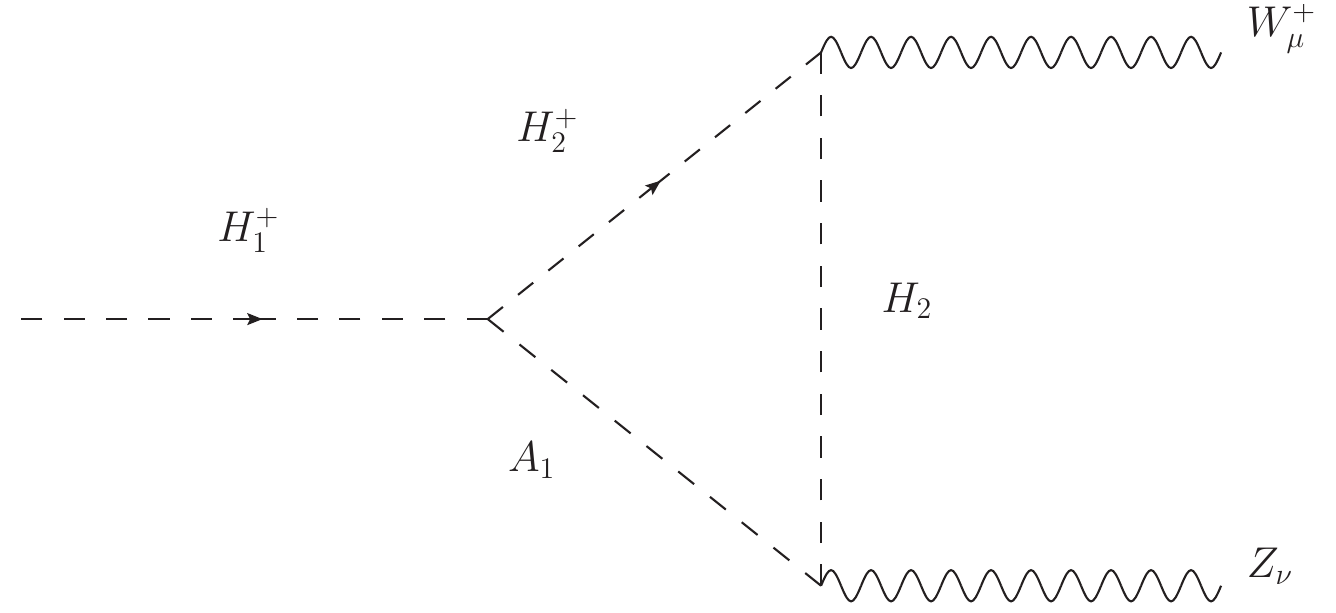}} 
\subfigure[$Z_1$]{
\includegraphics[height = 2 cm, width = 5.2 cm]{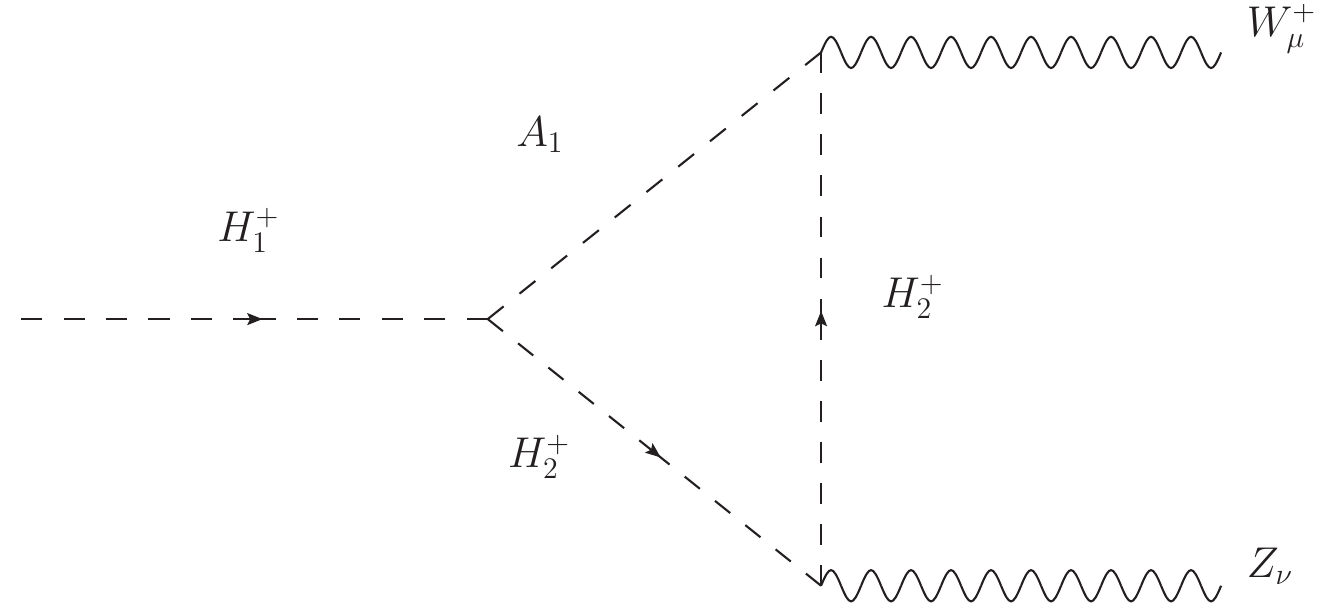}} \\
\subfigure[$X_2$]{
\includegraphics[height = 2 cm, width = 5.2 cm]{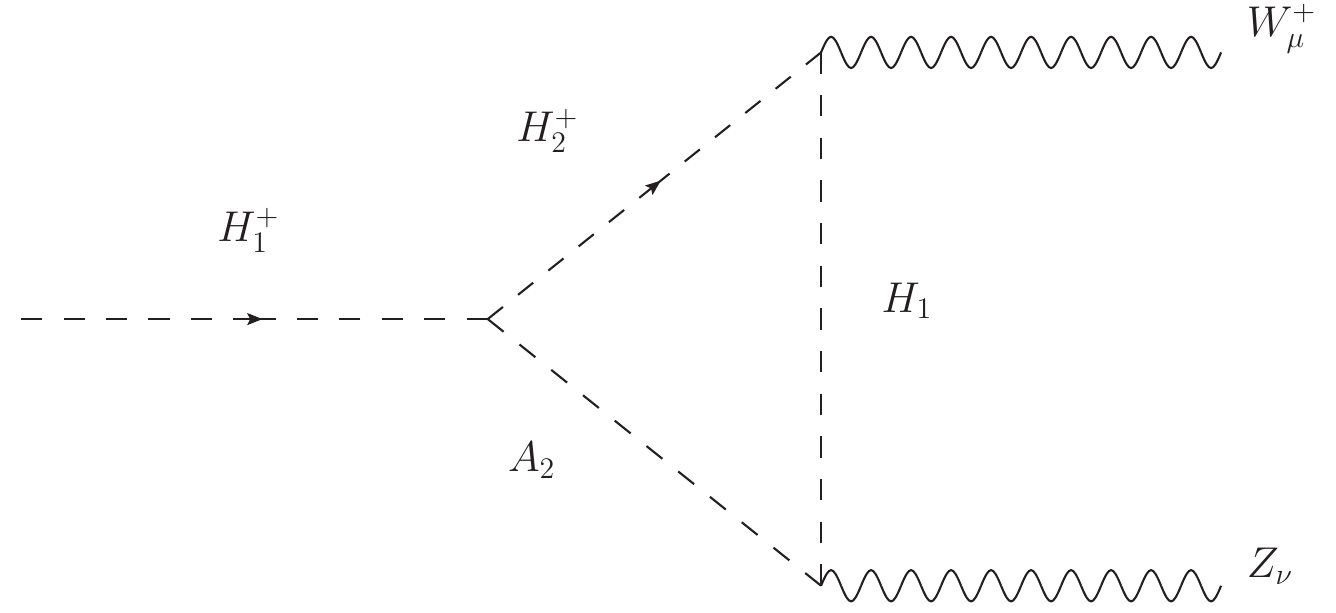}} 
\subfigure[$Y_2$]{
\includegraphics[height = 2 cm, width = 5.2 cm]{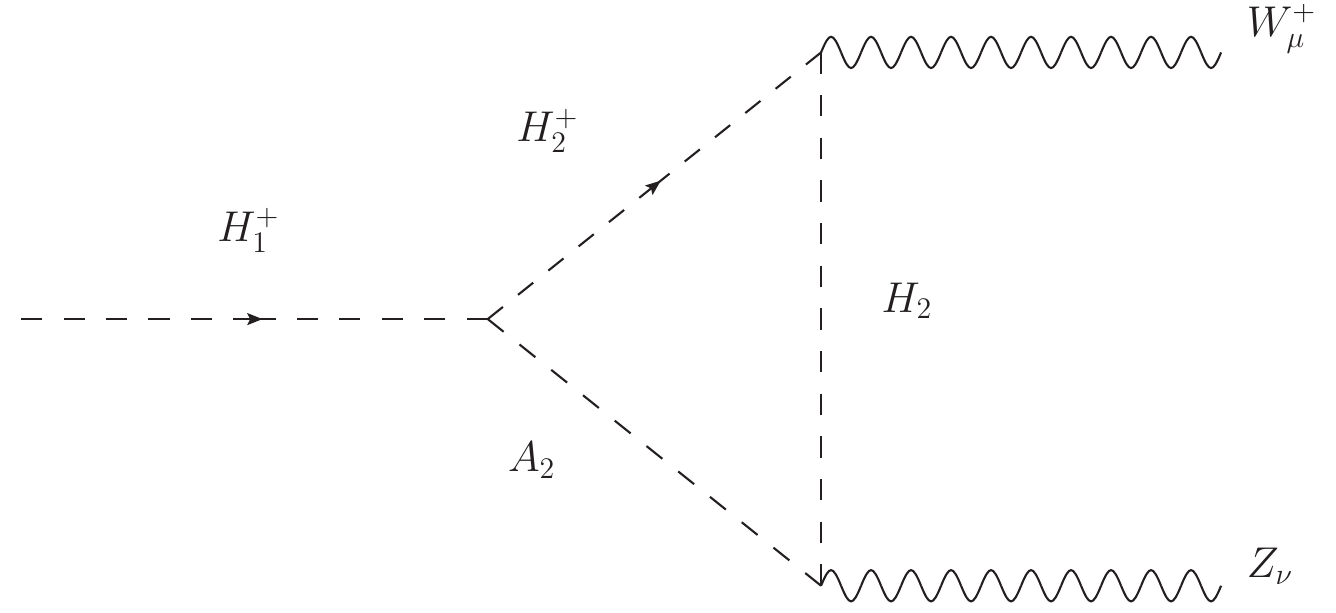}} 
\subfigure[$Z_2$]{
\includegraphics[height = 2 cm, width = 5.2 cm]{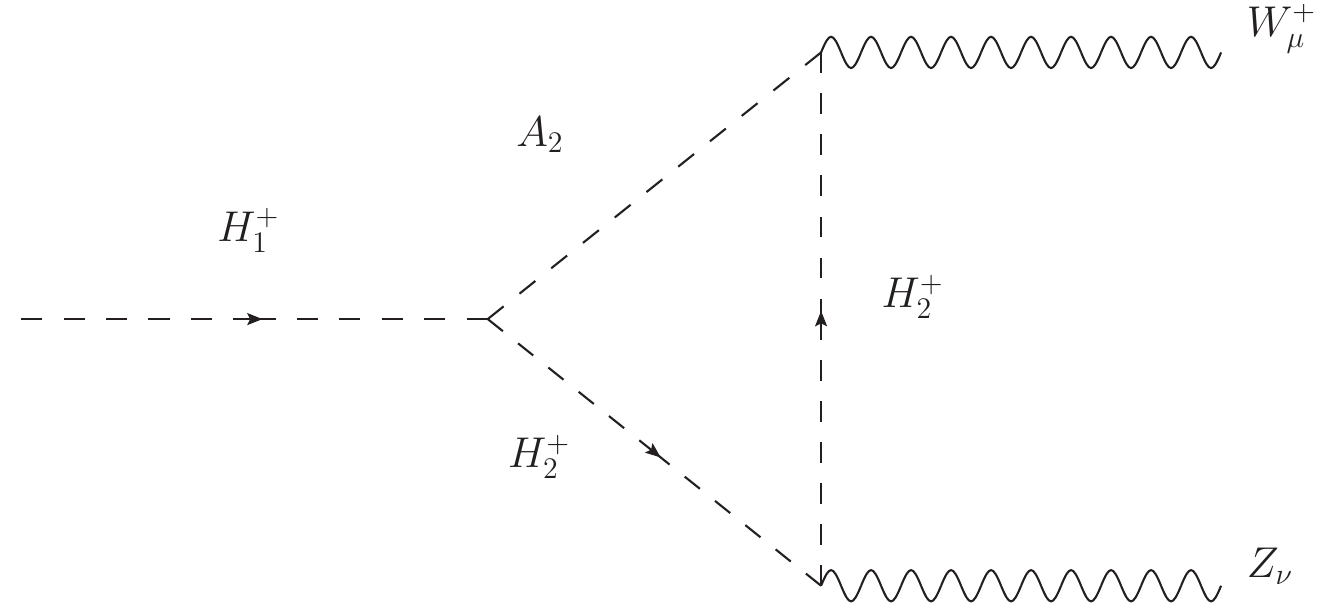}} \\
\subfigure[$X_3$]{
\includegraphics[height = 2 cm, width = 5.2 cm]{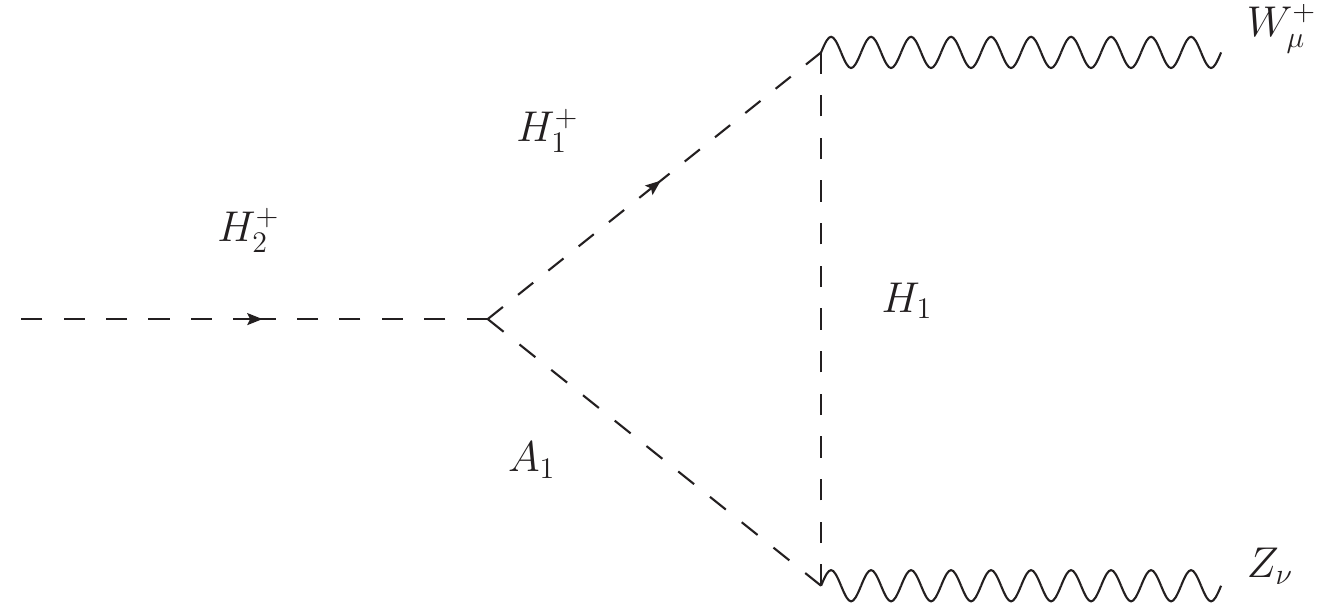}} 
\subfigure[$Y_3$]{
\includegraphics[height = 2 cm, width = 5.2 cm]{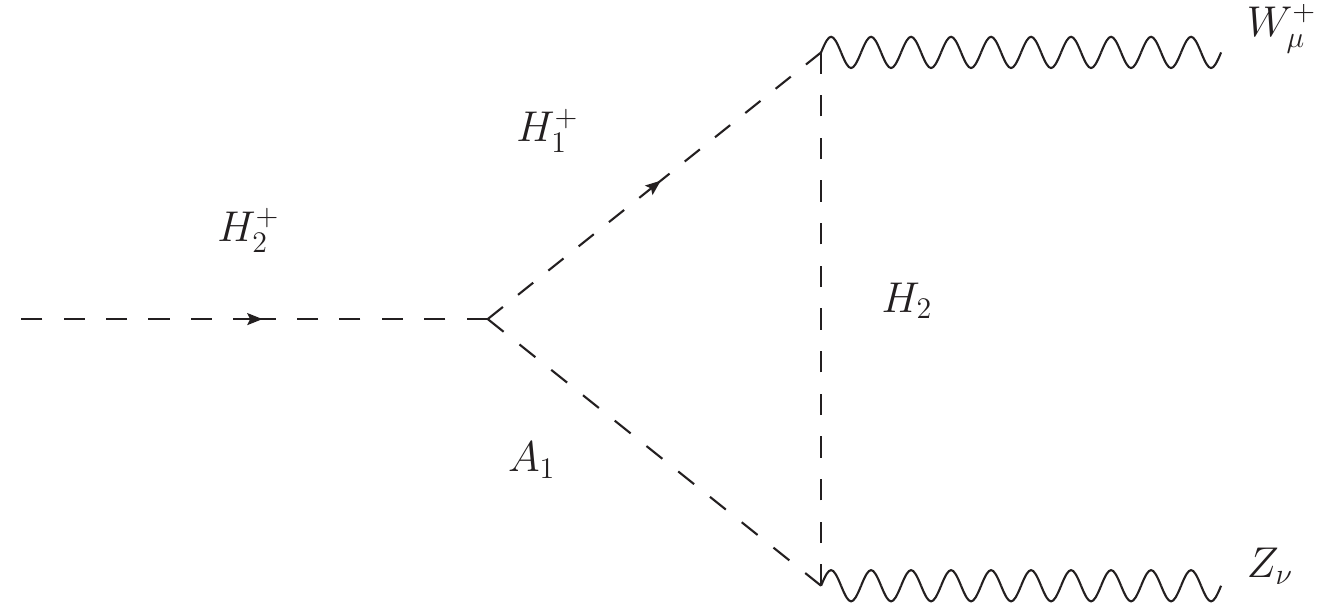}} 
\subfigure[$Z_3$]{
\includegraphics[height = 2 cm, width = 5.2 cm]{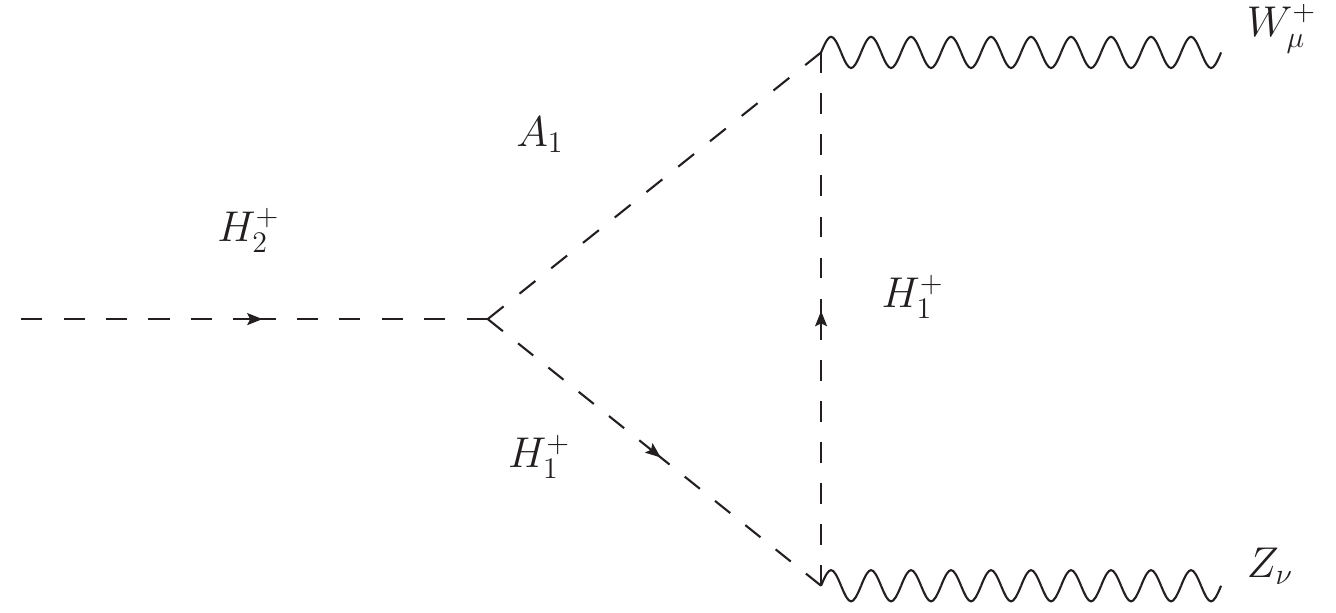}} \\
\subfigure[$X_4$]{
\includegraphics[height = 2 cm, width = 5.2 cm]{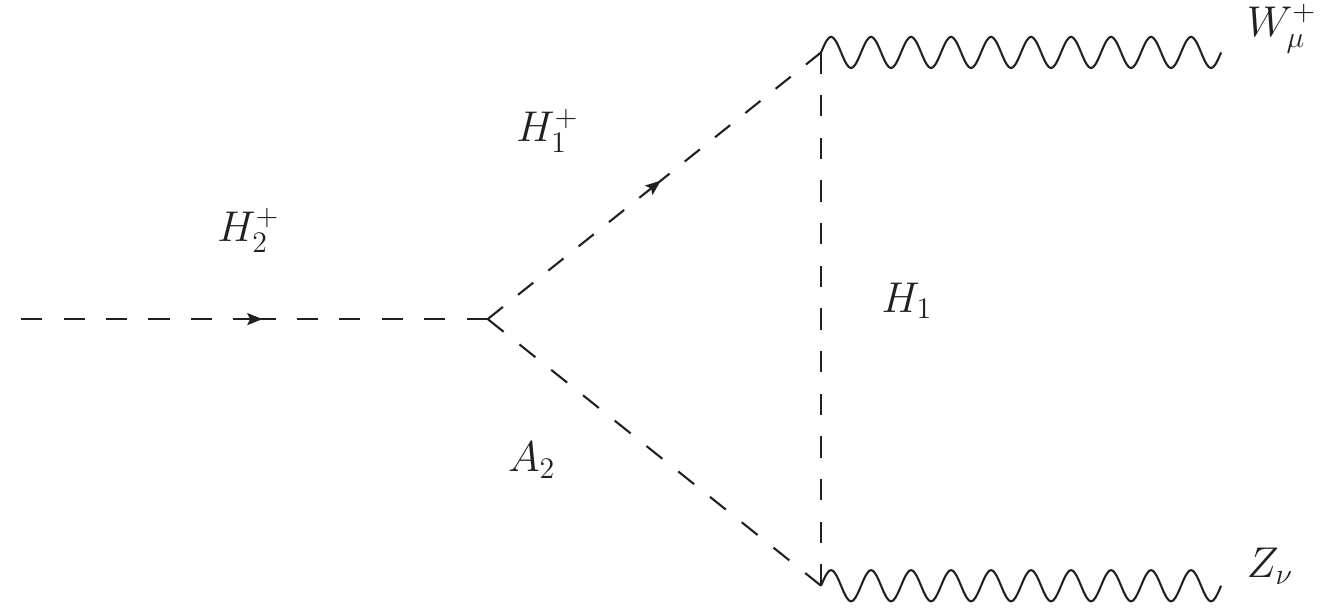}} 
\subfigure[$Y_4$]{
\includegraphics[height = 2 cm, width = 5.2 cm]{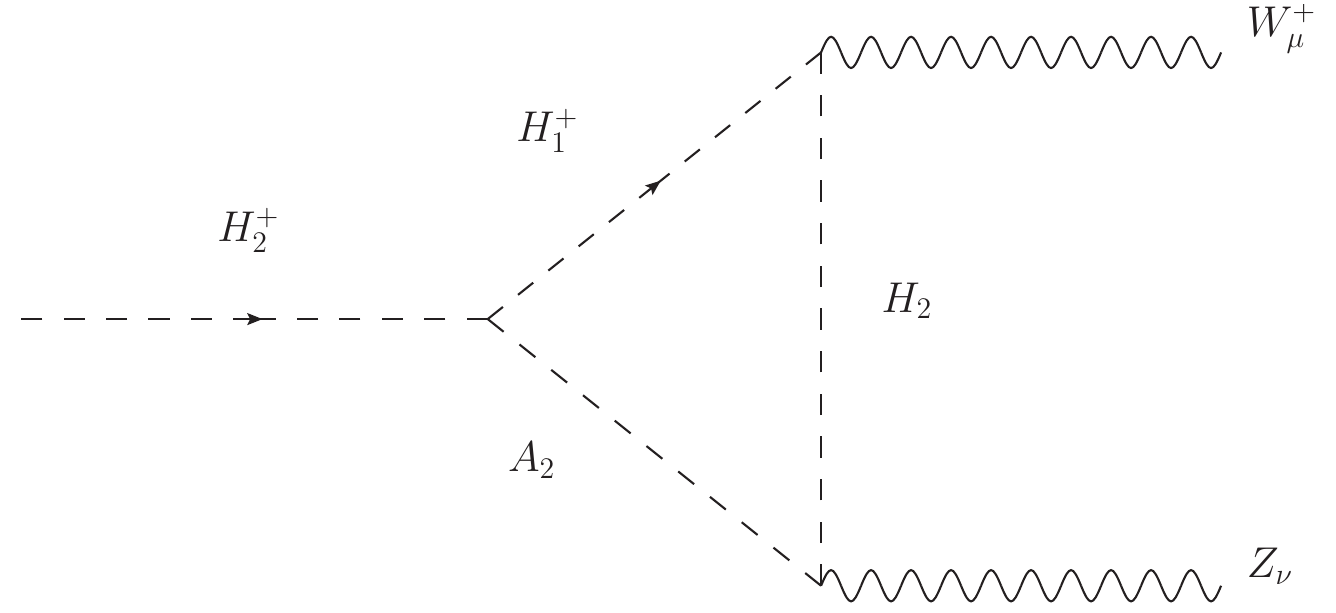}} 
\subfigure[$Z_4$]{
\includegraphics[height = 2 cm, width = 5.2 cm]{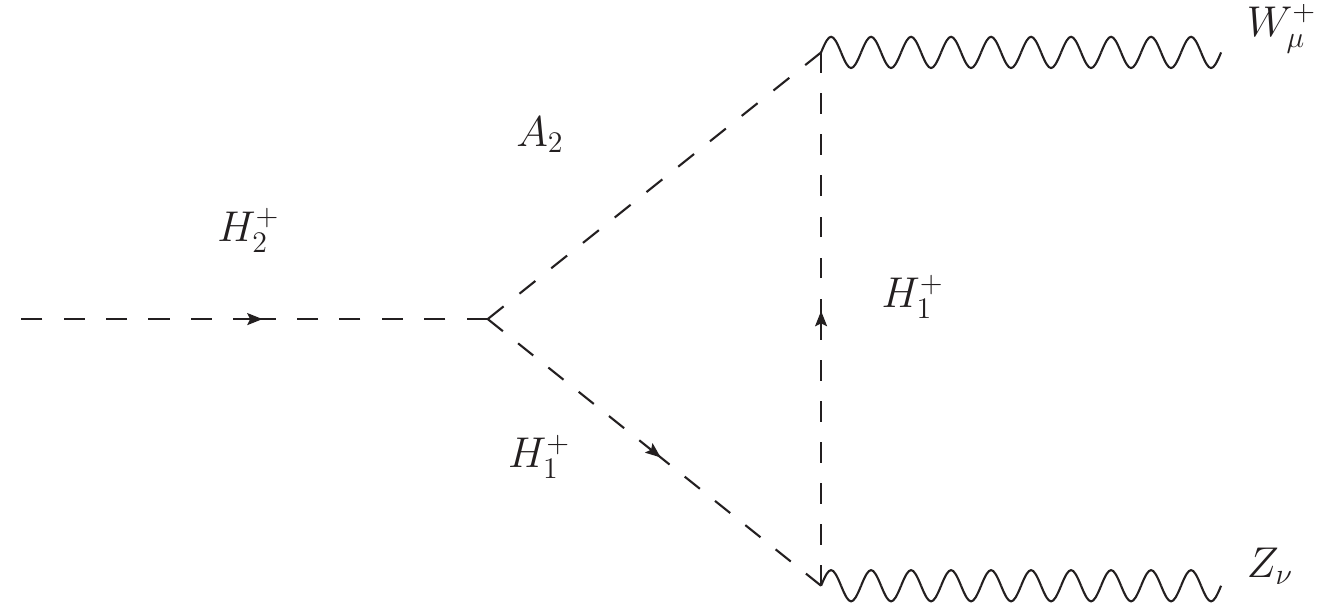}} 
} \\
\caption{}
\label{fig-6}
\end{figure}

\begin{center}
\begin{table}[htb!]
\resizebox{17 cm}{!}{
\begin{tabular}{|c|c|c|}\hline
Diagram no. & $ F_Z$ & $G_Z$ \\ \hline
$X_1$ & $ \left(-\frac{2 \lambda_{H_1^+ H_2^- A_1}}{16 \pi^2 v ~c_W}\right) (s_{\alpha_3+\delta_2}~ c_{\alpha_3+\delta_1}) C_{24}(H_2^+,H_1,A_1)$ & $ \left(-\frac{2 \lambda_{H_1^+ H_2^- A_1}}{16 \pi^2 v ~c_W}\right)~M_W^2 (s_{\alpha_3+\delta_2}~ c_{\alpha_3+\delta_1}) C_{1223}(H_2^+,H_1,A_1)$ \\ 
$Y_1$ & $ \left(-\frac{2 \lambda_{H_1^+ H_2^- A_1}}{16 \pi^2 v ~c_W}\right) (-c_{\alpha_3+\delta_2}~ s_{\alpha_3+\delta_1}) C_{24}(H_2^+,H_2,A_1)$ & $ \left(-\frac{2 \lambda_{H_1^+ H_2^- A_1}}{16 \pi^2 v ~c_W}\right)~M_W^2 (-c_{\alpha_3+\delta_2}~ s_{\alpha_3+\delta_1}) C_{1223}(H_2^+,H_2,A_1)$\\  
$Z_1$ & $(-\frac{2 c_{2W}}{c_W})~\frac{\lambda_{H_1^+ H_2^- A_1}}{16 \pi^2 v}~ s_{\delta_1 - \delta_2} ~ ~C_{24}(A_1,H_2^+, H_2^+)$ & $(-\frac{2 c_{2W}}{c_W})~\frac{\lambda_{H_1^+ H_2^- A_1}}{16 \pi^2 v}~ M_W^2~ s_{\delta_1 - \delta_2} ~ ~C_{1223}(A_1,H_2^+, H_2^+)$ \\  \hline \hline
$X_2$ & $ \left(-\frac{2 \lambda_{H_1^+ H_2^- A_2}}{16 \pi^2 v ~c_W}\right) (s_{\alpha_3+\delta_2}~ s_{\alpha_3+\delta_1}) C_{24}(H_2^+,H_1,A_2)$  & $ \left(-\frac{2 \lambda_{H_1^+ H_2^- A_2}}{16 \pi^2 v ~c_W}\right)M_W^2 (s_{\alpha_3+\delta_2}~ s_{\alpha_3+\delta_1}) C_{1223}(H_2^+,H_1,A_2)$\\ 
$Y_2$ & $ \left(-\frac{2 \lambda_{H_1^+ H_2^- A_2}}{16 \pi^2 v ~c_W}\right) (c_{\alpha_3+\delta_2}~ c_{\alpha_3+\delta_1}) C_{24}(H_2^+,H_2,A_2)$ & $ \left(-\frac{2 \lambda_{H_1^+ H_2^- A_2}}{16 \pi^2 v ~c_W}\right) M_W^2 (c_{\alpha_3+\delta_2}~ c_{\alpha_3+\delta_1}) C_{1223}(H_2^+,H_2,A_2)$\\  
$Z_2$ & $(\frac{2 c_{2W}}{c_W}) \frac{\lambda_{H_1^+ H_2^- A_2}}{16 \pi^2 v}~ c_{\delta_1 - \delta_2}~C_{24}(A_2,H_2^+, H_2^+)$  & $(\frac{2 c_{2W}}{c_W}) \frac{\lambda_{H_1^+ H_2^- A_2}}{16 \pi^2 v}~ M_W^2~ c_{\delta_1 - \delta_2}~C_{1223}(A_2,H_2^+, H_2^+)$ \\ \hline \hline
$X_3$ & $ \left(-\frac{2 \lambda_{H_2^+ H_1^- A_1}}{16 \pi^2 v ~c_W}\right) (c_{\alpha_3+\delta_2}~ c_{\alpha_3+\delta_1}) C_{24}(H_1^+,H_1,A_1)$ & $ \left(-\frac{2 \lambda_{H_2^+ H_1^- A_1}}{16 \pi^2 v ~c_W}\right) M_W^2 (c_{\alpha_3+\delta_2}~ c_{\alpha_3+\delta_1}) C_{1223}(H_1^+,H_1,A_1)$\\ 
$Y_3$ & $ \left(-\frac{2 \lambda_{H_2^+ H_1^- A_1}}{16 \pi^2 v ~c_W}\right) (s_{\alpha_3+\delta_2}~ s_{\alpha_3+\delta_1}) C_{24}(H_1^+,H_2,A_1)$ & $ \left(-\frac{2 \lambda_{H_2^+ H_1^- A_1}}{16 \pi^2 v ~c_W}\right) M_W^2 (s_{\alpha_3+\delta_2}~ s_{\alpha_3+\delta_1}) C_{1223}(H_1^+,H_2,A_1)$ \\  
$Z_3$ & $(\frac{2 c_{2W}}{c_W}) \frac{\lambda_{H_2^+ H_1^- A_1}}{16 \pi^2 v}~ c_{\delta_1 - \delta_2}~C_{24}(A_1,H_1^+, H_1^+)$  & $(\frac{2 c_{2W}}{c_W}) \frac{\lambda_{H_2^+ H_1^- A_1}}{16 \pi^2 v}~M_W^2~ c_{\delta_1 - \delta_2}~C_{1223}(A_1,H_1^+, H_1^+)$ \\  \hline  \hline
$X_4$ & $ \left(-\frac{2 \lambda_{H_2^+ H_1^- A_2}}{16 \pi^2 v ~c_W}\right) (c_{\alpha_3+\delta_2}~ s_{\alpha_3+\delta_1}) C_{24}(H_1^+,H_1,A_2)$ & $ \left(-\frac{2 \lambda_{H_2^+ H_1^- A_2}}{16 \pi^2 v ~c_W}\right) M_W^2 (c_{\alpha_3+\delta_2}~ s_{\alpha_3+\delta_1}) C_{1223}(H_1^+,H_1,A_2)$\\ 
$Y_4$ & $ \left(-\frac{2 \lambda_{H_2^+ H_1^- A_2}}{16 \pi^2 v ~c_W}\right) (-s_{\alpha_3+\delta_2}~ c_{\alpha_3+\delta_1}) C_{24}(H_1^+,H_2,A_2)$ & $ \left(-\frac{2 \lambda_{H_2^+ H_1^- A_2}}{16 \pi^2 v ~c_W}\right) M_W^2 (-s_{\alpha_3+\delta_2}~ c_{\alpha_3+\delta_1}) C_{1223}(H_1^+,H_2,A_2)$\\ 
$Z_4$ & $(\frac{2 c_{2W}}{c_W}) \frac{\lambda_{H_2^+ H_1^- A_2}}{16 \pi^2 v}~ s_{\delta_1 - \delta_2} ~C_{24}(A_2,H_1^+, H_1^+)$ & $(\frac{2 c_{2W}}{c_W}) \frac{\lambda_{H_2^+ H_1^- A_2}}{16 \pi^2 v}~M_W^2~ s_{\delta_1 - \delta_2} ~C_{1223}(A_2,H_1^+, H_1^+)$\\ 
\hline \hline
 \end{tabular}}
\caption{Corrected table}
\label{tab-6}
\end{table}
\end{center}

\subsubsection{$O^\prime$-type amplitudes}

\begin{figure}[htpb!]{\centering
\subfigure[$O_1^{'}$]{
\includegraphics[height = 3 cm, width = 6 cm]{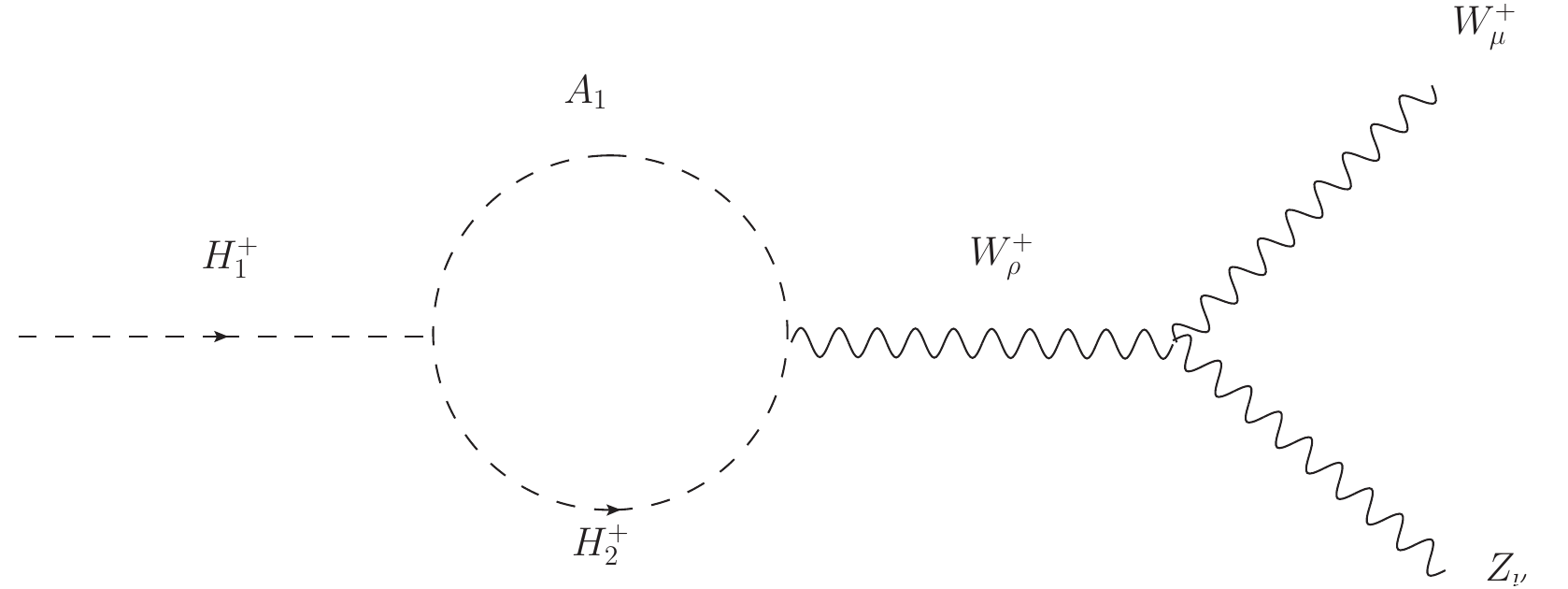}} 
\subfigure[$O_2^{'}$]{
\includegraphics[height = 3 cm, width = 6 cm]{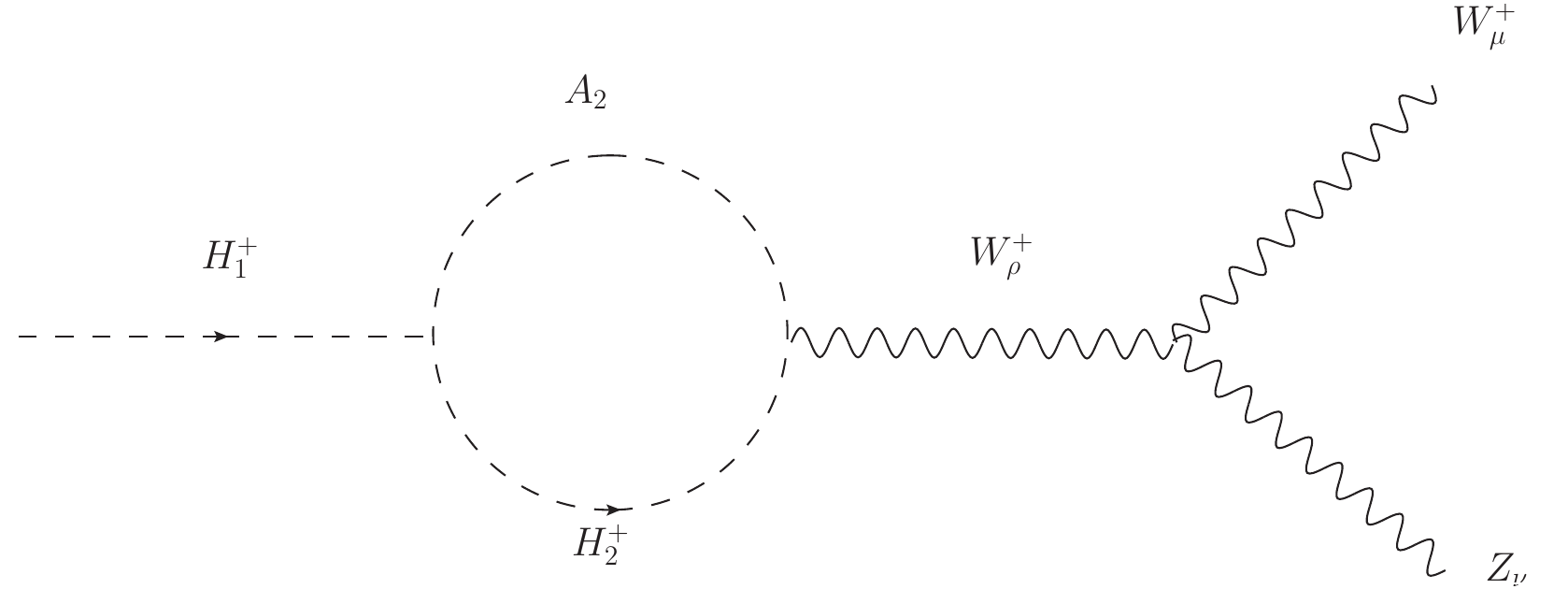}} \\
\subfigure[$O_3^{'}$]{
\includegraphics[height = 3 cm, width = 6 cm]{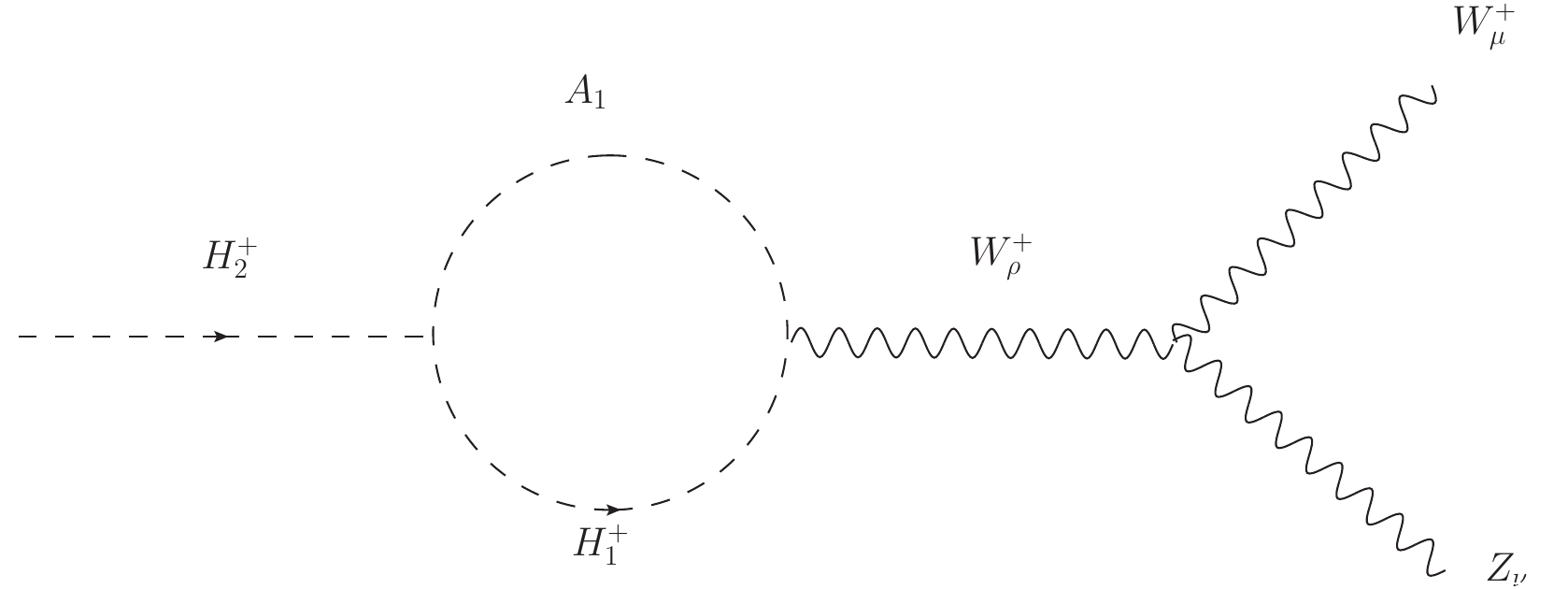}}
\subfigure[$O_4^{'}$]{
\includegraphics[height = 3 cm, width = 6 cm]{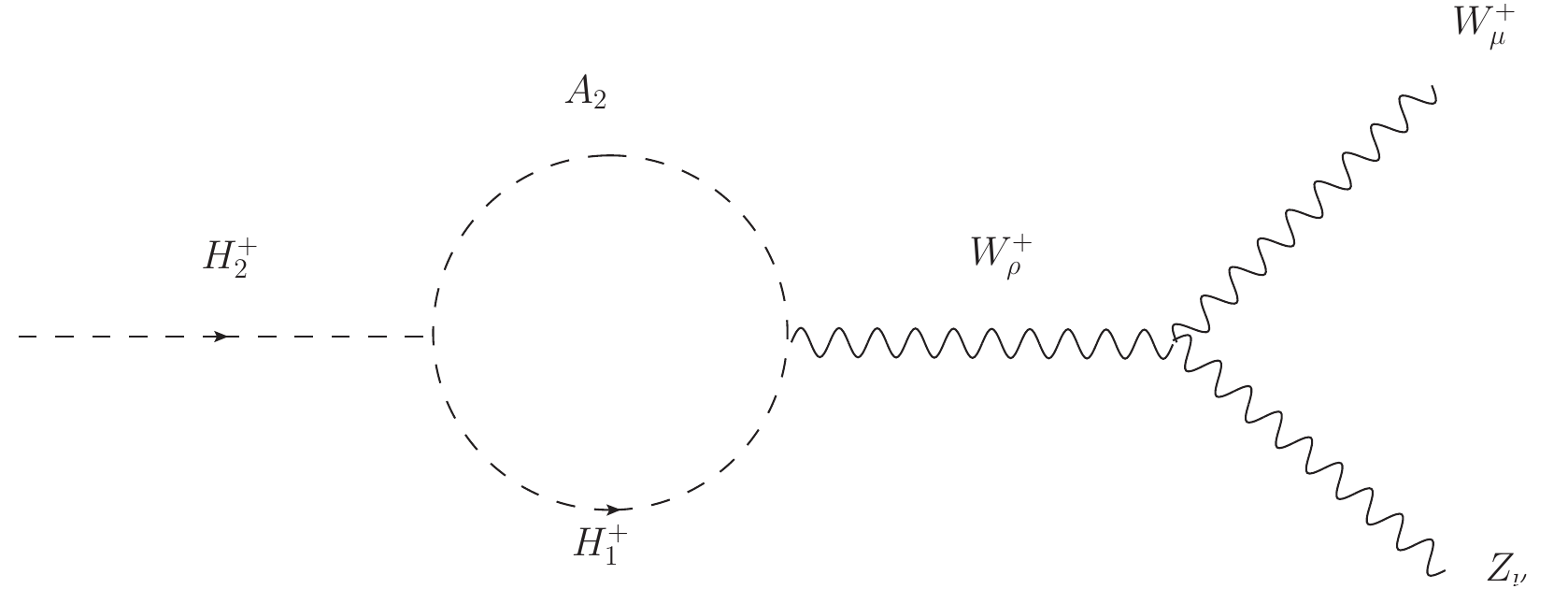}}
} \\
\caption{}
\label{fig-7}
\end{figure}

\begin{center}
\begin{table}[htb!]
\begin{tabular}{|c|c|}\hline
~~~~Diagram number ~~~~~& ~~~~$ F_Z$~~~~ \\ \hline \hline  
$O_1^{'}$ & $-\frac{1}{16 \pi^2 v}~\frac{s_W^2}{c_W}~ s_{\delta_1 - \delta_2} ~ \lambda_{H_1^+ H_2^- A_1}~[B_0(q^2,H_2^+, A_1)+2 B_1(q^2,H_2^+, A_1)]$ \\  
$O_2^{'}$ & $\frac{1}{16 \pi^2 v}~\frac{s_W^2}{c_W}~ c_{\delta_1 - \delta_2} ~ \lambda_{H_1^+ H_2^- A_2}~[B_0(q^2,H_2^+, A_2)+2 B_1(q^2, H_2^+, A_2)]$ \\ \hline \hline  
$O_3^{'}$ & $\frac{1}{16 \pi^2 v}~\frac{s_W^2}{c_W}~ c_{\delta_1 - \delta_2} ~ \lambda_{H_2^+ H_1^- A_1}~[B_0(q^2, H_1^+, A_1)+2 B_1(q^2,  H_1^+,A_1)]$ \\ 
$O_4^{'}$ & $\frac{1}{16 \pi^2 v}~\frac{s_W^2}{c_W}~ s_{\delta_1 - \delta_2} ~ \lambda_{H_2^+ H_1^- A_2}~[B_0(q^2, H_1^+, A_2)+2 B_1(q^2, H_1^+, A_2)]$ \\ \hline \hline
 \end{tabular}
\caption{}
\label{tab-5}
\end{table}
\end{center}

\subsection{Passarino-Veltman functions}
\besub
\bea
B_0(p^2;M_1^2, M_2^2) &=& \int \frac{d^d k}{i \pi^2} 
\frac{1}{(k^2 - M_1^2)((k + p)^2 - M_2^2)} \,, \\
p^\mu B_1(p^2;M_1^2, M_2^2) &=& \int \frac{d^d k}{i \pi^2} 
\frac{k^\mu}{(k^2 - M_1^2)((k + p)^2 - M_2^2)} .
\eea
\eesub
We also have
\besub
\bea
&& C_0(p_1^2,p_2^2,q^2;M_1^2,M_2^2,M_3^2) \nonumber \\
&& = \int \frac{d^d k}{i \pi^2} 
\frac{1}{(k^2 - M_1^2)((k + p_1)^2 - M_2^2)(k + q)^2 - M_3^2)}  \\
\nonumber \\
&& (p_1^\mu C_{11} + p_2^\mu C_{12})
 (p_1^2,p_2^2,q^2;M_1^2,M_2^2,M_3^2)  \nonumber \\
&& = \int \frac{d^d k}{i \pi^2} 
\frac{k^\mu}{(k^2 - M_1^2)((k + p_1)^2 - M_2^2)(k + q)^2 - M_3^2)}  \\
\nonumber \\
&& (p_1^\mu p_1^\nu C_{21} + p_2^\mu p_2^\nu C_{22} + p_1^\mu p_2^\nu C_{23}
 + g^{\mu\nu}C_{24}) 
(p_1^2,p_2^2,q^2;M_1^2,M_2^2,M_3^2)  \nonumber \\
&& = \int \frac{d^d k}{i \pi^2} 
\frac{k^\mu k^\nu}{(k^2 - M_1^2)((k + p_1)^2 - M_2^2)(k + q)^2 - M_3^2)}
\eea
\eesub
The Passarino-Veltman functions are expressed in terms of the Feynman parameter integrals as
\besub
\bea
B_0(p^2;M_1^2, M_2^2) &=& {\rm div} - \int_0^1 ~dx~ {\rm ln} \Delta_B \,, \\
B_1(p^2;M_1^2, M_2^2) &=& - \frac{{\rm div}}{2} + \int_0^1 ~ dx (1-x) {\rm ln} \Delta_B \,, \\
C_0(p_1^2,p_2^2,q^2;M_1^2,M_2^2,M_3^2) &=& - \int_0^1 ~ dx ~ \int_0^1 ~ dy \frac{y}{\Delta_C} \,, \\
C_{11}(p_1^2,p_2^2,q^2;M_1^2,M_2^2,M_3^2) &=& - \int_0^1 ~ dx ~ \int_0^1 ~ dy \frac{y(xy-1)}{\Delta_C} \,, \\
C_{12}(p_1^2,p_2^2,q^2;M_1^2,M_2^2,M_3^2) &=& - \int_0^1 ~ dx ~ \int_0^1 ~ dy \frac{y(y-1)}{\Delta_C} \,, \\
C_{21}(p_1^2,p_2^2,q^2;M_1^2,M_2^2,M_3^2) &=& - \int_0^1 ~ dx ~ \int_0^1 ~ dy \frac{y(1-xy)^2}{\Delta_C} \,, \\
C_{22}(p_1^2,p_2^2,q^2;M_1^2,M_2^2,M_3^2) &=& - \int_0^1 ~ dx ~ \int_0^1 ~ dy \frac{y(1-y)^2}{\Delta_C} \,, \\
C_{23}(p_1^2,p_2^2,q^2;M_1^2,M_2^2,M_3^2) &=& - \int_0^1  dx ~ \int_0^1  dy \frac{y(1-xy)(1-y)}{\Delta_C} \,, \\
C_{24}(p_1^2,p_2^2,q^2;M_1^2,M_2^2,M_3^2) &=& \frac{{\rm div}}{4} - \frac{1}{2}~ \int_0^1 ~ dx ~ \int_0^1 ~ dy ~y ~{\rm ln} \Delta_C \,.
\eea
\eesub
Where,
\besub
\bea
\Delta_B &=& -x(1-x)p^2 + x M_1^2 + (1-x) M_2^2 \,, \\
\Delta_C &=& y^2 (p_1 x + p_2)^2 + y [x (p_2^2 - q^2 + M_1^2 - M_2^2) + M_2^2 - M_3^2 - p_2^2] + M_3^2.
\eea
\eesub

\subsection{Perturbative unitarity}

This approach is based on the conservation of the total weak isospin $\sigma$ and total hypercharge Y in the high energy scalar-scalar scattering. In addition, the incoming and outgoing states should remain invariant under the discrete symmetry $\mathbb{Z}_2 \times \mathbb{Z}^{'}_2 \times \mathbb{Z}^{''}_2$ originally present in the scalar potential. The states acquire following different combinations which are either even or odd under the aforesaid symmetry : $(+ + +),(- - +), (+ - -)$ and $(- + -)$. We construct several scalar-scalar states out of the charged fields $\omega_i^\pm$ and neutral fields $n_i, n_i^*$ and tabulate them according to the weak isospin and its $z$-projection $\sigma_z$ , hypercharge and $\mathbb{Z}_2 \times \mathbb{Z}^{'}_2 \times \mathbb{Z}^{''}_2$ charge assignment. We have quoted the states with $Y=2$  and $Y=0$ in Table \ref{uni1} and Table \ref{uni2} respectively. From Table \ref{uni1}, it can be seen that the $(+ + +)$ scattering state with $Y = 2, \sigma = 0$ is absent owing to the Bose-Einstein symmetry of the identical particles.

\begin{table}[htpb!]
\centering
\begin{tabular}{ |c|c|c|c|c|c|c| } 
\hline
$Y$&$\sigma$ & $\sigma_z$ & (+ + +) & (- - + ) & (+ - -) & (- + -)\\ 
\hline \hline
& & 1 & $\begin{pmatrix}
 \omega_1^+ \omega_1^+ \\
 \omega_2^+ \omega_2^+ \\
 \omega_3^+ \omega_3^+ 
 \end{pmatrix}$ &  $\omega_1^+ \omega_2^+$ & $\omega_2^+ \omega_3^+$ & $\omega_1^+ \omega_3^+$\\  
2 &1 & 0 & $\begin{pmatrix}
 \omega_1^+ n_1 \\
 \omega_2^+ n_2 \\
 \omega_3^+ n_3 
 \end{pmatrix}$& $\frac{\omega_1^+ n_2 + \omega_2^+ n_1}{\sqrt{2}}$ &$\frac{\omega_2^+ n_3 + \omega_3^+ n_2}{\sqrt{2}}$ & $\frac{\omega_1^+ n_3 + \omega_3^+ n_1}{\sqrt{2}}$\\  
& & -1 & $\begin{pmatrix}
 n_1 n_1 \\
 n_2 n_2 \\
 n_3 n_3 
 \end{pmatrix}$ & $n_1 n_2$ & $n_2 n_3$ & $n_1 n_3$\\ 
&0 & 0 & absent & $\frac{(\omega_1^+ n_2 - \omega_2^+ n_1)}{\sqrt{2}}$ & $\begin{pmatrix}
 \frac{\omega_1^+ \omega_1^+ + n_1 n_1^* }{\sqrt{2}} \\
 \frac{\omega_2^+ \omega_2^+ + n_2 n_2^* }{\sqrt{2}}
\end{pmatrix}$ & $\begin{pmatrix}
 \frac{\omega_1^+ \omega_2^- + n_1 n_2^* }{\sqrt{2}} \\
 \frac{\omega_2^+ \omega_1^- + n_2 n_1^* }{\sqrt{2}}
\end{pmatrix}$\\  \hline
\end{tabular}
\caption{Scalar-scalar scattering states with $Y=2$.}
\label{uni1}
\end{table}

\begin{table}[htpb!]
\centering
\begin{tabular}{ |c|c|c|c|c|c|c| } 
\hline
$Y$&$\sigma$ & $\sigma_z$ & (+ + +) & (- - + ) & (+ - -) & (- + -)\\ 
\hline \hline
& & 1 & $\begin{pmatrix}
 \omega_1^+ n_1^* \\
 \omega_2^+ n_2^* \\
 \omega_3^+ n_3^* 
 \end{pmatrix}$ &  $\begin{pmatrix}
 \omega_1^+ n_2^* \\
 \omega_2^+ n_1^*  
 \end{pmatrix}$ & $\begin{pmatrix}
 \omega_2^+ n_3^* \\
 \omega_3^+ n_2^*  
 \end{pmatrix}$ & $\begin{pmatrix}
 \omega_1^+ n_3^* \\
 \omega_3^+ n_1^*  
 \end{pmatrix}$\\  
0 &1 & 0 & $\begin{pmatrix}
 \frac{\omega_1^+ \omega_1^+ - n_1 n_1^*}{\sqrt{2}} \\
  \frac{\omega_2^+ \omega_2^+ - n_2 n_2^*}{\sqrt{2}} \\
  \frac{\omega_3^+ \omega_3^+ - n_3 n_3^*}{\sqrt{2}}  
 \end{pmatrix}$& $\begin{pmatrix}
 \frac{\omega_1^+ \omega_2^- - n_1 n_2^*}{\sqrt{2}} \\
  \frac{\omega_2^+ \omega_1^- - n_2 n_1^*}{\sqrt{2}}  
 \end{pmatrix}$ &$\begin{pmatrix}
 \frac{\omega_2^+ \omega_3^- - n_2 n_3^*}{\sqrt{2}} \\
  \frac{\omega_3^+ \omega_2^- - n_3 n_2^*}{\sqrt{2}}  
 \end{pmatrix}$ & $\begin{pmatrix}
 \frac{\omega_3^+ \omega_1^- - n_3 n_1^*}{\sqrt{2}} \\
  \frac{\omega_1^+ \omega_3^- - n_1 n_3^*}{\sqrt{2}}  
 \end{pmatrix}$\\  
& & -1 & $\begin{pmatrix}
 \omega_1^- n_1 \\
 \omega_2^- n_2 \\
 \omega_3^- n_3 
 \end{pmatrix}$ &$\begin{pmatrix}
 \omega_1^- n_2 \\
 \omega_2^- n_1 \\ 
 \end{pmatrix}$  &$\begin{pmatrix}
 \omega_2^- n_3 \\
 \omega_3^- n_2 \\ 
 \end{pmatrix}$   & $\begin{pmatrix}
 \omega_3^- n_1 \\
 \omega_1^- n_3 \\ 
 \end{pmatrix}$ \\ 
&0 & 0 & $\begin{pmatrix}
 \frac{\omega_1^+ \omega_1^+ + n_1 n_1^*}{\sqrt{2}} \\
  \frac{\omega_2^+ \omega_2^+ + n_2 n_2^*}{\sqrt{2}} \\
  \frac{\omega_3^+ \omega_3^+ + n_3 n_3^*}{\sqrt{2}}  
 \end{pmatrix}$ & $\begin{pmatrix}
 \frac{\omega_1^+ \omega_2^- + n_1 n_2^*}{\sqrt{2}} \\
  \frac{\omega_2^+ \omega_1^- + n_2 n_1^*}{\sqrt{2}}  
 \end{pmatrix}$ & $\begin{pmatrix}
 \frac{\omega_2^+ \omega_3^- + n_2 n_3^*}{\sqrt{2}} \\
  \frac{\omega_3^+ \omega_2^- + n_3 n_2^*}{\sqrt{2}}  
 \end{pmatrix}$ & $\begin{pmatrix}
 \frac{\omega_1^+ \omega_3^- + n_1 n_3^*}{\sqrt{2}} \\
  \frac{\omega_3^+ \omega_1^- + n_3 n_1^*}{\sqrt{2}}  
 \end{pmatrix}$ \\ \hline \hline 
\end{tabular}
\caption{Scalar-scalar scattering states with $Y=0$.}
\label{uni2}
\end{table}

Corresponding scattering matrices for different two-scalar states with different hypercharges, total weak isospin and $\mathbb{Z}_2 \times \mathbb{Z}^{'}_2 \times \mathbb{Z}^{''}_2$ charges are given in Table \ref{uni-matrix}. These scattering matrices do not depend on $\sigma_z$. The elements of the scattering matrices are computed by extracting the coefficients of the scattering states out of the scalar potential and folding them with appropriate symmetry factors. Next we calculate the eigenvalues of the scattering matrices $e_i$'s and put the bound as : $|e_i| \geq 8 \pi$.

\begin{table}[htpb!]
\centering
	\resizebox{17cm}{!}{
\begin{tabular}{ |c|c|c|c|c|c| } 
\hline
$Y$&$\sigma$ & (+ + +) & (- - + ) & (+ - -) & (- + -)\\ 
\hline \hline 
$\pm$ 2 & 1 & $\begin{pmatrix}
\lambda_1 & \lambda_{12}^{''} & \lambda_{13}^{''}\\
\lambda_{12}^{''}& \lambda_2 & \lambda_{23}^{''}\\
\lambda_{13}^{''}& \lambda_{23}^{''} & \lambda_3
\end{pmatrix}$ &$(\lambda_{12} + \lambda_{12}^{''}) $ &$(\lambda_{23} + \lambda_{23}^{''}) $ & $(\lambda_{13} + \lambda_{13}^{''}) $\\ \hline
$\pm$ 2 & 0 & - & $(\lambda_{12} - \lambda_{12}^{''}) $ &$(\lambda_{23} - \lambda_{23}^{''}) $ & $(\lambda_{13} - \lambda_{13}^{''}) $  \\ \hline
0 & 1 & $\begin{pmatrix}
\lambda_1 & \lambda_{12}^{'} & \lambda_{13}^{'}\\
\lambda_{12}^{'}& \lambda_2 & \lambda_{23}^{'}\\
\lambda_{13}^{'}& \lambda_{23}^{'} & \lambda_3
\end{pmatrix}$&$\begin{pmatrix}
 \lambda_{12} & \lambda_{12}^{''}\\
\lambda_{12}^{''}& \lambda_{12}
\end{pmatrix} $ & $\begin{pmatrix}
 \lambda_{23} & \lambda_{23}^{''}\\
\lambda_{23}^{''}& \lambda_{23}  
\end{pmatrix}$& $\begin{pmatrix}
 \lambda_{13} & \lambda_{13}^{''}\\
\lambda_{13}^{''}& \lambda_{13} \end{pmatrix}$\\ \hline
0 & 0 &$\begin{pmatrix}
3 \lambda_1 & 2 \lambda_{12} + \lambda_{12}^{'} & 2 \lambda_{13} + \lambda_{13}^{'} \\
2 \lambda_{12} + \lambda_{12}^{'} & 3 \lambda_2 & 2 \lambda_{23} + \lambda_{23}^{'} \\
2 \lambda_{13} + \lambda_{13}^{'} & 2 \lambda_{23} + \lambda_{23}^{'} & 3 \lambda_3 
\end{pmatrix}$ &$\begin{pmatrix}
 \lambda_{12} + 2 \lambda_{12}^{'} & 3 \lambda_{12}^{'}\\
3 \lambda_{12}^{'}& \lambda_{12} + 2 \lambda_{12}^{'} \end{pmatrix}$ & $\begin{pmatrix}
 \lambda_{23} + 2 \lambda_{23}^{''} & 3 \lambda_{23}^{''}\\
3 \lambda_{23}^{''}& \lambda_{23} + 2 \lambda_{23}^{''} \end{pmatrix}$ &  $\begin{pmatrix}
 \lambda_{13} + 2 \lambda_{13}^{''} & 3 \lambda_{13}^{''}\\
3 \lambda_{13}^{''}& \lambda_{13} + 2 \lambda_{13}^{''} \end{pmatrix}$  \\ \hline
\end{tabular}}
\caption{Scattering matrices for two-scalar states.}
\label{uni-matrix}
\end{table}

\bibliographystyle{JHEP}
\bibliography{3HDMref} 

\end{document}